\documentclass{aa}  
\usepackage{linenoaa}

\usepackage{graphicx}
\usepackage[font=small, labelfont=bf, justification=justified, singlelinecheck=false]{caption}
\usepackage{subcaption}   
\usepackage{multirow} 
\usepackage{threeparttable}
\usepackage{booktabs}

\usepackage{txfonts}
\usepackage{array}
\usepackage{amsmath}
\usepackage{booktabs}
\usepackage{longtable} 
\usepackage{rotating}

\newcommand{\insertplot}[3]{%
    \begin{figure*}
        \centering
        \includegraphics[width=\textwidth]{#1}  
        \caption{#2}
        \label{#3}
    \end{figure*}
}
\newcommand{\obsA}{20}
\newcommand{\obsB}{47}

\usepackage[colorlinks=true, linkcolor=blue, citecolor=blue, filecolor=blue, urlcolor=blue]{hyperref}

\begin{document} 
   \title{The SOMA MM Survey. I.}

   \subtitle{An Astrochemical Census of Massive Protostars}

   \author{
        D. Gigli \inst{1,2}
        \and
        P. Gorai \inst{3,4}
        \and 
        C.Y. Law \inst{1}
        \and
        J. C. Tan \inst{2,5}
        \and
        M. Bonfand \inst{5}
        \and
        T. Rahman \inst{6}
        \and
        Y. Zhang \inst{7}
        \and
        K. Taniguchi \inst{8}
        \and
        R. Fedriani \inst{9}
        \and
        Z. Telkamp \inst{5}
        \and
        V. Rosero \inst{10}
        \and
        G. Cosentino \inst{11}
             }
   \institute{
    INAF, Osservatorio Astrofisico di Arcetri, Largo E. Fermi 5, I-50125 Firenze, Italy. \\  \email{diegogigli.dg@gmail.com}
    \and
    Department of Space, Earth \& Environment, Chalmers University of Technology, SE-412 96 Gothenburg, Sweden
    \and
    Rosseland Centre for Solar Physics, University of Oslo, PO Box 1029 Blindern, 0315 Oslo, Norway. \\
    \email{prasanta.astro@gmail.com}
    \and
    Institute of Theoretical Astrophysics, University of Oslo, PO Box 1029 Blindern, 0315 Oslo, Norway
    \and
    Department of Astronomy, University of Virginia, Charlottesville, VA 22904-4325, USA   
    \and
    Department of Physics, Kent State University, Kent, OH 44240, USA
    \and
    Department of Astronomy, Shanghai Jiao Tong University, 800 Dongchuan Road, Minhang, Shanghai 200240, People’s Republic of China
    \and
    National Astronomical Observatory of Japan, National Institutes of Natural Sciences, 2-21-1 Osawa, Mitaka, Tokyo 181-8588, Japan
    \and
    Instituto de Astrofísica de Andalucía, CSIC, Glorieta de la Astronomía s/n, E-18008 Granada, Spain
    \and
    Division of Physics, Mathematics, and Astronomy, California Institute of Technology, Pasadena, CA 91125, USA
    \and
    European Southern Observatory, Karl-Schwarzschild-Strasse 2, D-85748 Garching, Germany
   }
   
   \date{Received XXX; accepted XXX}

 
  \abstract
   {During the formation of massive stars, the dense gas experiences substantial chemical evolution, producing both simple and complex organic molecules (COMs), leading to sources known as "hot molecular cores". However, the extent to which chemical evolution depends on fundamental protostellar physical properties remains uncertain.
   }
   {Our goal is to investigate the chemical content of a large sample of well-studied massive protostars to study the potential correlation between chemical properties and protostellar physical properties.} 
   {We analyzed Atacama Compact Array (ACA) and Total Power (TP) 1.3~mm (Band 6) data toward 22 massive protostars selected from the SOMA Survey. The column densities, line widths, and excitation temperatures of the various molecular species were derived by modeling the detected emission lines under the assumption of local thermodynamic equilibrium (LTE) using \textsc{MADCUBA}. We compared the chemical properties with the protostellar properties derived from spectral energy distribution (SED) fitting. }
   {We present continuum images of our target sources, identifying 22 continuum sources. We extracted the spectra from each source and detected 35 molecular species, ranging from simple molecules (e.g., CO, SO, SiO) to complex organic molecules (COMs). Among the 22 sources, 7 exhibit a high degree of chemical complexity (i.e., $\gtrsim100$ transitions).
   Furthermore, we find variation of average excitation temperature ($T_\text{ex}$) across the sample, i.e., $T_\text{ex}>$100~K for 8 sources, $50\:{\rm K}<T_\text{ex}<100~$K for 4, and $T_\text{ex}<50$~K for the remainder.
   The measured parameters are compared with protostellar properties, such as the bolometric luminosity to envelope mass ratio ($L_{\rm bol}/M_{\rm env}$), an indicator of the evolutionary stage. We find a tentative correlation of line widths, $T_{\rm ex}$ and species column densities with $L_{\rm bol}/M_{\rm env}$.}
   {Our results confirm that sources with lower excitation temperatures ($T_{\rm ex} < 50$~K) trace lukewarm gas with relatively low chemical complexity. 
   The most chemically rich sources are associated with the highest excitation temperatures ($T_{\rm ex} > 100$~K), indicating the presence of typical hot cores where thermal desorption is efficient, resulting in line-rich spectra. Tentative trends are observed between the line widths, excitation temperatures, and column densities of various species with the luminosity-to-mass ratio. These data provide important constraints for chemodynamical models of massive protostellar cores.
}


   \maketitle

   \nolinenumbers
    %
\section{Introduction}\label{sect:1}

Massive stars ($m_\star\geq8 M_\odot$) play a crucial role in the evolution of the Universe, contributing to the enrichment of heavy elements, injection of energy into the interstellar medium (ISM), and regulation of star formation rates.
Despite their importance, the formation process of massive stars remains elusive, with several competing theories \citep[see, e.g.,][]{Tan2014}, including core accretion \citep{mckee2003formation, yorke2004theory} and competitive accretion \citep{bonnell2001competitive, grudic2022dynamics}. 
These different models predict distinct initial conditions and evolutionary pathways, which may then also influence the chemistry of the surrounding gas and dust.
In addition to their physical complexity, high-mass star-forming regions (HMSFRs) exhibit some of the most chemically rich environments observed in the ISM \citep{herbst2009complex, jorgensen2020astrochemistry}. 
To date, about 330 molecules have been detected in ISM or in circumstellar shells \citep[\url{https://cdms.astro.uni-koeln.de/classic/molecules;}][]{mcguire20222021}, many of which have been observed toward HMSFRs.
        
It is possible to outline evolutionary stages based on observational properties for HMSFRs. The birthplaces of stars are giant molecular clouds (GMCs; $M\gtrsim10^4\:M_\odot$). These cold and dense agglomerations of gas and dust fragment into smaller and denser clumps, which collapse under the influence of their gravity \citep[e.g.,][]{williams1999structure}. Infrared dark clouds (IRDCs) represent an early evolutionary phase of these clumps \citep[e.g.,][]{rathborne2006infrared}. These clouds are composed of quiescent cold and dense gas ($T\sim10-25~$K; $n_\text{H}\geq10^4~$cm$^{-3}$), often shaped in filamentary structures of sizes of $1-10$~pc. They are visible at (sub-)mm wavelengths in continuum and various line emissions \citep[e.g., N$_2$H$^+$, HNC, HCN, HCO$^+$, NH$_3$;][]{pillai2006ammonia,vasyunina2011chemistry,2022A&A...662A..39E} but show no (or weak) emission in the infrared. The ones lacking emission at 24 and 70~$\mu$m are colder and denser, and are believed to trace the earliest stages of high-mass star formation \citep{tan2013dynamics, sanhueza2013distinct, sanhueza2019alma}. IRDCs host (also) high-mass starless cores ($M\sim10-100~$M$_\odot$), with sizes of $\sim0.1\:$pc and peak densities $\geq10^6~$cm$^{-3}$ \citep[e.g.,][]{tan2013dynamics,2018ApJ...867...94K}. These starless cores undergo gravitational collapse and become high-mass protostellar objects \citep[HMPOs;][]{sridharan2002high, williams2004circumstellar, beuther2010high, duarte2013co}, characterized by high accretion rates ($\dot{M}\geq10^{-4}\:M_\odot\:$yr$^{-1}$) and a central protostar that heats the gas to higher temperatures ($\sim20-100$~K) \citep{gerner2014chemical}. It is important to note that not all cores within IRDCs will evolve into high-mass stars - most will form low- or intermediate-mass stars \citep[e.g.,][]{2018ApJ...862..105L}.

        
Massive protostellar cores exhibit strong emission in molecular lines and continuum at (sub-)mm and IR wavelengths,
in particular appearing as hot molecular cores \citep[HMCs;][]{cesaroni1997disk}, i.e., when the envelope of the protostar is heated to $T\geq100~$K. 
The HMCs are characterized by high densities ($n_\text{H}\gtrsim10^6~$cm$^{-3}$) over small scales ($<0.1~$pc) \citep{vanDishoeck1998,kurtz200hot} and high sensitivity studies have shown that HMCs are associated with weak radio continuum emission, with several tracing ionized jets \citep{rosero2019weak}. 
Due to the higher temperatures, the molecules formed in the ice layers of the dust grain surface desorb in the gas phase, revealing a plethora of molecular species, such as oxygen-bearing molecules (e.g., formaldehyde, H$_2$CO), nitrogen-bearing molecules (e.g., HC$_3$N; HNCO), sulfur-bearing molecules (e.g., SO; SO$_2$), or silicon-bearing molecules (e.g., SiO). 
Among these species, complex organic molecules (COMs; carbon-based species with $\geq6~$atoms) have also been identified \citep[e.g., methanol, CH$_3$OH; acetaldehyde, CH$_3$CHO; methyl formate, CH$_3$OCHO; ethanol, C$_2$H$_5$OH; dimethyl ether, CH$_3$OCH$_3$; acetone, CH$_3$COCH$_3$; vinyl cyanide, C$_2$H$_3$CN; ethyl cyanide, C$_2$H$_5$CN; formamide, NH$_2$CHO, methyl isocyanate, CH$_3$NCO;][]{belloche2009increased,rivilla2017formation,belloche2017rotational,belloche2019re,pagani2019complexity,colzi2021guapos,Gorai2024}. 
In the late stages of massive star formation, the strong UV protostellar radiation can dissociate and ionize the envelope gas, forming hyper/ultra-compact H\textsc{II} regions \citep[HCH\textsc{II}/UCH\textsc{II}, $R<0.01~$pc/$R<0.1~$pc;][]{van2004hot, cesaroni2010structure, kuiper2018first}. These ionized regions expand with time and can be studied at cm wavelengths via their free-free emission.

Recent observations with high-sensitivity facilities, such as ALMA, have further refined our understanding of molecular emission in HMSFRs \citep{csengeri2018search,liu2020atoms,mininni2020guapos}. 
Interpreting large spectral datasets remains challenging due to the wide variety of molecular species and the diversity of physical conditions. Expanding molecular surveys and studying the chemical inventories of HMSFRs provide valuable insight into their chemical and physical conditions, which is essential for understanding their formation and evolution.
        
To characterize and study the astrochemical inventories in massive star formation and to test different formation theories, we have initiated various line-surveys of a sample of massive protostars selected from the SOFIA Massive (SOMA) Star Formation survey (PI: J. C. Tan). The SOMA survey has studied over 50 high- and intermediate-mass star-forming regions across various evolutionary stages, masses, and environments with the SOFIA-FORCAST instrument, covering wavelengths ranging from about 7 to 40 $\mu$m. Protostellar properties have been investigated in several SOMA papers focused on SOFIA observations and spectral energy distribution (SED) analysis \citep{DeBuizer2017,Liu2019,Liu2020,Fedriani2023,Telkamp2025}.
For the line survey observations, a variety of different interferometric arrays (ALMA, SMA, VLA, and NOEMA) and single-dish telescopes (e.g., Yebes-40m, IRAM-30m, APEX) have been utilized. Results from the Yebes-40m observations have been presented by \citet{taniguchi2024sofia} and \citet{gorai2025sofia}. 
        
Here, as part of these line survey observations, we study the astrochemical inventories of 20 SOMA target regions using ALMA Atacama Compact Array (ACA) and Total Power (TP) observations at 1.3~mm (Band 6). We compare observational results with SED-derived protostellar properties to explore the correlation between chemical complexity and physical characteristics. The paper is organized as follows. We describe the observations in Sect.~\ref{sect:2} and the analysis methods in Sect.~\ref{sect:3}. We present and discuss the results in Sect.~\ref{sect:4}. Finally, we provide our conclusions in Sect.~\ref{sect:5}.
\section{Observations and data reduction}\label{sect:2}

The ALMA data used in this study are part of the Cycle 9 project (2021.2.00177.S, PI: Y. Zhang) with 7M (ACA) and TP observations.  We summarize the general information, including the angular resolution, spectral bandwidth, and spectral resolution, in Table \ref{table:info_obs}. Overall, \obsB\,SOMA regions have TP observations. Of these, \obsA\,also have ACA observations, and these regions are the focus of this paper. All observations have passed the QA2 quality check. 
    
        The ACA and TP array observations were reduced and imaged by the CASA-PIPELINE \citep[ver.6.5.4, ][]{mcmullin2007casa}. 
        We combine the ACA and TP observations with the FEATHER function using the default parameters, resulting in products (hereafter ACA+TP) that will be used for both line identifications and spectral analysis for this work.
    
        \begin{table}[h!]
            \centering
            \caption{General information of the interferometric ACA and single-dish TP observations.}
            \label{table:info_obs}
            \begin{tabular}{cccc}
                \hline\hline
                \noalign{\smallskip}
                Config. & Angular & MRS\tablefootmark{b} & Spectral \\
                 & resolution\tablefootmark{a} & &  resolution \\
                & ($''$) & ($''$) & (km~s$^{-1}$; kHz)\\
                \noalign{\smallskip}
                \hline
                \noalign{\smallskip}
                ACA & $\sim 6$ & $\sim32$ & $\sim 0.3$; 244\\
                TP & $\sim 30$ & $\sim412$ & $\sim 0.3$; 244\\
                ACA+TP & $\sim 6$ & $\sim32$ & $\sim 0.3$; 244\\
                \noalign{\smallskip}
                \hline
            \end{tabular}
            \tablefoot{
                \tablefoottext{a}{The synthesized beam of each observation is presented in Table~\ref{tab:observations}}
                \tablefoottext{b}{Maximum recoverable scale.}
            }
        \end{table}
        \begin{table*}
            \centering
            \caption{List of the 22 continuum sources detected in the 20 regions observed, with coordinates, heliocentric distances, local standard of rest (LSR) velocities, bolometric luminosities, envelope and protostellar masses.}
            \label{tab:sources}
            \begin{tabular}{lcccccccc}
                \hline
                \hline
                \noalign{\smallskip}
                Source\tablefootmark{a} & R.A. (ICRS) \tablefootmark{b} & Dec. (ICRS) \tablefootmark{b} & $d$\tablefootmark{c} & $v_\text{LSR}$\tablefootmark{d} & $L_\text{bol}$\tablefootmark{c} & $M_\text{env}$\tablefootmark{c} & $m_\star$\tablefootmark{c} & $N_{\text{H}_2}$\tablefootmark{e}\\
                \noalign{\smallskip}
                 & (hh:mm:ss) & ($\circ$:$'$:$''$) & (kpc) & (km/s) & (L$_\odot$) & (M$_\odot$) & (M$_\odot$) & (cm$^{-2}$)\\
                \noalign{\smallskip}
                \hline
                \noalign{\smallskip}
                AFGL 5180 a & 06:08:53.38 & 21:38:11.80 & 1.8 & 10.0 & ... & ... & ... & $2.1-6.8\times10^{22}$\\
                AFGL 5180 b & 06:08:53.38 & 21:38:30.50 & 1.8 & 10.0 & ... & ... & ... & $1.3-4.3\times10^{22}$\\
                G010.62-00.38 & 18:10:28.69 & -19:55:49.58 & 4.9 & -4.0 & $4.9\substack{+3.6\\ -2.1}\times 10^5$ & $291\substack{+96\\ -72}$ & $41\substack{+26\\ -16}$ & $1.7-5.6\times10^{23}$\\
                G011.94-00.62 & 18:14:01.10 & -18:53:24.39 & 3.8 & 40.0 & $2.4\substack{+2.2\\ -1.1}\times 10^5$ & $232\substack{+205\\ -109}$ & $27\substack{+16\\ -10}$ & $3.0-9.8\times10^{22}$\\
                G012.81-00.20 & 18:14:13.78 & -17:55:44.01 & 2.4 & 30.5 & $7.6\substack{+6.5\\ -3.5}\times 10^5$ & $210\substack{+90\\ -63}$ & $62\substack{+33\\ -22}$ & $1.7-5.4\times10^{23}$\\
                G045.47+00.05 & 19:14:25.67 & 11:09:24.90 & 8.4 & 61.5 & $3.5 \substack{+2.4 \\ - 1.4} \times 10^5$ & $207\substack{+58 \\ -45}$ & $29\substack{+12 \\ -9}$ & $2.7-5.4\times10^{22}$\\
                G049.27-00.34 & 19:23:06.72 & 14:20:11.81 & 5.5 & 67.8 & $1.0\substack{+0.5 \\ -0.3} \times 10^5$ & $244\substack{+221 \\ -116}$ & $25\substack{+5 \\ -4}$ & $1.1-3.6\times10^{22}$\\
                G049.37-00.30 & 19:23:10.82 & 14:26:39.89 & 5.4 & 50.6 & $3.6\substack{+4.7\\ -2.0} \times 10^5$ & $281\substack{+145\\ -95}$ & $34\substack{+27\\ -15}$ & $0.4-1.2\times10^{23}$\\
                G058.77+00.65 &  19:38:49.13 & 23:08:40.13 & 3.3 & 32.0 & $2.5\substack{+5.5 \\ -1.7} \times 10^4$ & $178\substack{+149 \\ -81}$ & $13\substack{+11 \\ -6} $ & $0.9-2.9\times10^{22}$\\
                G061.48+00.09 a & 19:46:49.07 & 25:12:47.30 & 2.2 & 21.9 & $ 0.9\substack{+2.5\\ -0.7} \times 10^5$ & $63\substack{+69 \\ -33}$ & $19\substack{+20 \\ -10}$& $2.5-8.1\times10^{22}$\\
                G061.48+00.09 b & 19:46:48.26 & 25:12:49.50 & 2.2 & 21.9 & $ 0.4\substack{+1.6\\ -0.3} \times 10^5$ & $41\substack{+75 \\ -26}$ & $13\substack{+19\\ -8}$ & $2.1-6.7\times10^{22}$\\
                G305.20+0.21 & 13:11:13.77 & -62:34:40.98 & 4.1  & -40.0 & $1.5\substack{+3.3 \\ -1.0} \times 10^5$ & $92\substack{+92 \\ -46}$ & $25\substack{+24 \\ -12}$ & $0.8-2.6\times10^{23}$\\
                G305.80-0.24 & 13:16:43.13 & -62:58:32.09 & 4.0 & -33.2 & $1.8\substack{+1.1\\ -0.7} \times 10^5$ &$282\substack{+154\\ -100} $ &$22\substack{+7.0\\ -5.3} $ & $0.7-2.3\times10^{23}$\\
                G309.92+0.48 & 13:50:41.76 & -61:35:10.10 & 5.5 & -56.8 & $6.6\substack{+4.8 \\ -2.8} \times 10^5$ & $240\substack{+71 \\ -55}$ & $49\substack{+32 \\ -19}$ & $0.5-1.7\times10^{23}$\\
                G317.43-00.56 & 14:51:37.60 & -60:00:19.80 & 14.2 & 27.4 & $4.7\substack{+4.1\\ -2.2} \times 10^5$ &$315\substack{+112 \\ -83}$ &$38\substack{+25 \\ -15}$ & $1.5-4.7\times10^{22}$\\
                G318.95-0.20 & 15:00:55.27 & -58:58:53.00 & 2.4 & -34.5 & $0.9\substack{+3.8\\ -0.8} \times 10^5$ & $127\substack{+94 \\ -54}$& $19\substack{+30 \\ -12}$& $0.4-1.2\times10^{23}$\\
                G337.40-0.40 & 16:38:50.50 & -47:28:00.86 &  3.1 & -40.9 & $2.4\substack{+1.7\\ -1.0} \times 10^5$ & $305\substack{+156 \\ -103}$ & $23\substack{+6 \\ -5} $ & $0.6-1.9\times10^{23}$\\
                MMS 6/OMC3  & 05:35:23.42 & -05:01:30.57 & 0.4 & 11.0 & $4.7\substack{+4.2\\ -2.2} \times 10^2$ & $2.7\substack{+4.3\\ -1.6}$ & $2.3\substack{+1.8\\ -1.0}$& $0.6-2.0\times10^{23}$\\
                NGC 2071 & 05:47:04.64 & 00:21:47.44 & 0.4 & 9.5 & $6.9\substack{+12 \\ -4.4} \times 10^2$ & $4.2\substack{+4.7 \\ -2.2}$ & $1.7\substack{+2.2 \\ -1.0}$ & $0.3-1.1\times10^{22}$\\
                OMC-2/3 MM7 & 05:35:26.52 & -05:03:54.90 & 0.4 & 5.5 & $5.5\substack{+22\\ -4.4} \times 10^2$ & $2.8\substack{+4.8 \\ -1.8}$&$2.3\substack{+3.7 \\ -1.4}$& $0.8-2.7\times10^{22}$\\
                OMC1-S 137-408  & 05:35:13.70 & -05:24:08.72 & 0.4 & 9.0 & ... & ... & ... & $1.2-3.9\times10^{23}$\\
                W51 e2; North/IRS2 & 19:23:42.00 & 14:30:36.00 & 5.4 & 60.0 & ... & ... & ... & $0.6-2.1\times10^{23}$\\
                \noalign{\smallskip}   
                \hline
            \end{tabular}
            \tablefoot{
                \tablefoottext{a}{The sources detected in the same region are labeled with a letter after the name.}
                \tablefoottext{b}{Coordinates of the intensity peaks of the 225.7~GHz continuum maps (Fig.~\ref{fig:continuum}).}
                \tablefoottext{c}{Parameter estimated from the SED fitting performed in \cite{Telkamp2025} (for G058.77, G045.47, G049.27, G305.20, G309.92, and NGC 2071) and Rahman et al. in prep. (for the remaining sources).}
                \tablefoottext{d}{The Local Standard of Rest (LSR) velocities used to center the spectra.}
                \tablefoottext{e}{Range of values estimated in this work (Sect.~\ref{sect:maps}).}
            }
        \end{table*}
        
\section{Methods}\label{sect:3}

In this work, we will focus only on the \obsA\,regions that have both ACA and TP observations, as the goal of the study is to characterize the chemical inventories of the continuum sources identified within the ALMA field-of-view (FoV) with a diameter of $\sim44^{\prime\prime}$. The main steps in our study include: source identification, noise estimation, and spectra extraction (Sect.~\ref {sect:rms}); derivation of H$_2$ column densities from the continuum maps (Sect.~\ref {sect:lineidentification}); line analysis (Sect.~\ref{sect:lineidentification}); and SED analysis (Sect.~\ref{sect:sedfit}). In the following, we describe each step in detail.
    
\subsection{Source identification and spectra extraction}\label{sect:rms}

Source identification was performed based on the continuum maps via two criteria: (1) intensity peak required to be $> 5\sigma$; (2) distance between different peaks required to be larger than the synthesized beam of the observation. The $1\sigma$ noise level is estimated from the rms noise of the map. A sigma-clipping algorithm has been used through the \texttt{sigma$\_$clip} function from the module \texttt{astropy.stats}, which allows us to mask the outliers of a distribution (e.g., the distribution of intensities of the map), and then calculate the rms noise of the masked data. The rms noise levels of each continuum map are shown in Table~\ref{tab:observations}.
            
We extracted the spectra from a circular beam of $5.1''$ centered on the maximum intensity. The diameter of the extracted region is the average minor axis of the beam of the data cubes (Table~\ref{tab:observations}). The spectra were extracted from the same circular beam for each region to obtain comparable parameters from the spectral analysis. The rms noise of the spectra was determined using the same approach as for the continuum maps, after on-source spectra were extracted from the data cubes for each spectral window. The average noise for each source is shown in Table~\ref{tab:observations}.

\subsection{H$_2$ column density calculations}\label{sect:maps}

We extracted the flux to estimate the molecular hydrogen column density, $N_{\text{H}_2}$, from the dust emission maps. We extracted the flux using the same $5.1''$ beam as for the spectral extraction and, assuming optically thin conditions, applied the following equation \citep{battersby2011characterizing}:
\begin{equation}
N_{\text{H}_2}=\frac{\gamma F_\nu}{\Omega_\text{S} \kappa_\nu B_\nu(T_\text{d})\mu_{\text{H}_2} m_\text{H}},
\end{equation}
where $\gamma$ is the gas-to-dust ratio, $F_\nu$ is the continuum flux density at frequency $\nu$, $\Omega_S$ is the source solid angle, $\kappa_\nu$ is the dust mass opacity, $T_\text{d}$ is the dust temperature, $\mu_{\text{H}_2}$ is the mass per $\rm H_2$ molecule in units of $m_{\rm H}$, which is assumed to be 2.8. 
The H$_2$ column density has been estimated with a constant gas-to-dust ratio ($\gamma=100$), and assuming three different values of $T_\text{d}$ (i.e., 50, 100, and 150~K). The dust mass opacity has been derived at the observed frequencies from the relation $\kappa_{\nu}/\kappa_{\nu_0}=(\nu/\nu_0)^{\beta}$, assuming $\kappa_{\nu_0}=0.899~$cm$^2$ g$^{-1}$ at $\nu_0 = 230~$GHz, obtained by \cite{ossenkopf1994dust}, for thin ice mantles at a gas density of $10^6~$cm$^{-3}$. The spectral index $\beta$ adopted is 1.5. The H$_2$ column densities are summarized in Table~\ref{tab:sources}.

\subsection{Line identification}\label{sect:lineidentification}

The spectra were fitted using the MAdrid Data CUBe Analysis \citep[\textsc{MADCUBA}\footnote{\textsc{MADCUBA} is a software developed in the Madrid Center of Astrobiology (CAB, CSIC-INTA), which enables the visualization and analysis of single spectra and data cubes: \url{https://cab.inta-csic.es/madcuba/}};][]{martin2019spectral} software. The transitions were identified using the Spectral Line Identification and LTE Modelling (SLIM) tool of \textsc{MADCUBA}, which makes use of the Jet Propulsion Laboratory \citep[JPL;][]{pickett1998submillimeter} and Cologne Database for Molecular Spectroscopy \citep[CDMS;][]{muller2001cologne} catalogs. The lines were fitted using the AUTOFIT function of SLIM. This function produces the Gaussian synthetic spectrum that best matches the observed spectrum, assuming local thermodynamic equilibrium (LTE) conditions. It estimates the physical parameters, i.e., total molecular column density ($N_\text{tot}$), full-width at half-maximum (FWHM), radial systemic velocity of the source ($v_\text{LSR}$), excitation temperature ($T_\text{ex}$), and angular size of the emission ($\theta_S$) and their associated uncertainties through a non-linear least-squares fitting procedure. AUTOFIT assumes a constant set of these parameters for all transitions of a given species. 
We assume that the source emission $\theta_S$ fills the beam, which is supported by the continuum maps shown in Fig.~\ref{fig:continuum}. In the case of single-line detections, where the excitation temperature cannot be constrained, $T_\text{ex}$ is fixed to 100~K, a typical value for hot cores. If the fitting algorithm converges, FWHM and $v_\text{LSR}$ were left free.

\subsection{SED fitting}\label{sect:sedfit}

The spectral energy distribution (SED) analysis of the sample was performed using the latest version (0.9.11) of \texttt{sedcreator}, an open-source Python package described in \cite{Telkamp2025}, which provides a comprehensive description of the methods employed \citep[also see ][]{Fedriani2023}. This package is hosted in both GitHub\footnote{https://github.com/fedriani/sedcreator} and PyPi\footnote{https://pypi.org/project/sedcreator/} (the documentation can be accessed at this URL: \url{https://sedcreator.readthedocs.io/}). This package has two main classes, SedFluxer and SedFitter. SedFluxer performs aperture photometry on a given image, coordinates, and aperture size using functions from \cite{bradley2020}. SedFitter fits observations to a grid of models following the \cite{zhang2018radiation} radiative transfer models. The results of the SED fitting for our sample, obtained in \cite{Telkamp2025} and in Rahman et al. in prep., are summarized in Sect.~\ref{sect:sourceprop}.

\section{Results and discussion}\label{sect:4}

\subsection{Sources physical properties}\label{sect:sourceprop}

The SED fitting results provide the physical properties of our protostar sample. The sources are characterized by bolometric luminosities, $L_\text{bol}$, between $4.7\times10^2$ and $7.6\times10^5~L_\odot$, envelope masses, $M_\text{env}$, between $3$ and $305\:M_\odot$, and current protostellar masses, $m_\star$, between 2 and $62~M_\odot$. The values of bolometric luminosity, envelope, and current protostellar mass for the whole sample are summarized in Table~\ref{tab:sources}. The results for G58.77, G45.47, G49.27, G305.20, G309.92, and NGC 2071 have been presented in \cite{Telkamp2025}. Those for AFGL~5180 have also been reported by \cite{Telkamp2025}; however, given the lower spatial resolution of the maps used for the SED fitting, it is not possible to disentangle the contribution from the two different continuum sources detected in this paper (AFGL 5180 a and AFGL 5180 b), so in this case, no results have been presented or used. The properties of the remaining sources are taken from a companion study (Rahman et al., in prep.) and will be described in detail in a separate publication. No SED parameters have yet been estimated for OMC1-S 
and W51 e2, so analysis of these regions is deferred to a future study. 

\subsection{Description of the dust continuum maps}

\begin{figure*}
        \centering
        \begin{minipage}[c]{0.03\textwidth}
            \centering
            \includegraphics[height=0.15\textheight]{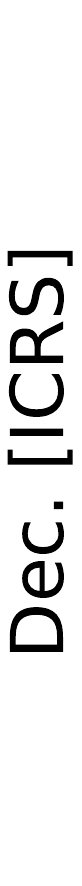}
        \end{minipage}
        \begin{minipage}[c]{0.92\textwidth}
            \centering
            \begin{subfigure}{0.24\textwidth}\centering
                \includegraphics[width=\textwidth]{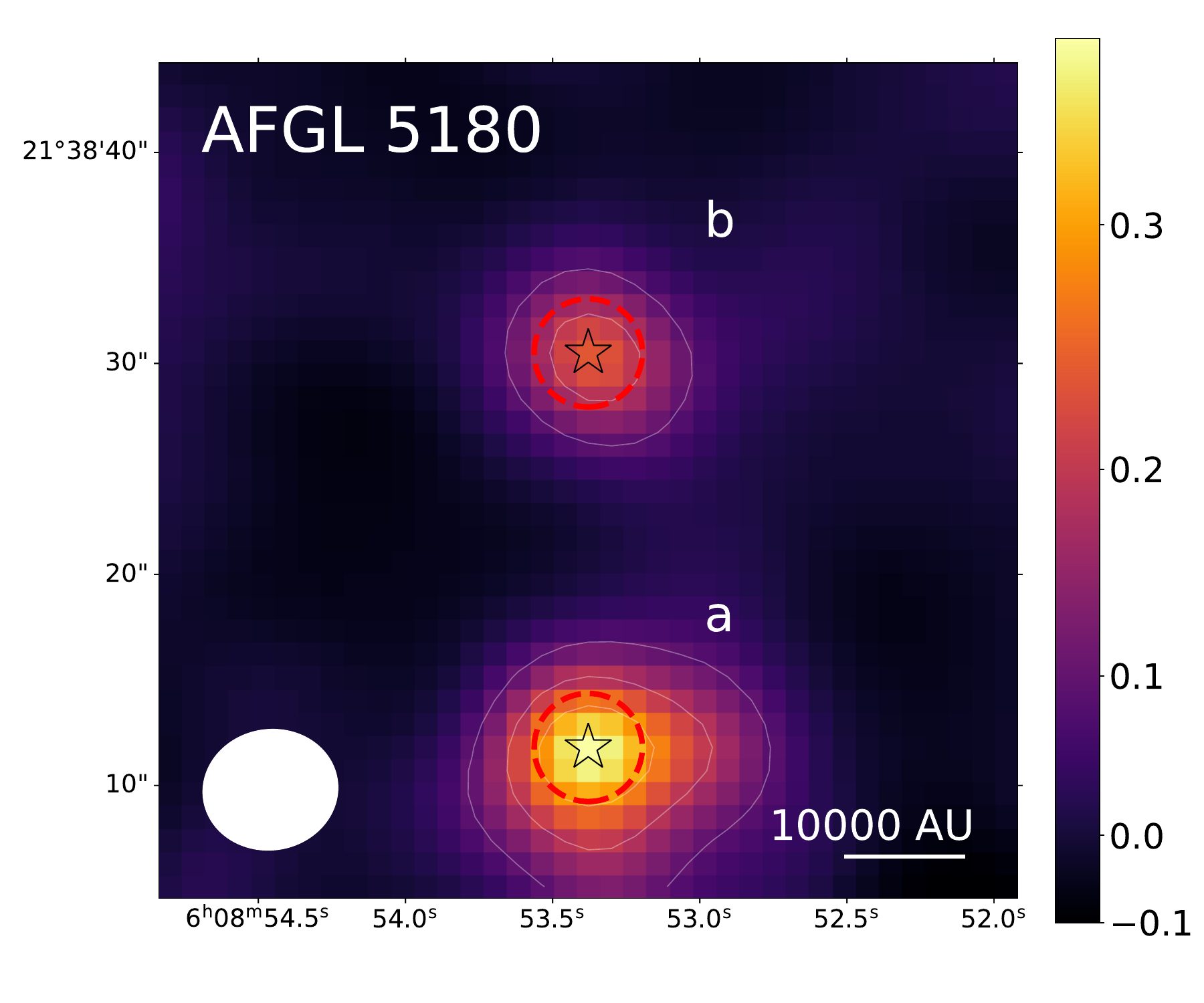}
            \end{subfigure}
            \begin{subfigure}{0.24\textwidth}\centering
                \includegraphics[width=\textwidth]{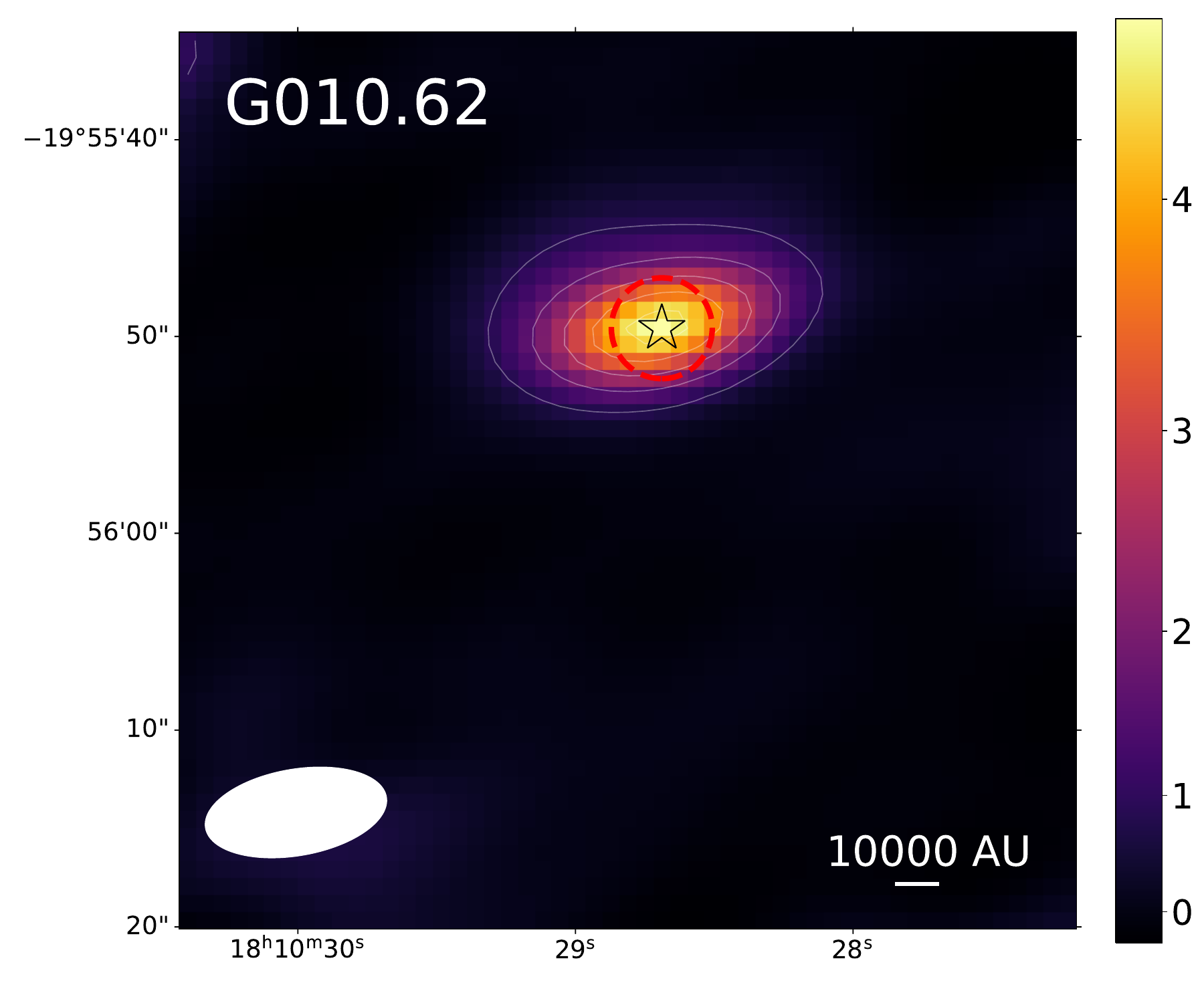}
            \end{subfigure}
            \begin{subfigure}{0.24\textwidth}\centering
                \includegraphics[width=\textwidth]{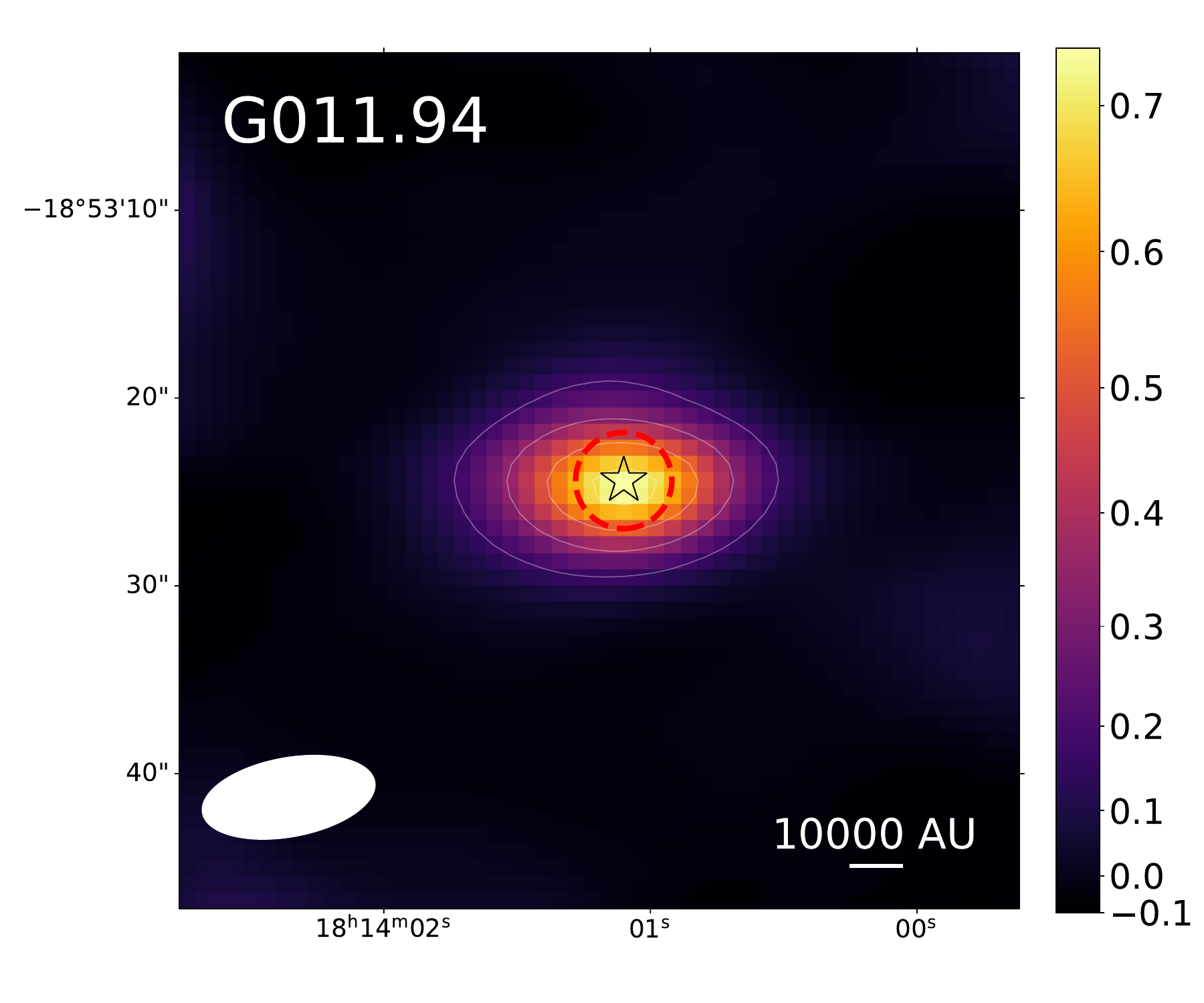}
            \end{subfigure}
            \begin{subfigure}{0.24\textwidth}\centering
                \includegraphics[width=\textwidth]{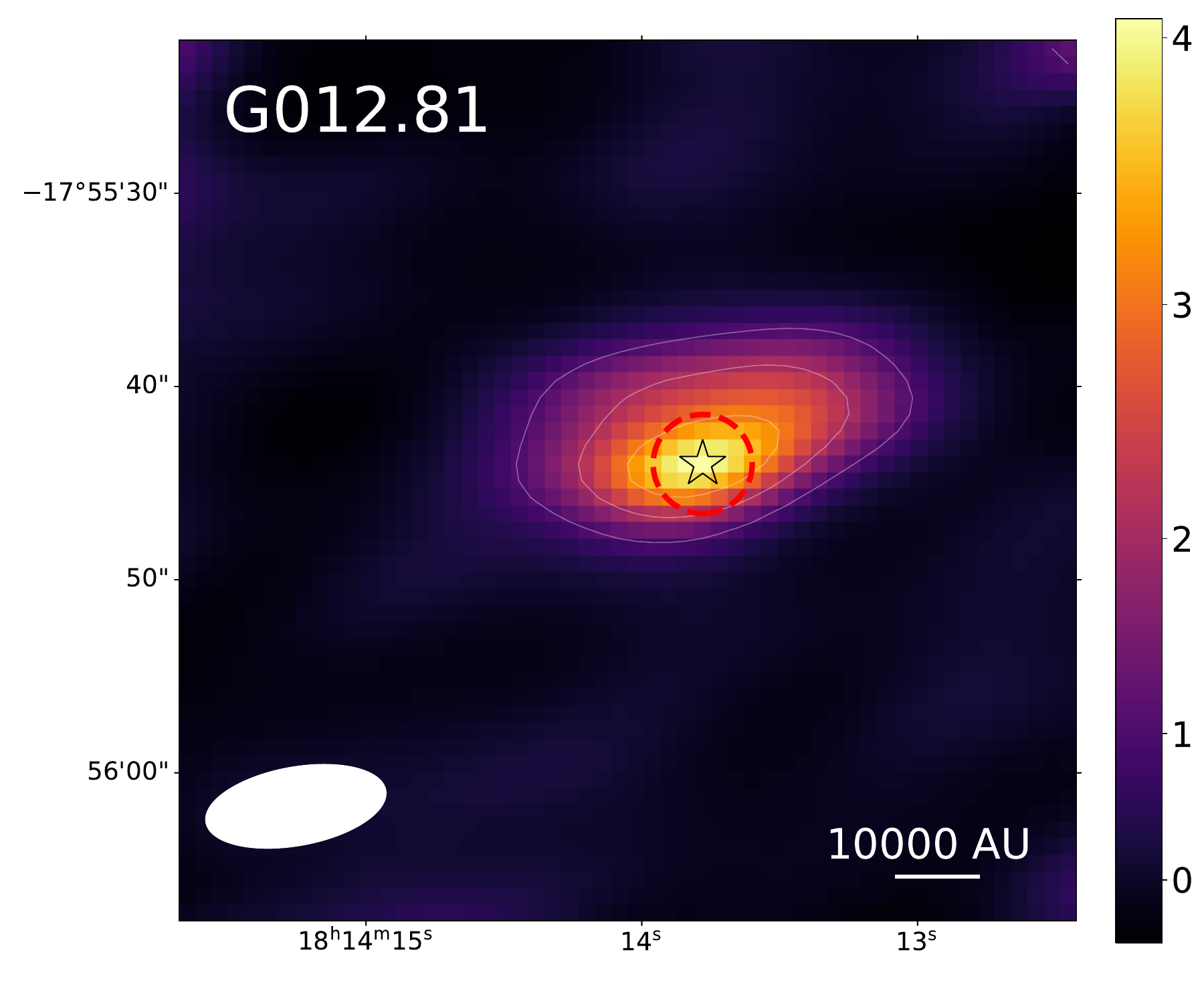}
            \end{subfigure}
            \begin{subfigure}{0.24\textwidth}\centering
                \includegraphics[width=\textwidth]{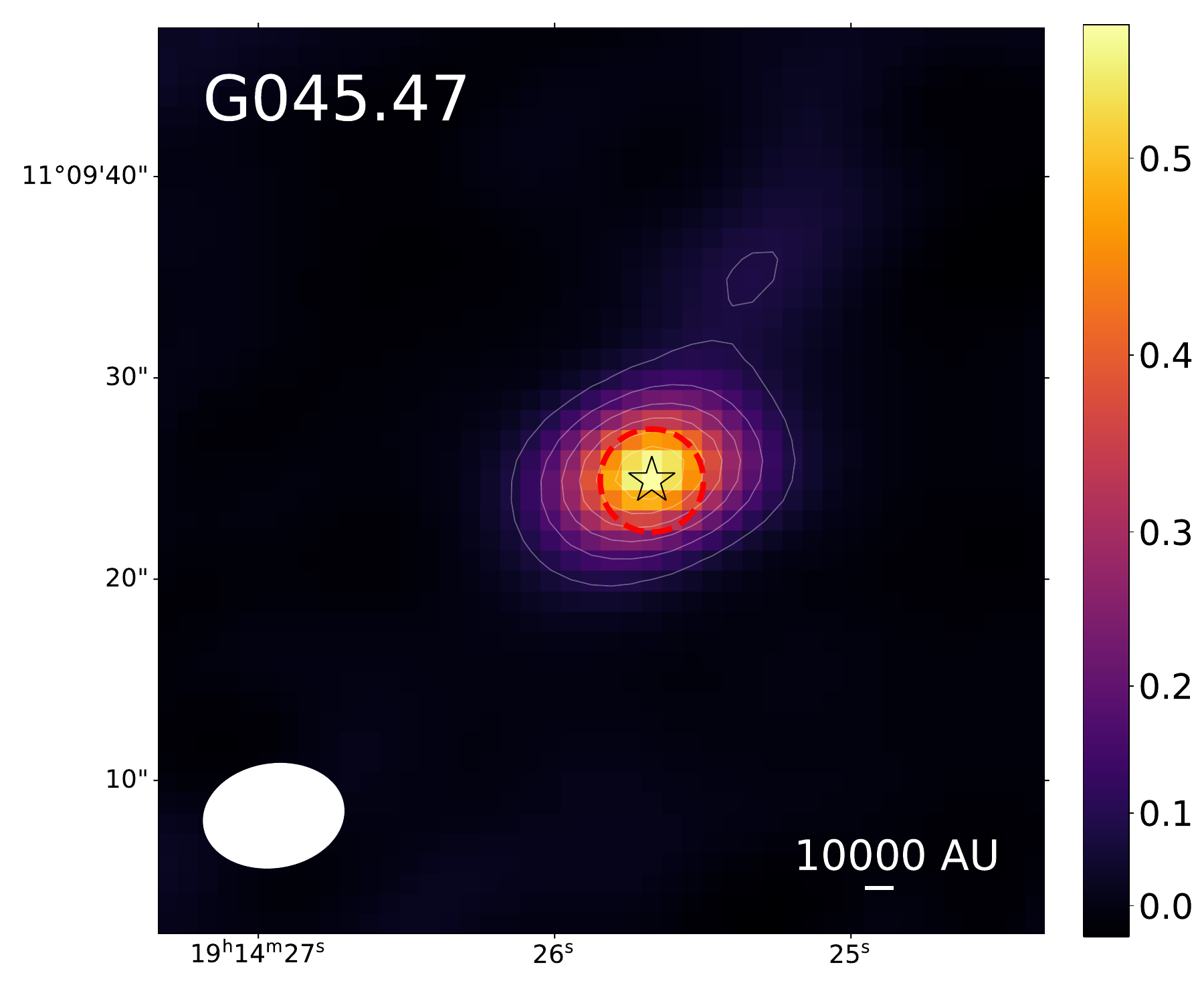}
            \end{subfigure}
            \begin{subfigure}{0.24\textwidth}\centering
                \includegraphics[width=\textwidth]{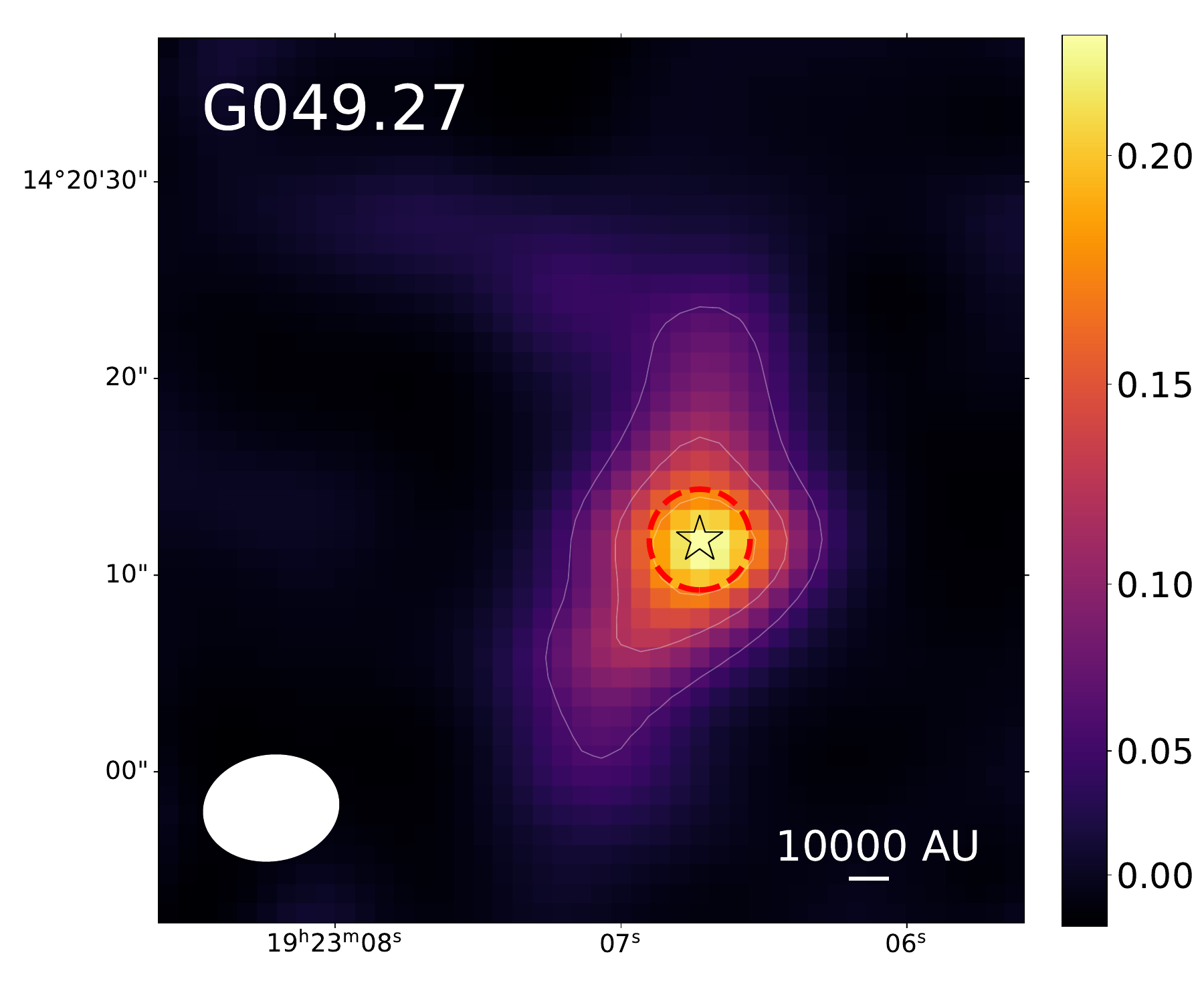}
            \end{subfigure}
            \begin{subfigure}{0.24\textwidth}\centering
                \includegraphics[width=\textwidth]{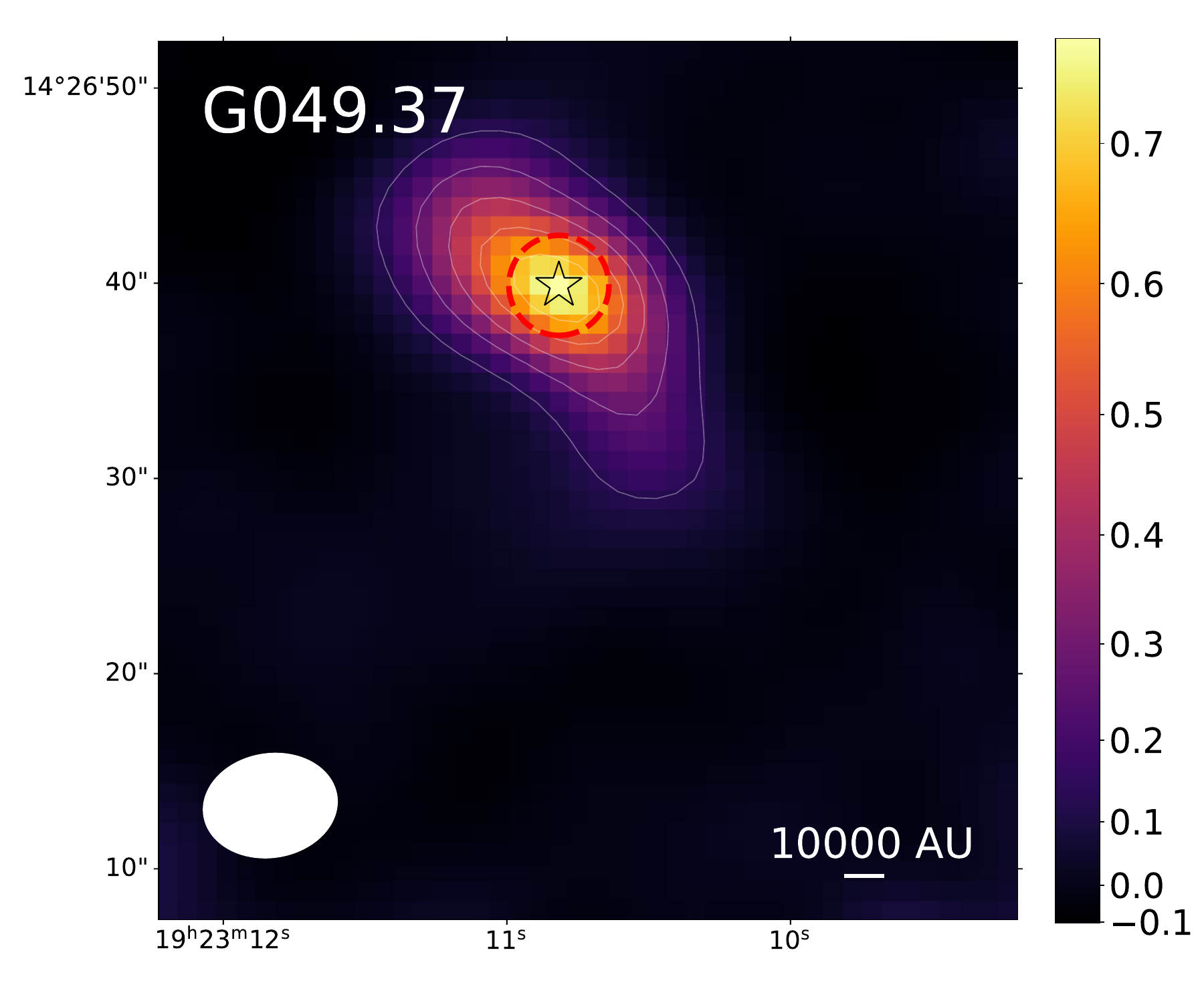}
            \end{subfigure}
            \begin{subfigure}{0.24\textwidth}\centering
                \includegraphics[width=\textwidth]{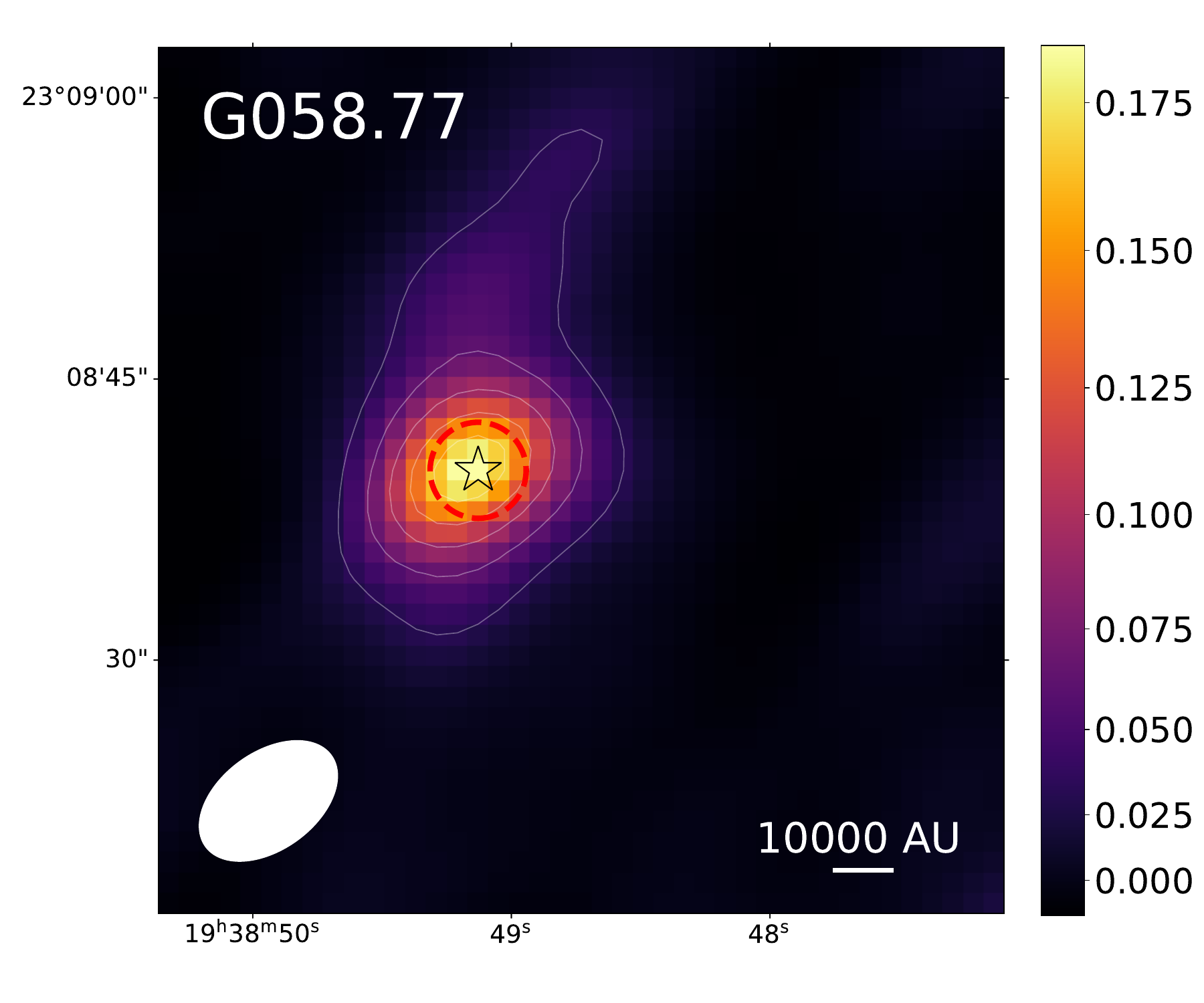}
            \end{subfigure}
            \begin{subfigure}{0.24\textwidth}\centering
                \includegraphics[width=\textwidth]{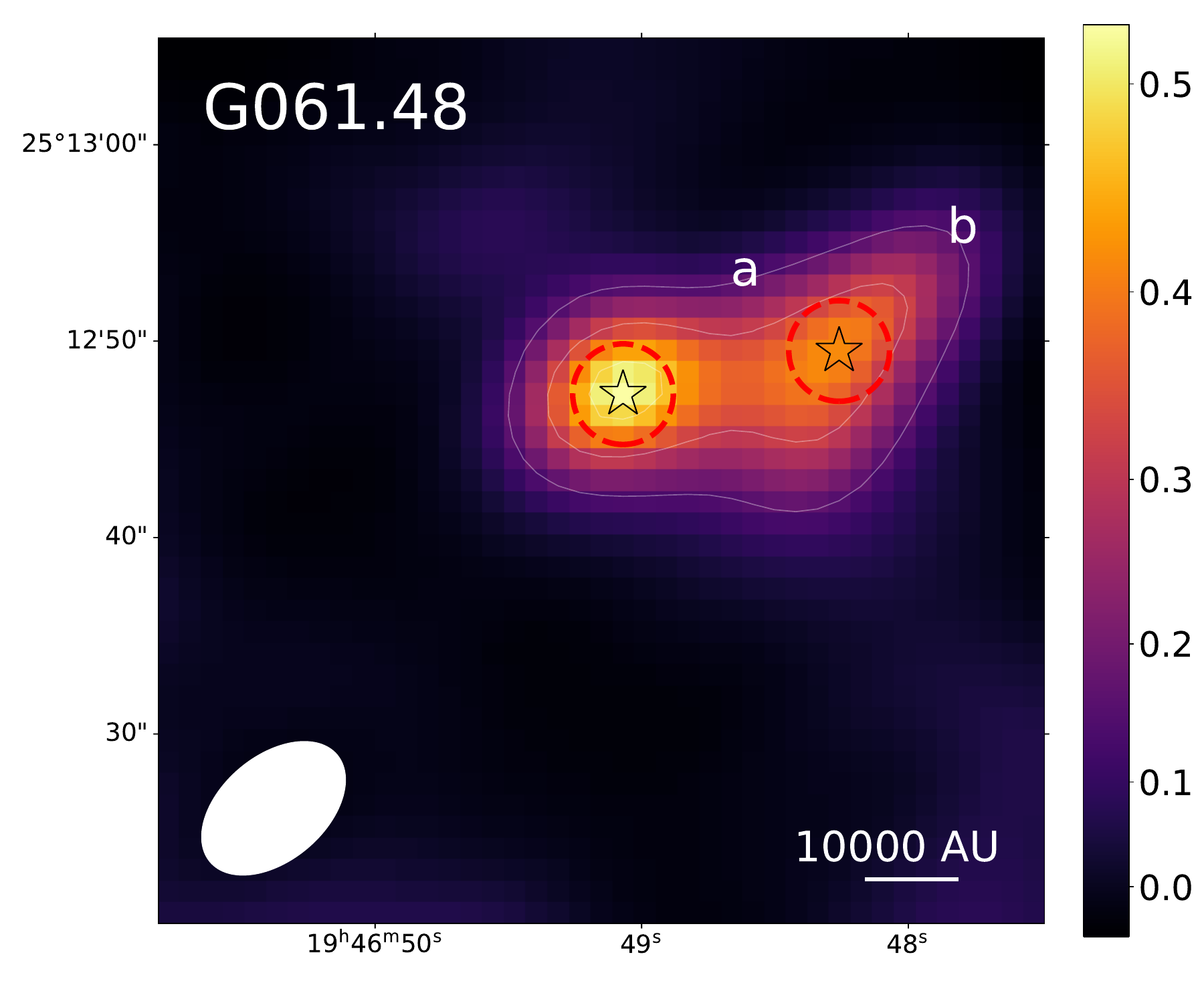}
            \end{subfigure}
            \begin{subfigure}{0.24\textwidth}\centering
                \includegraphics[width=\textwidth]{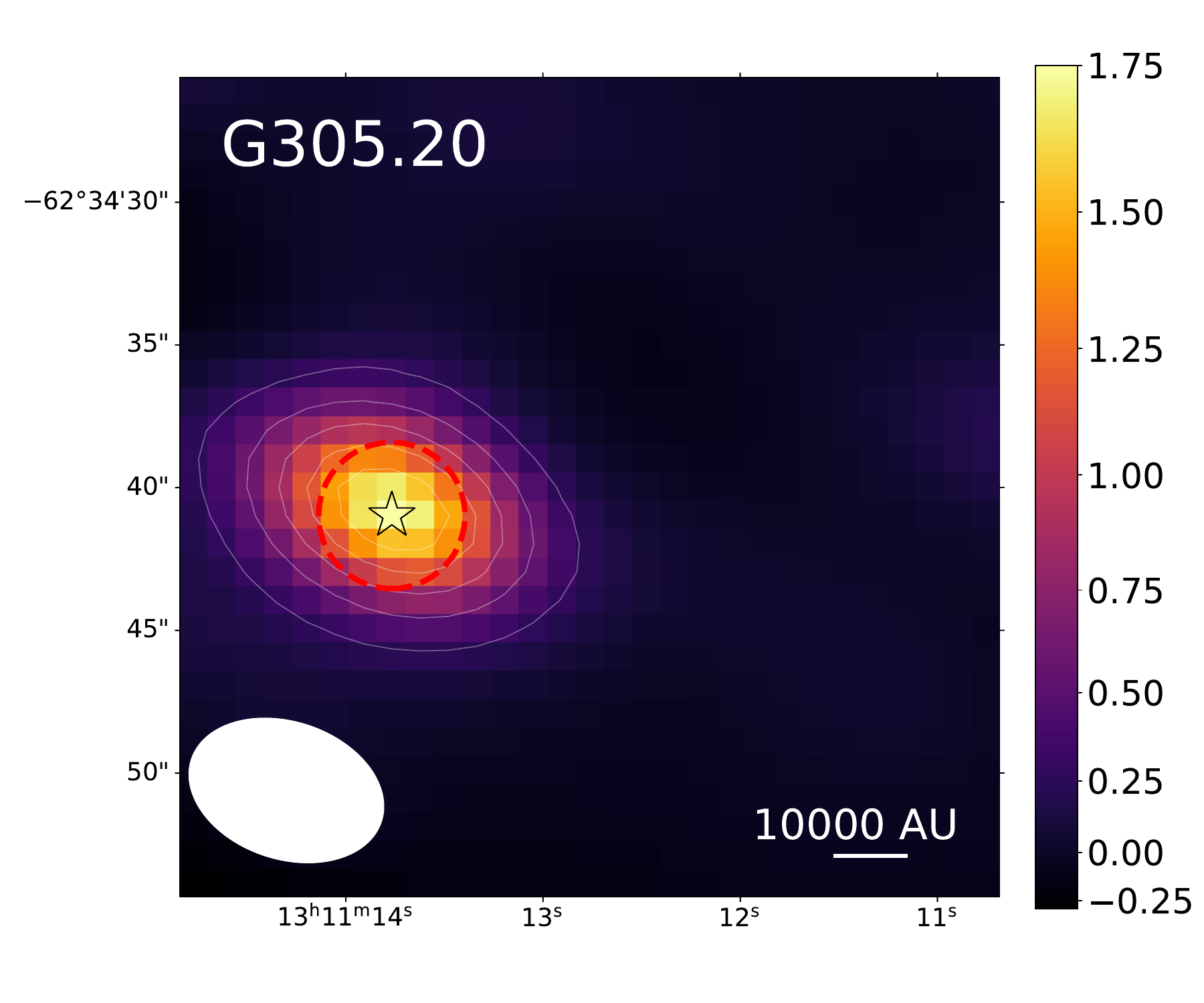}
            \end{subfigure}
            \begin{subfigure}{0.24\textwidth}\centering
                \includegraphics[width=\textwidth]{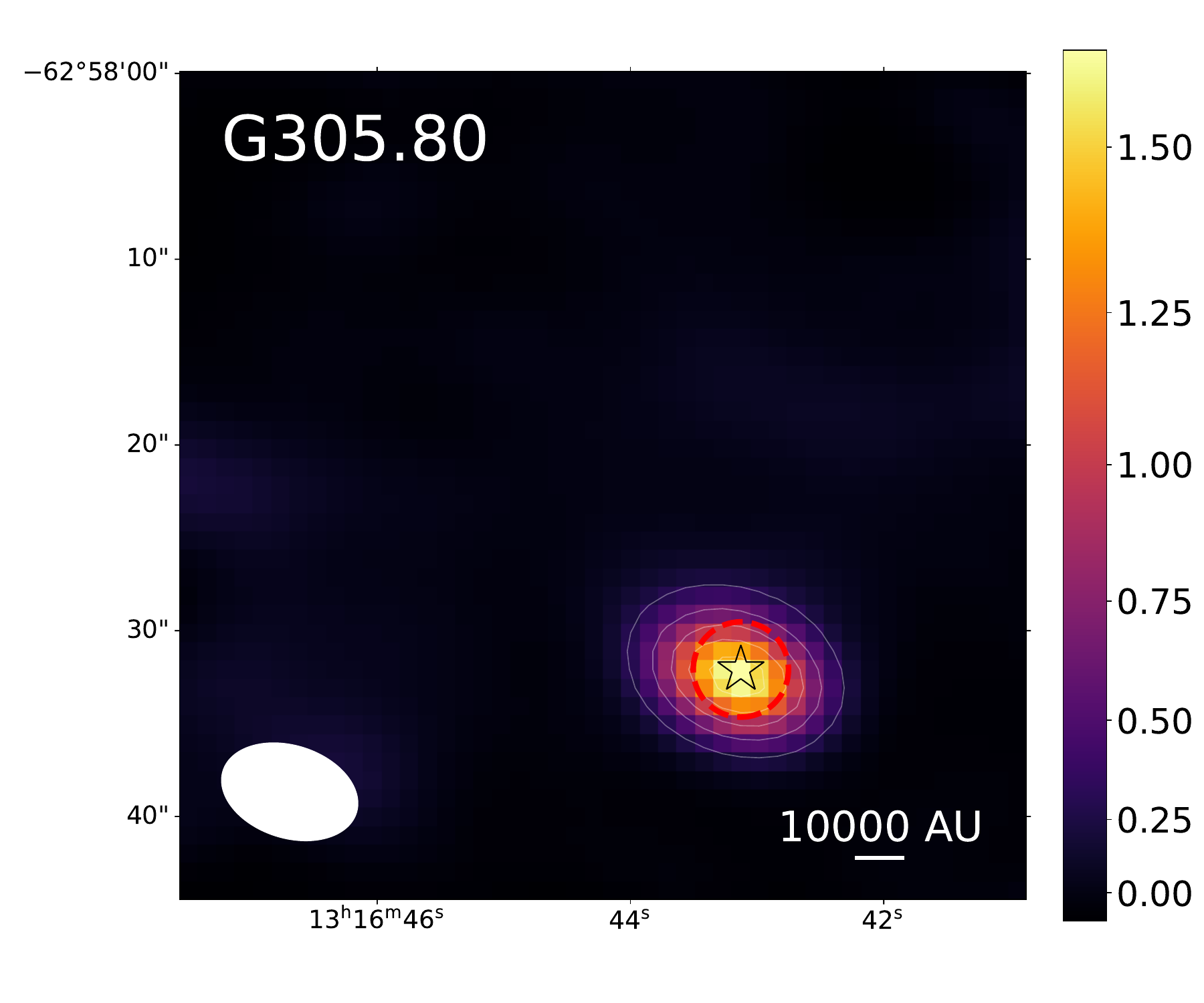}
            \end{subfigure}
            \begin{subfigure}{0.24\textwidth}\centering
                \includegraphics[width=\textwidth]{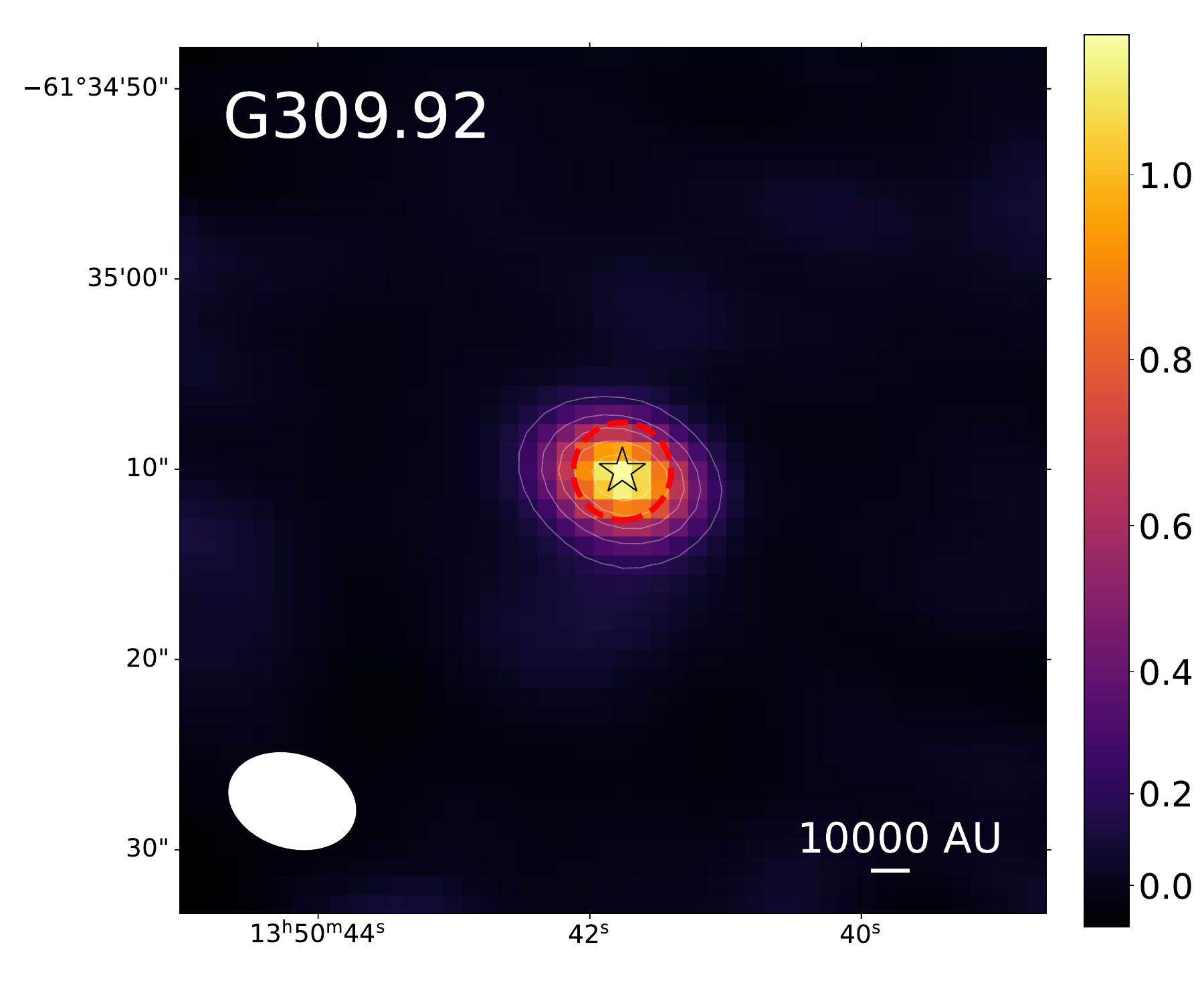}
            \end{subfigure}
            \begin{subfigure}{0.24\textwidth}\centering
                \includegraphics[width=\textwidth]{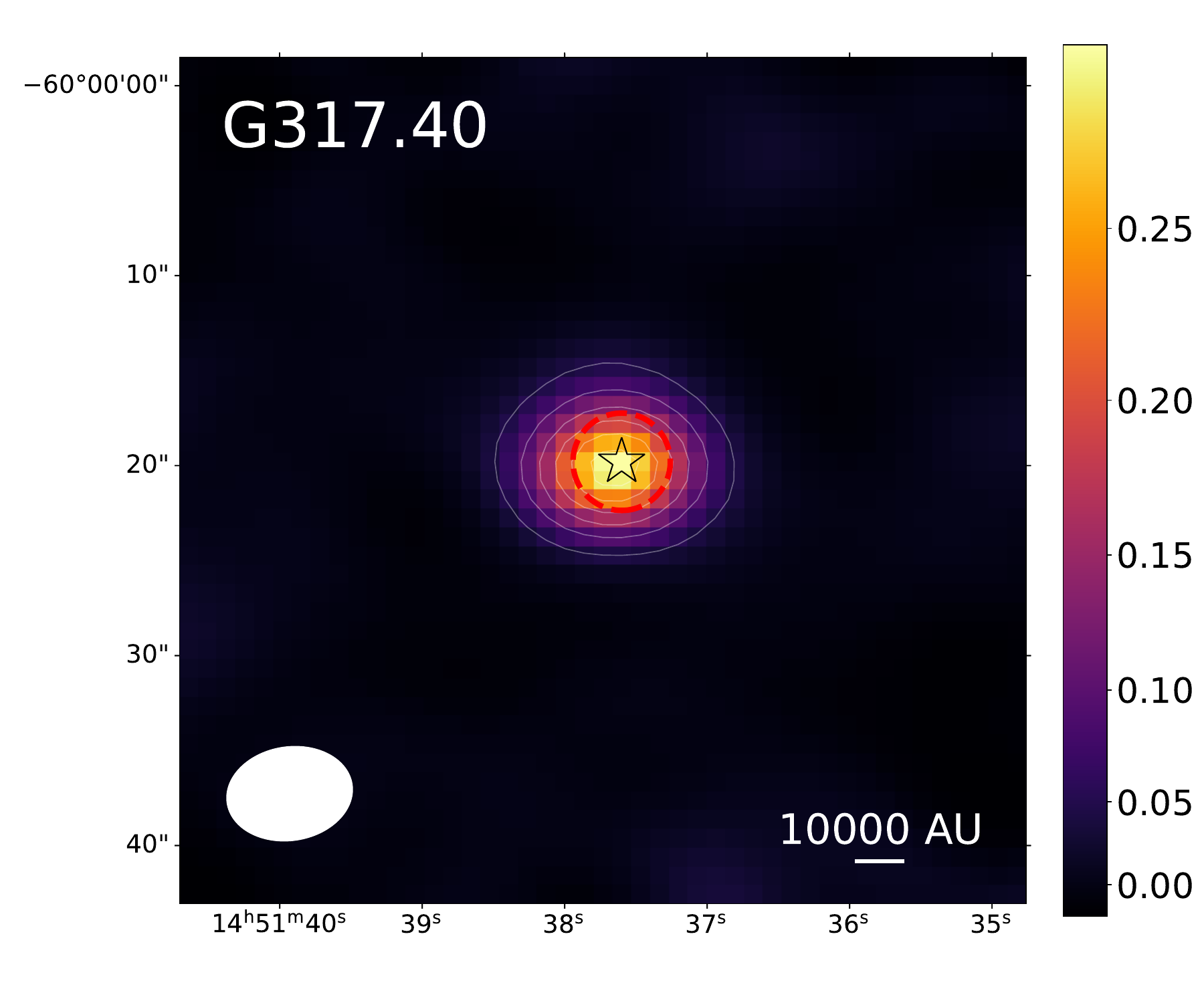}
            \end{subfigure}
            \begin{subfigure}{0.24\textwidth}\centering
                \includegraphics[width=\textwidth]{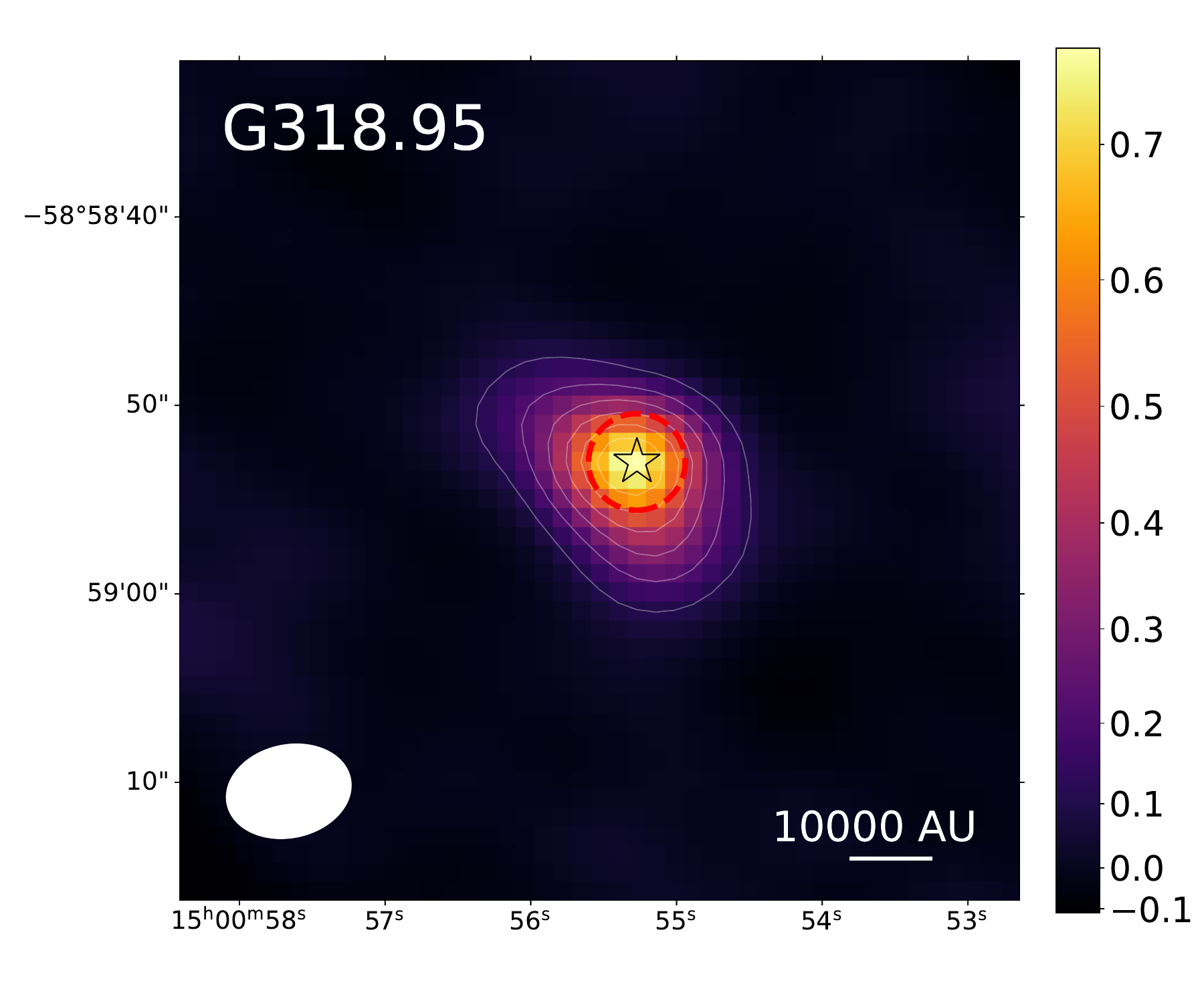}
            \end{subfigure}
            \begin{subfigure}{0.24\textwidth}\centering
                \includegraphics[width=\textwidth]{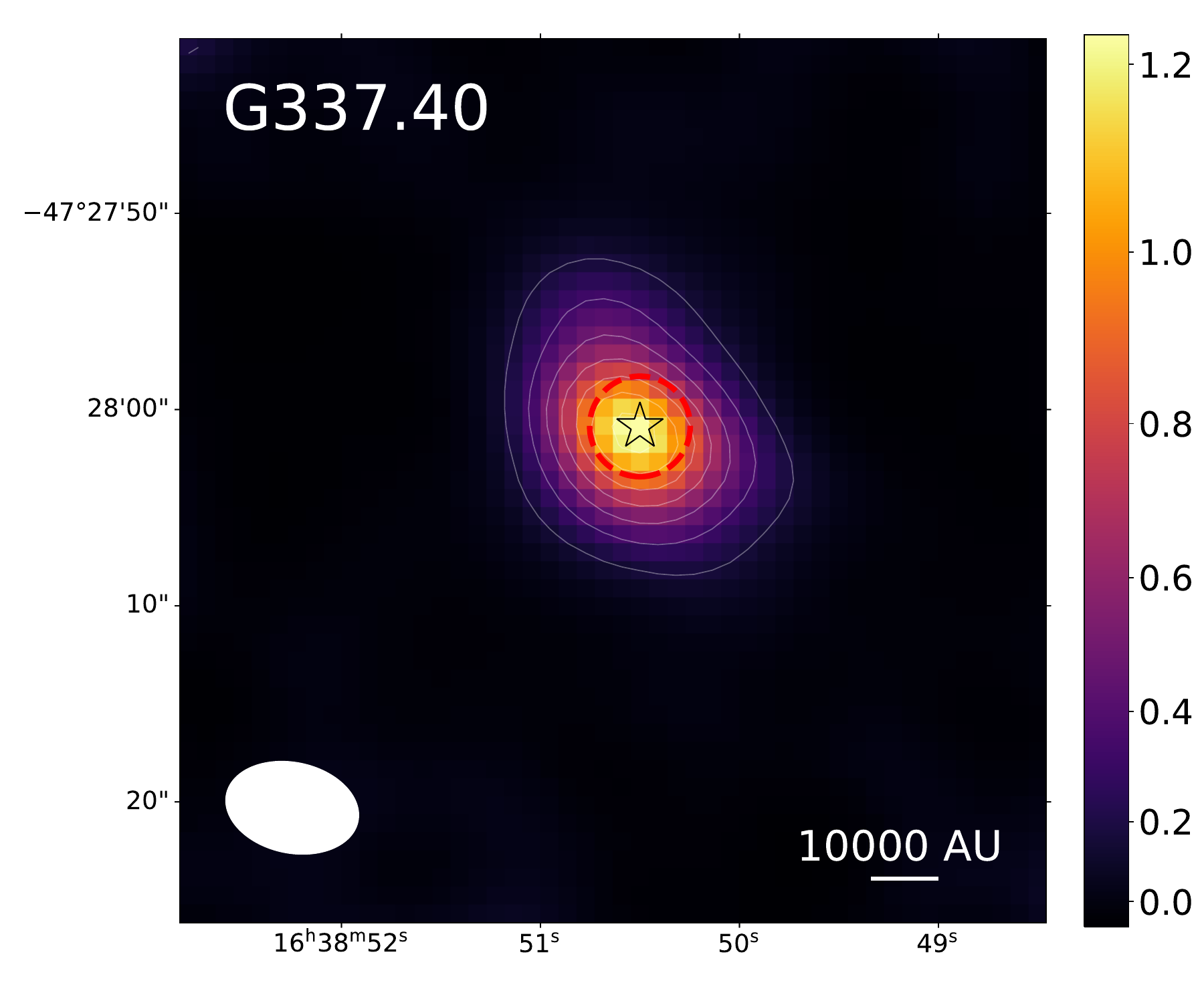}
            \end{subfigure}
            \begin{subfigure}{0.24\textwidth}\centering
                \includegraphics[width=\textwidth]{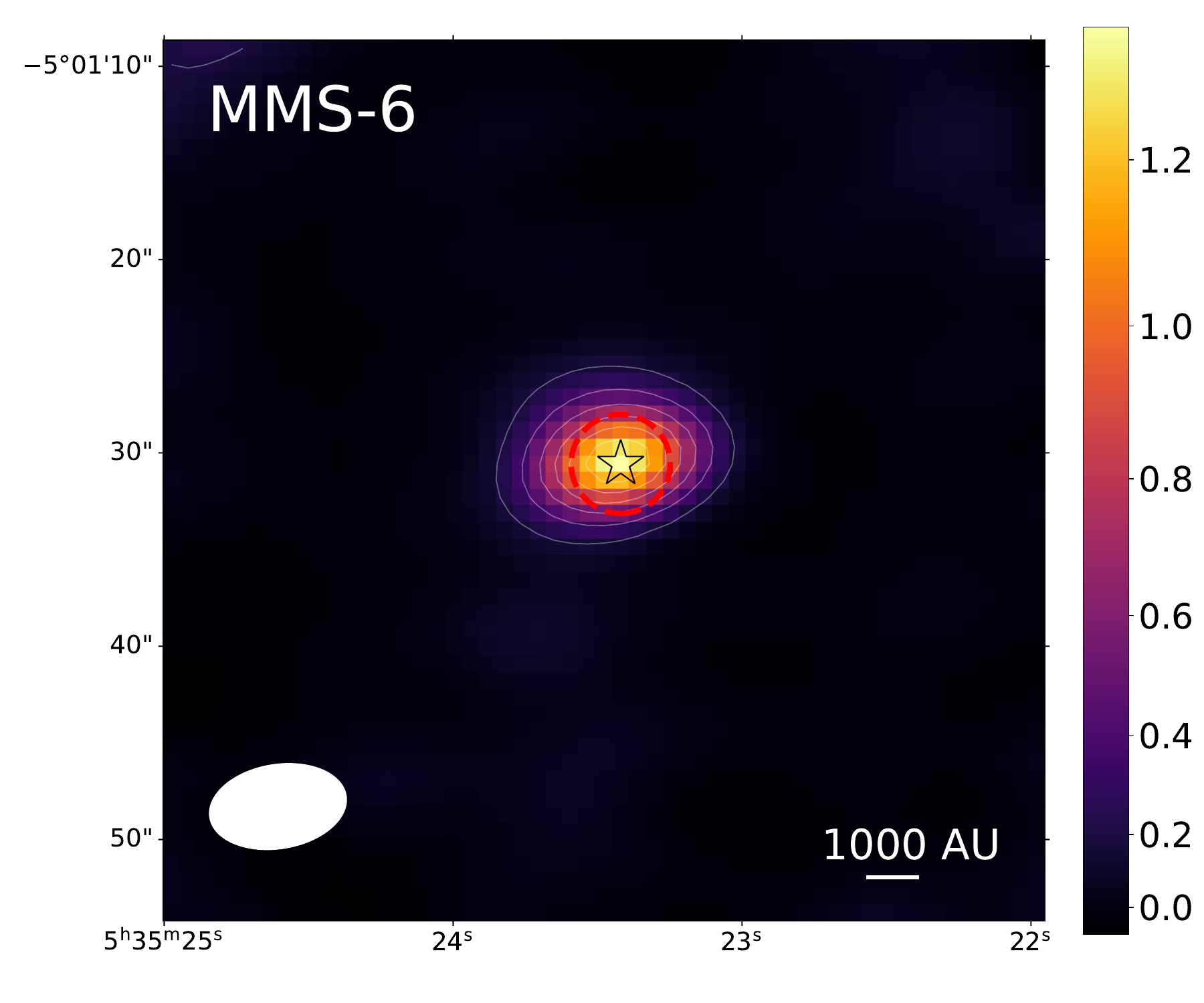}
            \end{subfigure}
            \begin{subfigure}{0.24\textwidth}\centering
                \includegraphics[width=\textwidth]{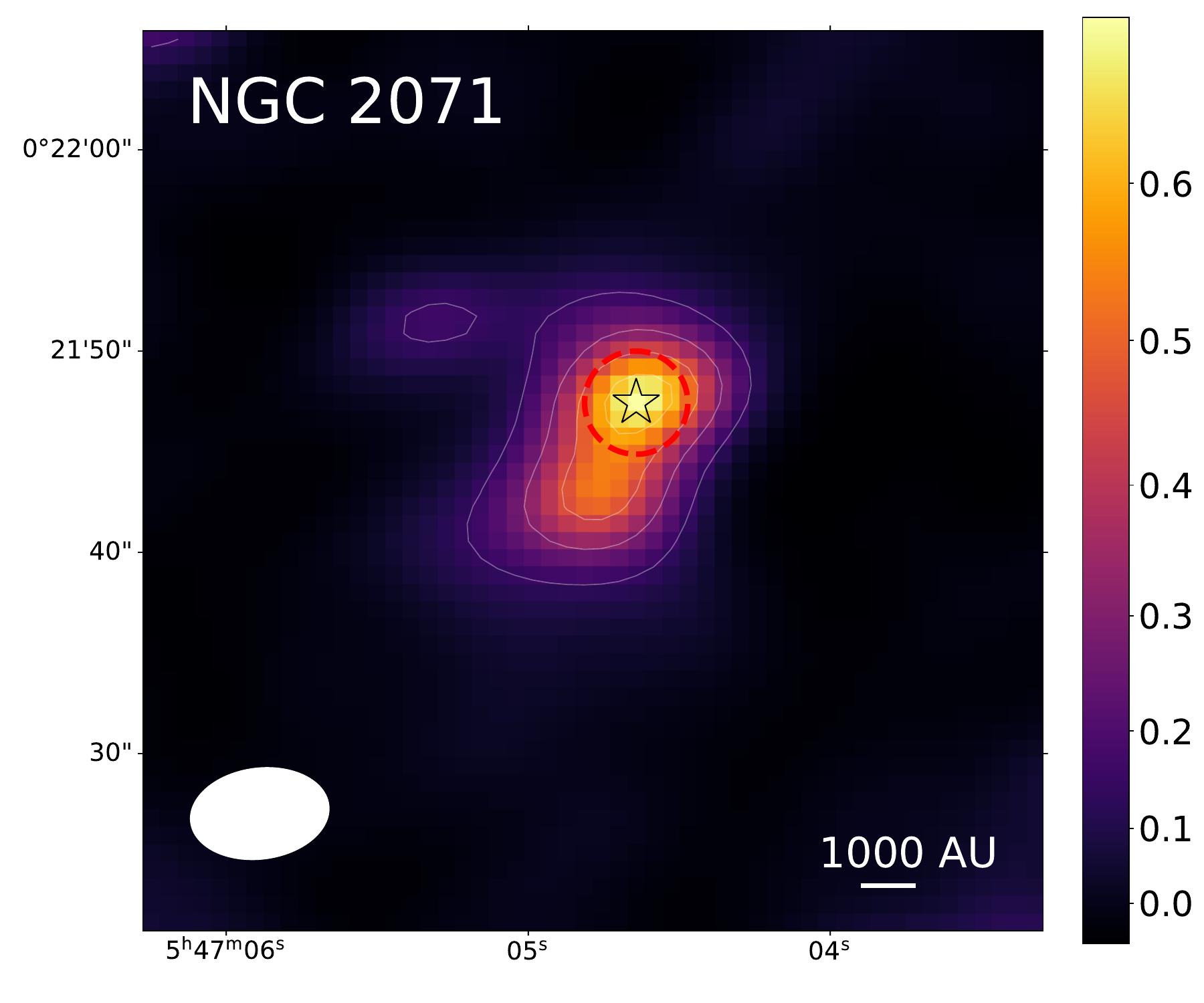}
            \end{subfigure}
            \begin{subfigure}{0.24\textwidth}\centering
                \includegraphics[width=\textwidth]{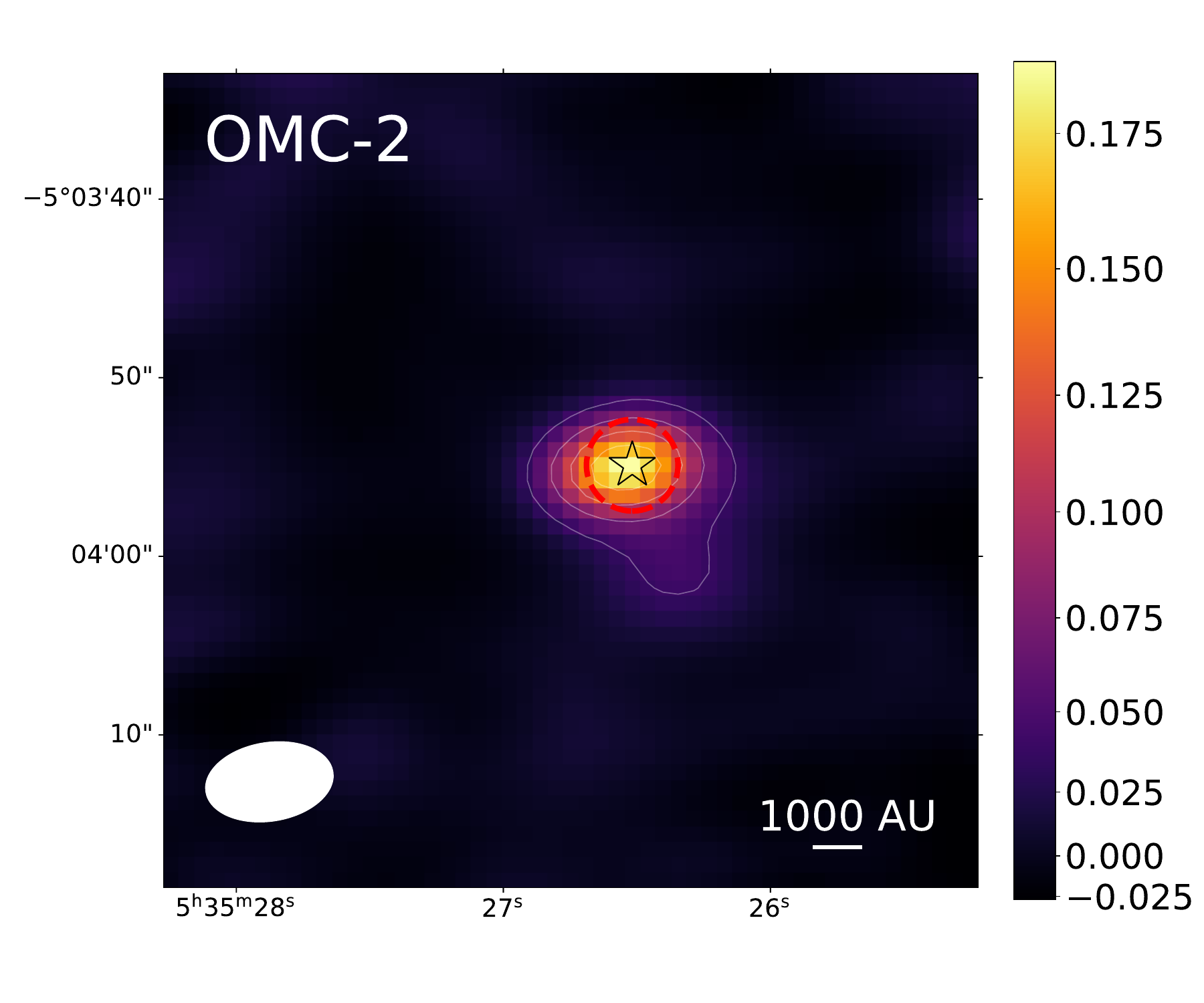}
            \end{subfigure}
            \begin{subfigure}{0.24\textwidth}\centering
                \includegraphics[width=\textwidth]{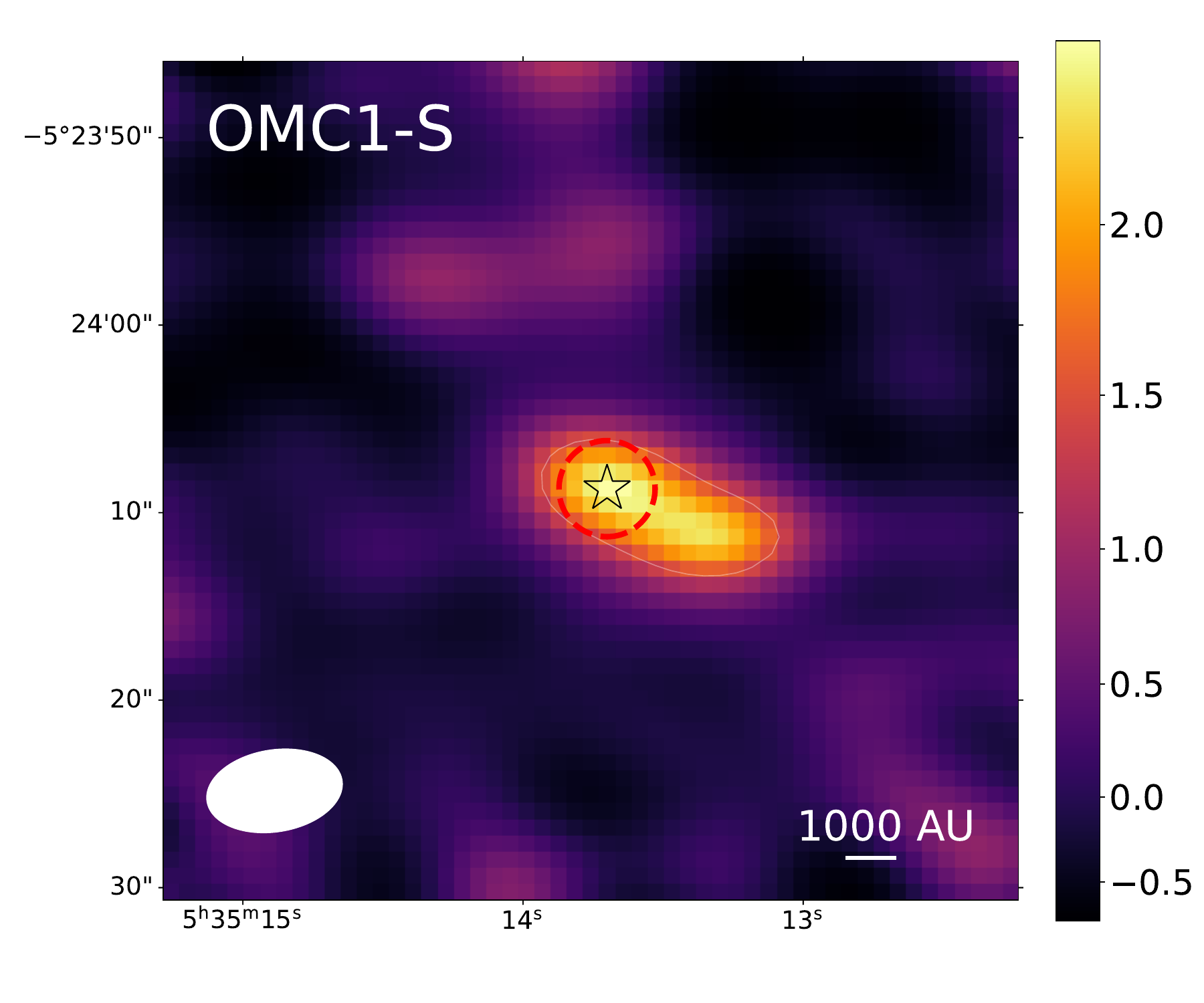}
            \end{subfigure}
            \begin{subfigure}{0.24\textwidth}\centering
                \includegraphics[width=\textwidth]{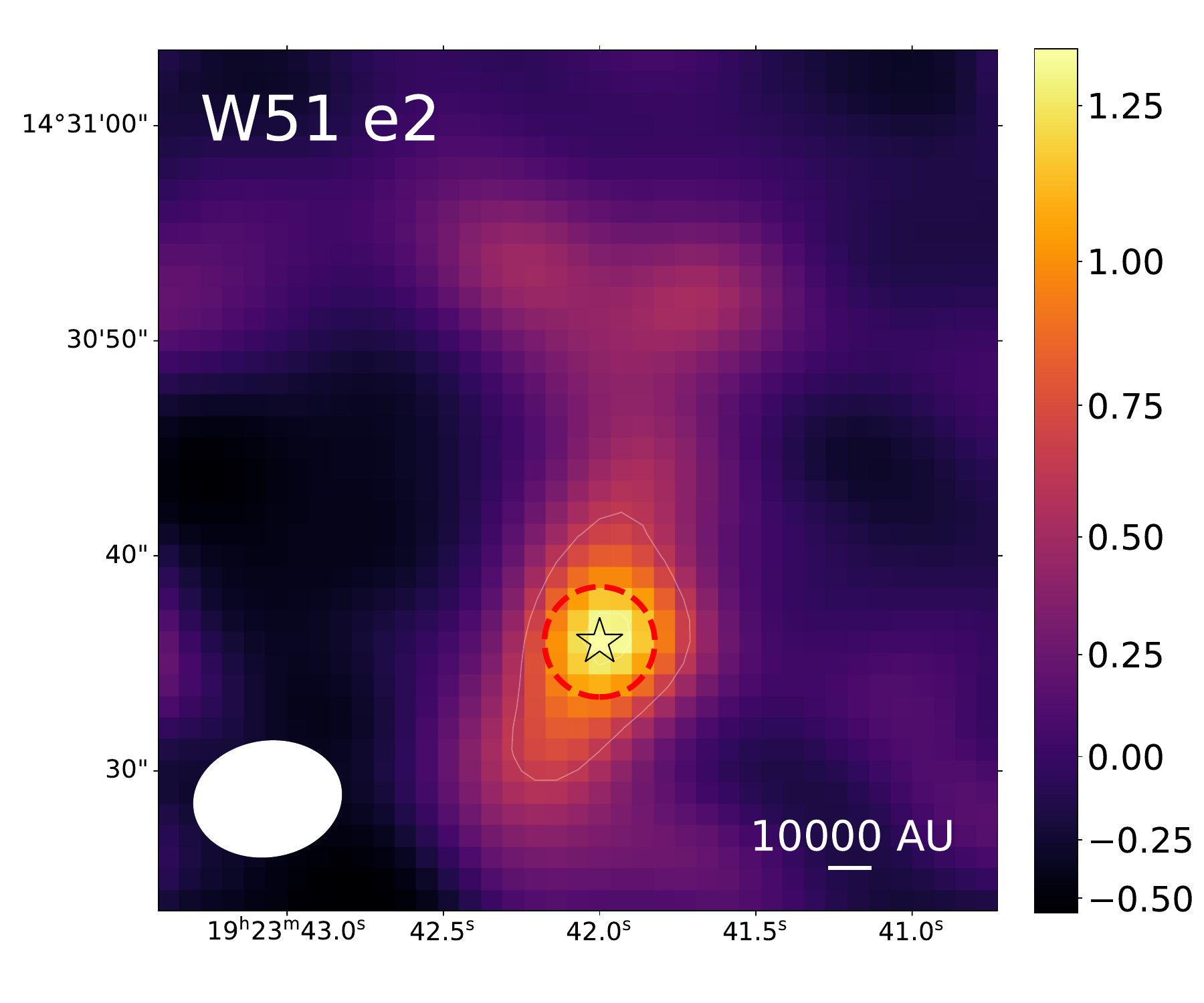}
            \end{subfigure}
        \end{minipage}
        \begin{minipage}[c]{0.03\textwidth}
            \centering
            \includegraphics[height=0.15\textheight]{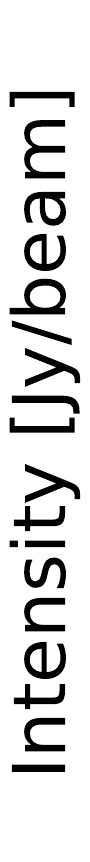}
        \end{minipage}
        \vspace{1mm}
        \includegraphics[width=0.15\textwidth]{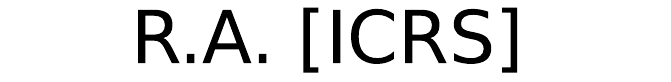}
        \caption{The 1.33~mm continuum emission maps of the SOMA MM sample obtained with ALMA at $\sim225.7~$GHz.
        Contour levels start at 3$\sigma$ (1$\sigma$ level is shown in Table~\ref{tab:observations} for each source) and steps are in 3$\sigma$.
        The red-dashed circles represent the aperture size from which we extracted the spectra.
        The black stars show the peaks of emission.
        The name of the region is shown in the upper-left corner of each map, and a letter is added as a label if more than one source is detected in the same region.
        The ellipses at the bottom-left corner of the images show the synthesized beam (see Table~\ref{tab:observations}), and a linear scale is displayed at the bottom right.}
        \label{fig:continuum}
    \end{figure*}

The dust continuum maps, gathered with ACA set-up at the frequency of $225.7$~GHz (1.33~mm), provide an overview of the shape of the sources and their environment. Fig.~\ref{fig:continuum} shows the 1.33 mm continuum maps of the 20 high-mass star-forming regions. We detected 22 continuum sources in total. The regions named AFGL 5180 and G061.48 are the only two where we identified two sources in the FoV of our data. 

\subsection{Detection summary of molecular species}\label{sect:detections}

Table~\ref{tab:detectionsummary} summarizes the species detection for the whole analysis: it shows the molecular species that have been detected at a significance level of $3\sigma$ for at least one source and shows which species have been detected for each source. A species is considered detected if at least one well-fitted, uncontaminated line is observed above the $3\sigma$ threshold. We present all the ACA+TP spectra in Appendix~\ref{app:spectra_SOMA}. Figure~\ref{fig:stats} displays the detection statistics: we detected 35 molecular species in total, with 21 main isotopologues and 14 rarer isotopologues (e.g., D, $^{13}$C, $^{18}$O, $^{15}$N, $^{34}$S). We identified seven relatively line-rich sources, G337.40, G318.95, G309.92, G305.80, G305.20, G010.62, and OMC1-S, characterized by the detection of $\gtrsim100$ transitions in the spectral windows of the sample.
            
Lines exhibiting complex profiles, such as non-Gaussian shapes, (self-)absorption features, or high-velocity components, were not fitted, since reliable estimation of the fit parameters is not achievable in these cases (in Table~\ref{tab:detectionsummary} these lines are labeled). Some sources show multiple velocity components: 2 components for G45.47 (H$_2$CO), G58.77 ($^{13}$CO), and NGC~2071 (C$^{18}$O, H$_2$CO, CH$_3$OH, SO, SiO); 3 components for G61.48 b ($^{13}$CO, C$^{18}$O, H$_2$CO, SO). However, these different components remain spatially unresolved in the current observations, considering the synthesized beam size of the observations. Hence, a single beam is used to extract the spectra. Therefore, for these emissions, the parameters obtained from \textsc{MADCUBA} are averaged over the number of different features observed. These features can be related to different sub-sources inside the beam or distinct velocity components of the same source (i.e., outflows), and they have been distinguished from a self-absorption effect by assessing the consistency of the $v_\text{LSR}$ of the component through the different molecules for each source, or the association of a double-component emission to an outflow.
            
Recombination lines have been observed in the sample: H$_{30}\alpha$ has been detected in G10.62, G11.94, G12.81, G45.47, G49.37, G61.48 a, G61.48 b, G317.40, OMC1-S, and W51 e2; He$_{30}\alpha$ has been detected in a subsample of the previous list (e.g., G11.94, G12.81, W51 e2). The presence of recombination lines indicates the presence of ionized gas, which is evidence of more evolved sources, already harboring H\textsc{II} regions. The occurrence of recombination lines in our sample shows no dependence on $L_{\mathrm{bol}}$, $M_{\mathrm{env}}$, $m_{\star}$, or $L_{\mathrm{bol}}/M_{\mathrm{env}}$, as they are detected in sources spanning a broad range of these parameters.
The line width of the recombination lines is higher ($\sim2-20$ times) compared to that of the rotational emission of molecules. The $v_\text{LSR}$ of these lines is also generally different (from -11 to +8 km/s w.r.t. the average molecular $v_\text{LSR}$), indicating that these emissions trace gas with very different kinematics with respect to the molecular emission.

    \begin{figure}
                \centering
                \begin{subfigure}{\hsize}
                    \includegraphics[width=\hsize]{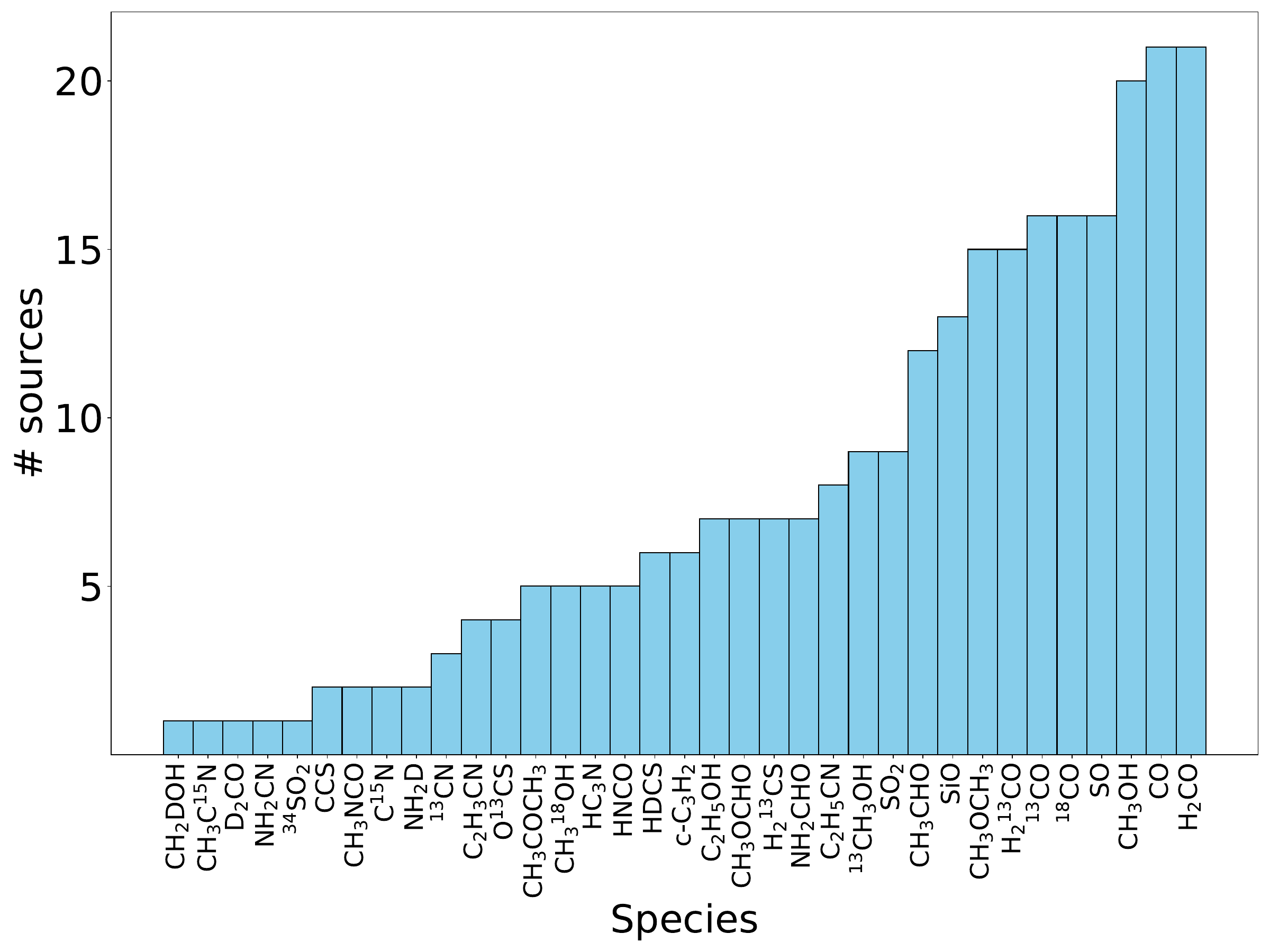}

                \end{subfigure}
                \begin{subfigure}{\hsize}
                    \includegraphics[width=\hsize]{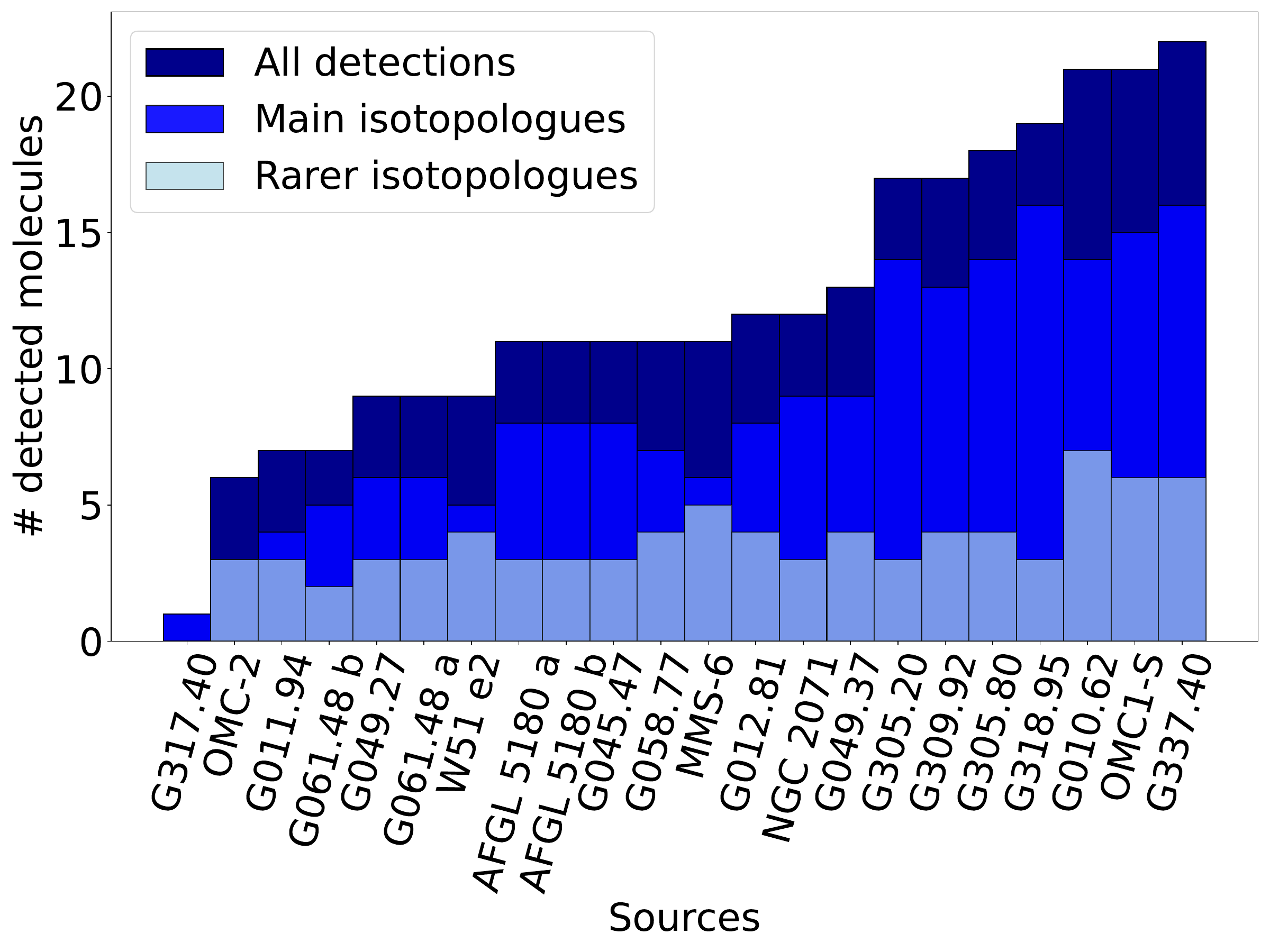}
                \end{subfigure}
                \begin{subfigure}{\hsize}
                    \includegraphics[width=\hsize]{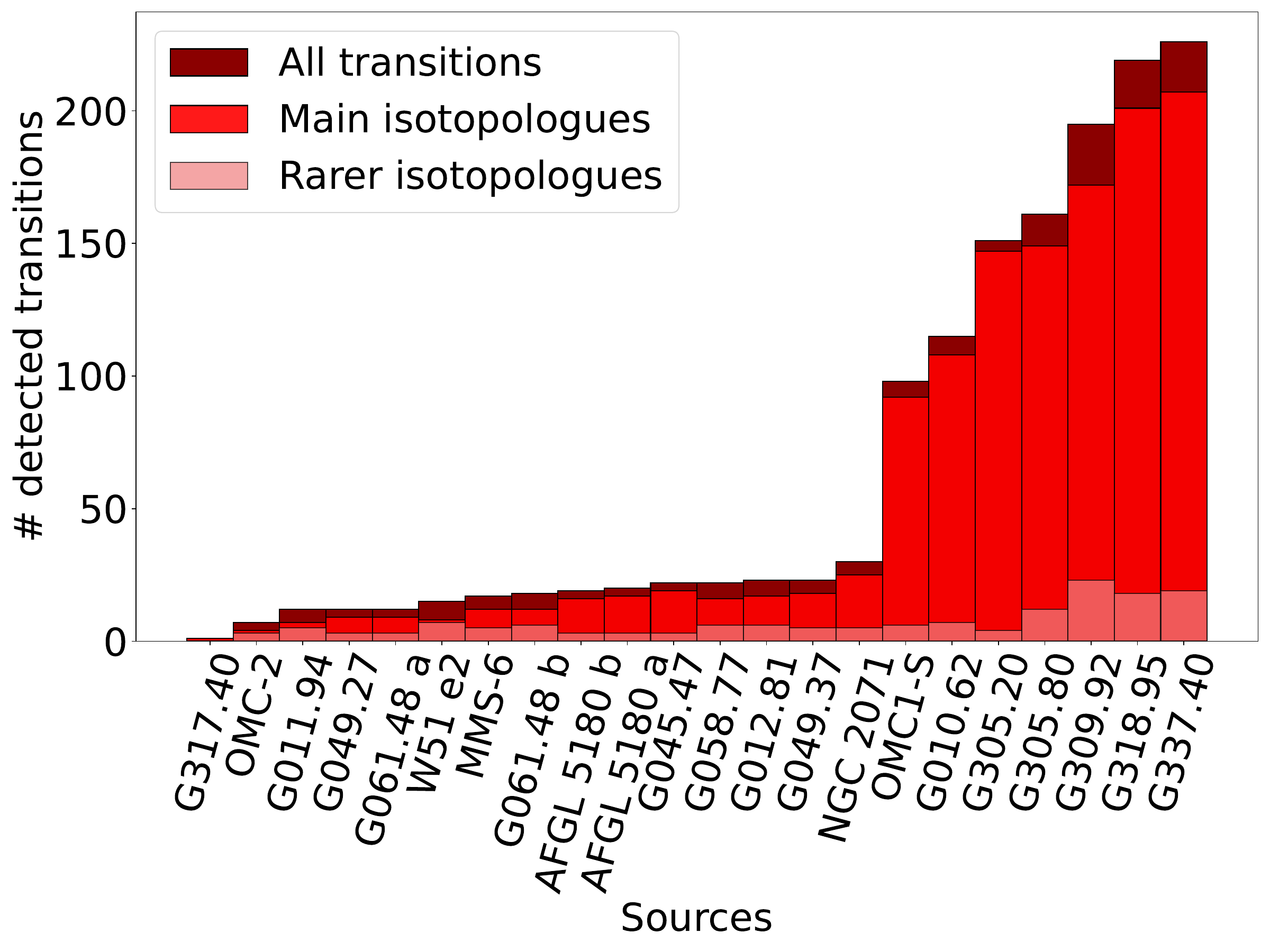}
                \end{subfigure}
                \caption{\textit{Top:} Number of sources with detection for each species over the 22 sources studied. \textit{Middle:} Number of detected molecules for each source. \textit{Bottom:} Number of detected transitions for each source, including separate values of the main and rarer isotopologues.}
                \label{fig:stats}
            \end{figure}

    \begin{sidewaystable*}
                \centering
                \small
                \setlength{\tabcolsep}{3pt}
                \caption{Summary of the detected species observed in the sample.}
                \label{tab:detectionsummary}
                \begin{tabular}{l@{\hspace{0.5cm}}cccccccccccccccccccccccccccccccccccccccccc} 
                \hline\hline
                \noalign{\smallskip}
                Source & \multicolumn{15}{c}{O-bearing molecules} & \multicolumn{1}{c}{} & \multicolumn{11}{c}{N-bearing molecules} & \multicolumn{1}{c}{} & \multicolumn{7}{c}{S-bearing molecules} & \multicolumn{1}{c}{} & \multicolumn{1}{c}{Si-bearing} & \multicolumn{1}{c}{} & Carbon & \multicolumn{1}{c}{} & \multicolumn{2}{c}{Recomb.}\\
                & & & & & & & & & & & & & & & & & & & & & & & & & & & & & & & & & & & & & \multicolumn{1}{c}{ molecules} & & chains & & \multicolumn{2}{c}{lines}\\
                \cline{2-16}\cline{18-28}\cline{30-36} \cline{38-38}\cline{40-40}\cline{42-43}
                \noalign{\smallskip}
                & \rotatebox{90}{CO} & \rotatebox{90}{$^{13}$CO} & \rotatebox{90}{C$^{18}$O} & \rotatebox{90}{H$_2$CO} & \rotatebox{90}{H$_2^{13}$CO} & \rotatebox{90}{D$_2$CO} & \rotatebox{90}{CH$_3$OH} & \rotatebox{90}{$^{13}$CH$_3$OH} & \rotatebox{90}{CH$_2$DOH} & \rotatebox{90}{CH$_3$$^{18}$OH} & \rotatebox{90}{CH$_3$CHO} & \rotatebox{90}{CH$_3$OCHO} & \rotatebox{90}{CH$_3$OCH$_3$} & \rotatebox{90}{CH$_3$COCH$_3$} & \rotatebox{90}{C$_2$H$_5$OH} & \multicolumn{1}{c}{} & \rotatebox{90}{$^{13}$CN} & \rotatebox{90}{C$^{15}$N} & \rotatebox{90}{HNCO} & \rotatebox{90}{CH$_3$C$^{15}$N} & \rotatebox{90}{C$_2$H$_3$CN} & \rotatebox{90}{C$_2$H$_5$CN} & \rotatebox{90}{CH$_3$NCO} & \rotatebox{90}{NH$_2$CHO} & \rotatebox{90}{NH$_2$CN} & \rotatebox{90}{NH$_2$D} & \rotatebox{90}{HC$_3$N} & \multicolumn{1}{c}{} & \rotatebox{90}{SO} & \rotatebox{90}{SO$_2$} & \rotatebox{90}{$^{34}$SO$_2$} & \rotatebox{90}{CCS} & \rotatebox{90}{O$^{13}$CS} & \rotatebox{90}{HDCS} & \rotatebox{90}{H$_2^{13}$CS} & \multicolumn{1}{c}{} & \rotatebox{90}{SiO} & \multicolumn{1}{c}{} & \rotatebox{90}{\textit{c}-C$_3$H$_2$} & \multicolumn{1}{c}{} & \rotatebox{90}{H$_{30}\alpha$} & \rotatebox{90}{He$_{30}\alpha$}\\
                \noalign{\smallskip}
                \hline
                \noalign{\smallskip}
                AFGL 5180 a & Y\tablefootmark{a} & Y\tablefootmark{a} & Y & Y & Y & n & Y & n & n & n & Y & n & Y & n & n & \multicolumn{1}{c}{} & n & n & n & n & n & n & n & n & n &  &  & \multicolumn{1}{c}{} & Y & n & n & n & n & Y & n & \multicolumn{1}{c}{} & Y & \multicolumn{1}{c}{} & n & \multicolumn{1}{c}{} & n & n\\
                AFGL 5180 b & Y\tablefootmark{a} & Y\tablefootmark{a} & Y & Y & Y & n & Y & n & n & n & Y & n & Y & n & n & \multicolumn{1}{c}{} & n & n & n & n & n & n & n & n & n &  &  & \multicolumn{1}{c}{} & Y & n & n & n & n & Y & n & \multicolumn{1}{c}{} & Y & \multicolumn{1}{c}{} & n & \multicolumn{1}{c}{} & n & n\\
                G010.62 & Y\tablefootmark{a} & Y\tablefootmark{a} & Y & Y & Y & n & Y & Y & n & n & Y & Y & Y & n & Y & \multicolumn{1}{c}{} & n & n & n & n & Y & Y & n & Y & n &  &  & \multicolumn{1}{c}{} & Y & Y & Y & Y & Y & n & Y & \multicolumn{1}{c}{} & Y & \multicolumn{1}{c}{} & n & \multicolumn{1}{c}{} & Y & n\\
                G011.94 & Y\tablefootmark{a} & Y\tablefootmark{a} & Y\tablefootmark{a} & Y\tablefootmark{a} & n & n & Y & n & n & n & n & n & n & n & n & \multicolumn{1}{c}{} & Y & n & n & n & n & n & n & n & n &  &  & \multicolumn{1}{c}{} & Y & n & n & n &  & n & n & \multicolumn{1}{c}{} & n & \multicolumn{1}{c}{} & n & \multicolumn{1}{c}{} & Y & Y\\
                G012.81 & Y\tablefootmark{a} & Y\tablefootmark{a} & Y\tablefootmark{a} & Y\tablefootmark{a} & n & n & Y & Y & n & n & Y & n & Y & n & Y & \multicolumn{1}{c}{} & Y & n & n & n & n & n & n & n & n &  &  & \multicolumn{1}{c}{} & Y\tablefootmark{a} & n & n & n & n & n & n & \multicolumn{1}{c}{} & Y & \multicolumn{1}{c}{} & n & \multicolumn{1}{c}{} & Y & Y\\
                G045.47 & Y\tablefootmark{a} & Y\tablefootmark{a} & Y & Y\tablefootmark{b} & Y & n & Y & n & n & n & n & n & Y & n & n & \multicolumn{1}{c}{} & n & n & n & n & n & n & n & n & n &  &  & \multicolumn{1}{c}{} & Y & Y & n & n & n & Y & n & \multicolumn{1}{c}{} & Y & \multicolumn{1}{c}{} & n & \multicolumn{1}{c}{} & Y & n\\
                G049.27 & Y\tablefootmark{a} & Y\tablefootmark{a} & Y & Y & Y & n & Y & n & n & n & n & n & n & n & n & \multicolumn{1}{c}{} & n & n & n & n & n & n & n & n & n &  &  & \multicolumn{1}{c}{} & Y & Y & n & n & n & n & n & \multicolumn{1}{c}{} & Y & \multicolumn{1}{c}{} & n & \multicolumn{1}{c}{} & n & n\\
                G049.37 & Y\tablefootmark{a} & Y\tablefootmark{a} & Y\tablefootmark{a} & Y & Y & n & Y & n & n & n & Y & n & Y & n & n & \multicolumn{1}{c}{} & n & Y & n & n & n & n & n & n & n &  &  & \multicolumn{1}{c}{} & Y & n & n & n & n & Y & n & \multicolumn{1}{c}{} & Y & \multicolumn{1}{c}{} & Y & \multicolumn{1}{c}{} & Y & n\\
                G058.77 & Y\tablefootmark{a} & Y & Y & Y\tablefootmark{b} & Y & n & Y & n & n & n & Y & n & Y & n & n & \multicolumn{1}{c}{} & n & Y & n & n & n & n & n & n & n & n &  & \multicolumn{1}{c}{} & Y & n & n & n & n & n & n & \multicolumn{1}{c}{} & Y & \multicolumn{1}{c}{} & n & \multicolumn{1}{c}{} & n & n\\
                G061.48 a & Y\tablefootmark{a} & Y & Y & Y & Y & n & Y & n & n & n & n & n & n & n & n & \multicolumn{1}{c}{} & n & n & n & n & n & n & n & n & n & n &  & \multicolumn{1}{c}{} & Y & n & n & n & n & n & n & \multicolumn{1}{c}{} & Y & \multicolumn{1}{c}{} & Y & \multicolumn{1}{c}{} & Y & n\\
                G061.48 b & Y\tablefootmark{a} & Y\tablefootmark{b} & Y\tablefootmark{b} & Y\tablefootmark{b} & n & n & Y & n & n & n & n & n & n & n & n & \multicolumn{1}{c}{} & n & n & n & n & n & n & n & n & n & n &  & \multicolumn{1}{c}{} & Y\tablefootmark{b} & n & n & n & n & n & n & \multicolumn{1}{c}{} & Y & \multicolumn{1}{c}{} & n & \multicolumn{1}{c}{} & Y & n\\
                G305.20 & Y\tablefootmark{a} &  &  & Y &  & n & Y & Y & n & Y & Y & Y & Y & Y & Y & \multicolumn{1}{c}{} & n & n & n & n & Y & Y & n & Y & n & n & Y & \multicolumn{1}{c}{} &  & Y & n & n &  & n & Y & \multicolumn{1}{c}{} &  & \multicolumn{1}{c}{} & Y & \multicolumn{1}{c}{} & n & n\\
                G305.80 & Y\tablefootmark{a} &  &  & Y & Y & n & Y & Y & n & Y & Y & Y & Y & Y & Y & \multicolumn{1}{c}{} & n & n & Y & n & Y & Y & n & Y & n & n & Y & \multicolumn{1}{c}{} &  & Y & n & n &  & n & Y & \multicolumn{1}{c}{} &  & \multicolumn{1}{c}{} & n & \multicolumn{1}{c}{} & n & n\\
                G309.92 &  &  &  & Y &  & n & Y & Y & n & Y & Y & Y & Y & Y & Y & \multicolumn{1}{c}{} & n & n & Y & n & n & Y & Y & Y & n & n & Y & \multicolumn{1}{c}{} &  & Y & n & n & Y & n & Y & \multicolumn{1}{c}{} &  & \multicolumn{1}{c}{} & n & \multicolumn{1}{c}{} & n & n\\
                G317.40 & Y\tablefootmark{a} &  &  & n & n & n & n & n & n & n & n & n & n & n & n & \multicolumn{1}{c}{} & n & n & n & n & n & n & n & n & n & n &  & \multicolumn{1}{c}{} &  & n & n & n & n & n & n & \multicolumn{1}{c}{} &  & \multicolumn{1}{c}{} & n & \multicolumn{1}{c}{} & Y & n\\
                G318.95 & Y\tablefootmark{a} &  &  & Y &  & n & Y & Y & Y & Y & Y & Y & Y & Y & Y & \multicolumn{1}{c}{} & n & n & Y & n & n & Y & n & Y & Y & Y & Y & \multicolumn{1}{c}{} &  & Y & n & n &  & n & Y & \multicolumn{1}{c}{} &  & \multicolumn{1}{c}{} & n & \multicolumn{1}{c}{} & n & n\\
                G337.40 & Y\tablefootmark{a} &  &  & Y & Y & n & Y & Y & n & Y & Y & Y & Y & Y & Y & \multicolumn{1}{c}{} & n & n & Y & Y & Y & Y & Y & Y & n & Y & Y\tablefootmark{a} & \multicolumn{1}{c}{} &  & Y & n & n & Y & n & Y & \multicolumn{1}{c}{} &  & \multicolumn{1}{c}{} & n & \multicolumn{1}{c}{} & n & n\\
                MMS-6 & Y\tablefootmark{a} & Y & Y & Y & Y & Y & Y & Y & n & n & n & n & Y & n & n & \multicolumn{1}{c}{} & n & n & n & n & n & n & n & n & n & n &  & \multicolumn{1}{c}{} & Y & n & n & n & n & n & n & \multicolumn{1}{c}{} & n & \multicolumn{1}{c}{} & Y & \multicolumn{1}{c}{} & n & n\\
                NGC 2071 & Y\tablefootmark{a} & Y\tablefootmark{a} & Y\tablefootmark{b} & Y\tablefootmark{b} & Y & n & Y\tablefootmark{b} & n & n & n & n & n & Y & n & n & \multicolumn{1}{c}{} & n & n & n & n & n & Y & n & n & n & n &  & \multicolumn{1}{c}{} & Y\tablefootmark{b} & n & n & n & n & Y & n & \multicolumn{1}{c}{} & Y\tablefootmark{b} & \multicolumn{1}{c}{} & Y & \multicolumn{1}{c}{} & n & n\\
                OMC-2 & Y\tablefootmark{a} & Y & Y & Y & Y & n & n & n & n & n & n & n & n & n & n & \multicolumn{1}{c}{} & n & n & n & n & n & n & n & n & n & n &  & \multicolumn{1}{c}{} & Y & n & n & n & n & n & n & \multicolumn{1}{c}{} & n & \multicolumn{1}{c}{} & n & \multicolumn{1}{c}{} & n & n\\
                OMC1-S & Y\tablefootmark{a} & Y\tablefootmark{a} & Y\tablefootmark{a} & Y & Y & n & Y & Y & n & n & Y & Y & Y & n & n & \multicolumn{1}{c}{} & n & n & Y & n & n & Y & n & Y & n & n &  & \multicolumn{1}{c}{} & Y & Y & n & Y & Y & Y & Y & \multicolumn{1}{c}{} & Y & \multicolumn{1}{c}{} & Y & \multicolumn{1}{c}{} & Y & n\\
                W51 e2 & Y\tablefootmark{a} & Y\tablefootmark{a} & Y\tablefootmark{a} & Y\tablefootmark{a} & Y & n & Y & n & n & n & n & n & n & n & n & \multicolumn{1}{c}{} & Y & n & n & n & n & n & n & n & n & n &  & \multicolumn{1}{c}{} & Y\tablefootmark{a} & n & n & n & n & n & n & \multicolumn{1}{c}{} & Y & \multicolumn{1}{c}{} & n & \multicolumn{1}{c}{} & Y & Y\\
                \noalign{\smallskip}
                \cline{2-16}\cline{18-28}\cline{30-36} \cline{38-38}\cline{40-40}\cline{42-43}
                \noalign{\smallskip}
                & \rotatebox{90}{21/21} & \rotatebox{90}{16/16} & \rotatebox{90}{16/16} & \rotatebox{90}{21/22} & \rotatebox{90}{15/19} & \rotatebox{90}{1/22} & \rotatebox{90}{20/22} & \rotatebox{90}{9/22} & \rotatebox{90}{1/22} & \rotatebox{90}{5/22} & \rotatebox{90}{12/22} & \rotatebox{90}{7/22} & \rotatebox{90}{15/22} & \rotatebox{90}{5/22} & \rotatebox{90}{7/22} & \multicolumn{1}{c}{} & \rotatebox{90}{3/22} & \rotatebox{90}{2/22} & \rotatebox{90}{5/22} & \rotatebox{90}{1/22} & \rotatebox{90}{4/22} & \rotatebox{90}{8/22} & \rotatebox{90}{2/22} & \rotatebox{90}{7/22} & \rotatebox{90}{1/22} & \rotatebox{90}{2/14} & \rotatebox{90}{5/5} & \multicolumn{1}{c}{} & \rotatebox{90}{16/16} & \rotatebox{90}{9/22} & \rotatebox{90}{1/22} & \rotatebox{90}{2/22} & \rotatebox{90}{4/18} & \rotatebox{90}{6/22} & \rotatebox{90}{7/22} & \multicolumn{1}{c}{} & \rotatebox{90}{13/16} & \multicolumn{1}{c}{} & \rotatebox{90}{6/22} & \multicolumn{1}{c}{} & \rotatebox{90}{10/22} & \rotatebox{90}{3/22}\\
                \noalign{\smallskip}
                \hline
                \end{tabular}
                \tablefoot{
                    The species with emission above the significance level of $3\sigma$ are labeled with "Y", while the ones under that level are labeled with "n". The boxes are left empty for those species with no transitions in the spectral windows of the source.\\
                    \tablefoottext{a}{The line shows non-Gaussian features (i.e., self-absorbed lines, high-velocity wings), and the fit with \textsc{MADCUBA} is not performed.}
                    \tablefoottext{b}{Multiple velocity features present (fitted with multiple Gaussians).}
                }
            \end{sidewaystable*}
        
\subsection{Correlations between molecular species}

In this section, we investigate the correlation between column densities and line widths, derived from the line fitting analysis (Sect.~\ref{sect:lineidentification}), for different molecular species. Strong correlations and high Pearson coefficients are expected for pairs of species with a close chemical link. For column densities, such correlations may arise not only from direct chemical relationships, but also from shared formation timescales or similar temperature conditions within the source. Correlations in line widths may indicate the type of kinematic component, e.g., infall envelope, disk, or outflow, where a given species resides.

\begin{figure*}
                \centering
                \begin{minipage}{0.025\textwidth}
                    \centering
                    \includegraphics[height=0.09\textheight]{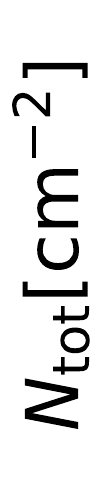} 
                \end{minipage}%
                \begin{minipage}{0.97\textwidth}
                    \begin{subfigure}{0.247\textwidth}
                        \centering
                        \includegraphics[width=\textwidth]{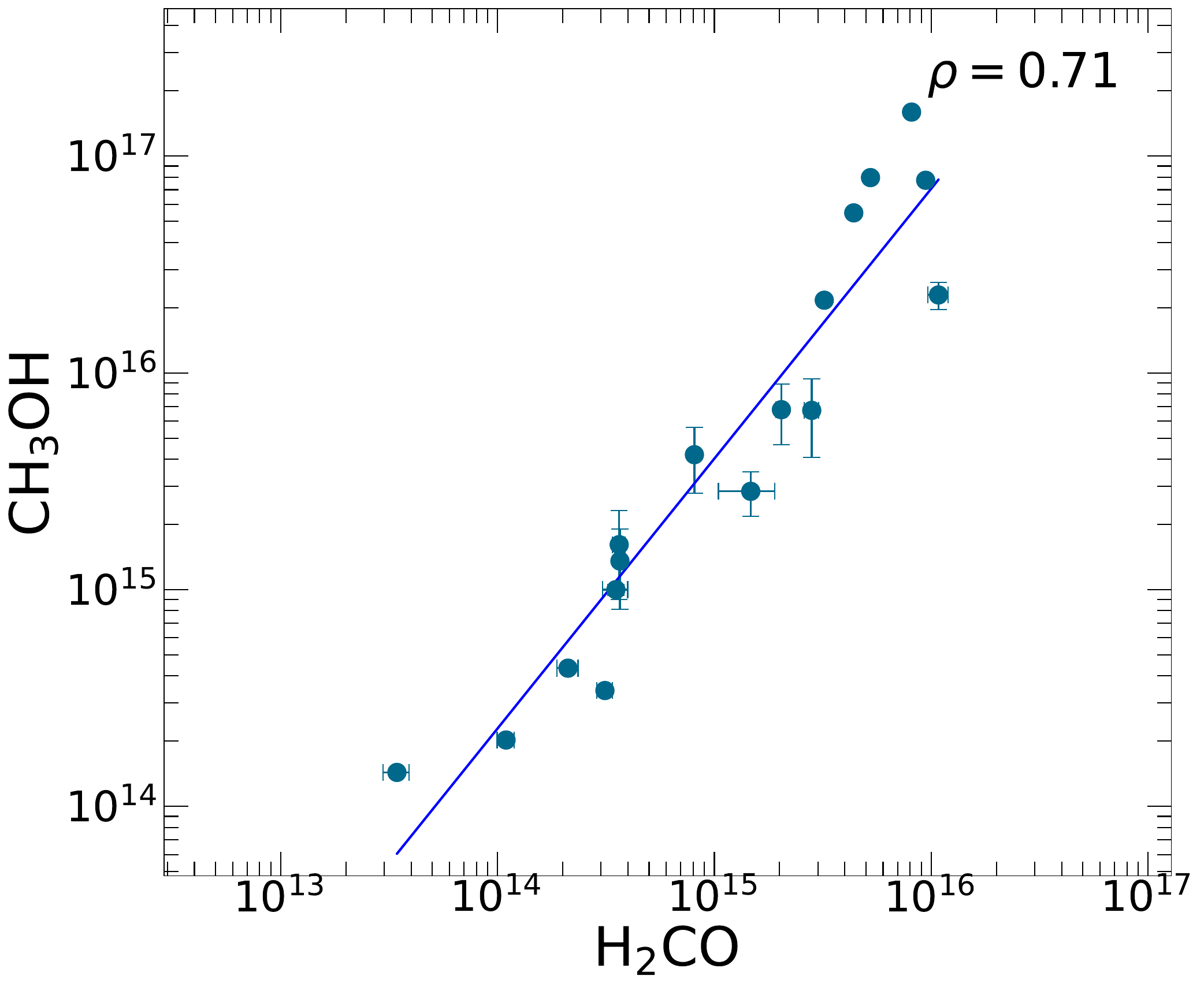}
                    \end{subfigure}
                    \begin{subfigure}{0.247\textwidth}
                        \centering
                        \includegraphics[width=\textwidth]{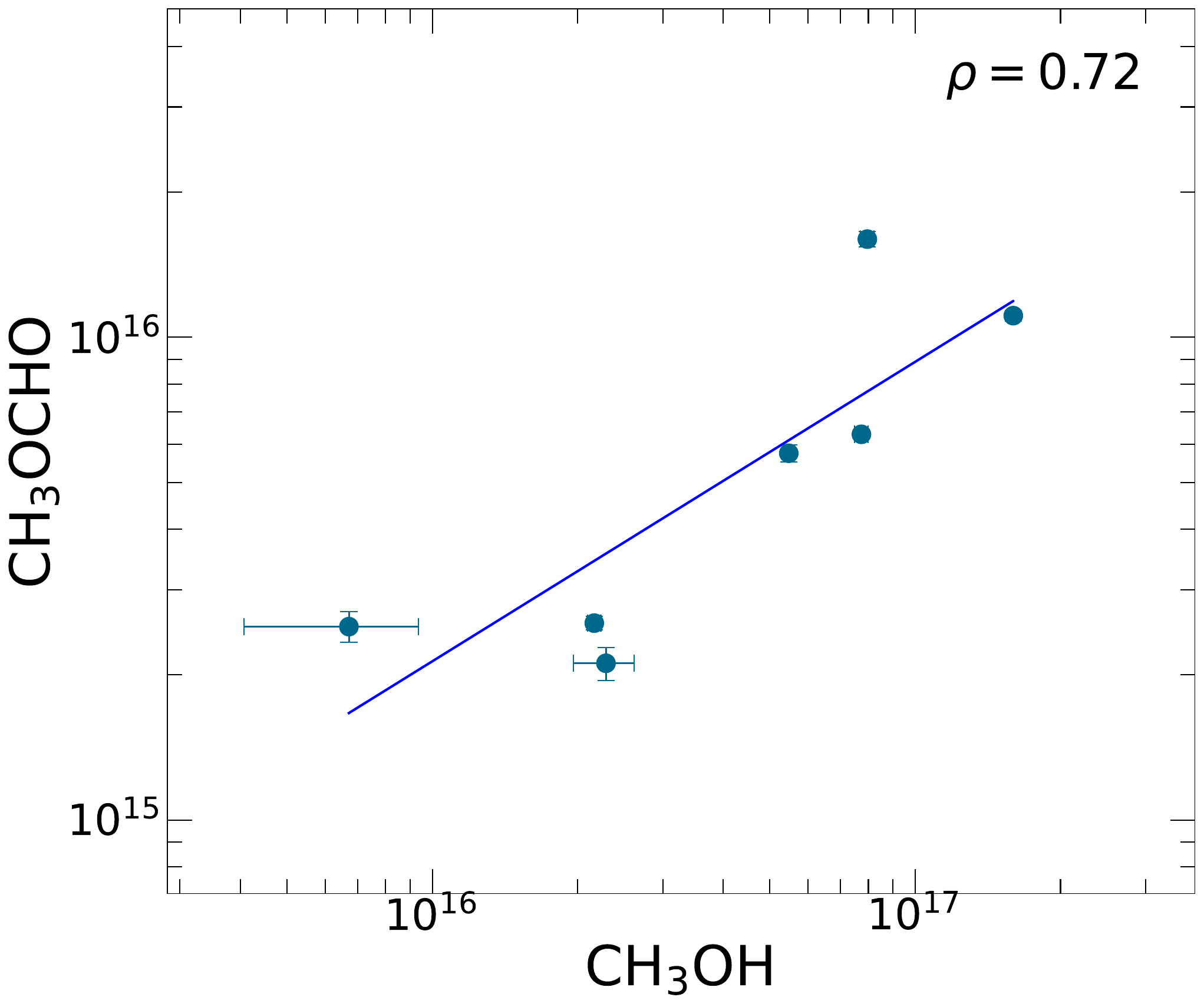}
                    \end{subfigure}
                    \begin{subfigure}{0.247\textwidth}
                        \centering
                        \includegraphics[width=\textwidth]{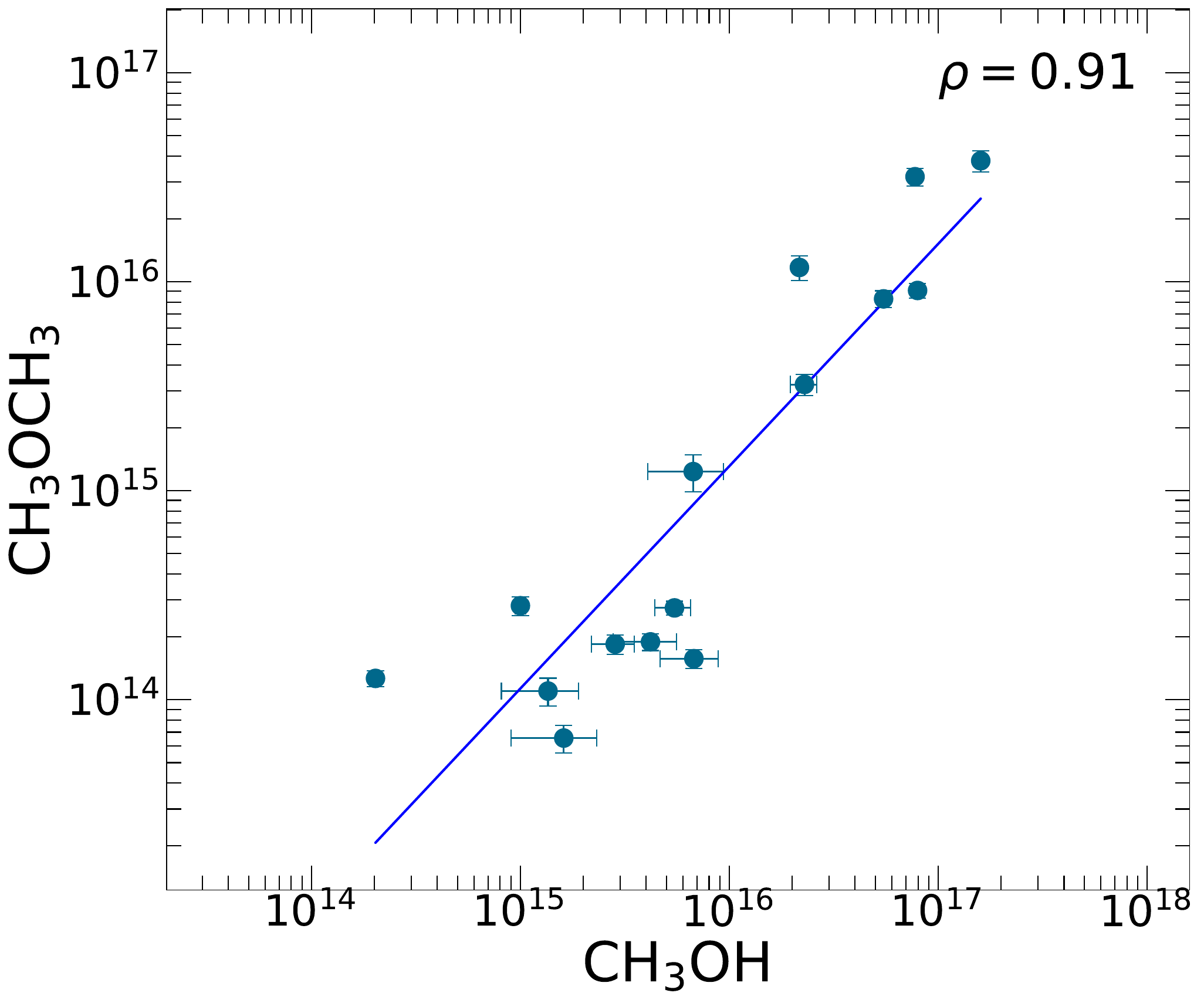}
                    \end{subfigure}
                    \begin{subfigure}{0.247\textwidth}
                        \centering
                        \includegraphics[width=\textwidth]{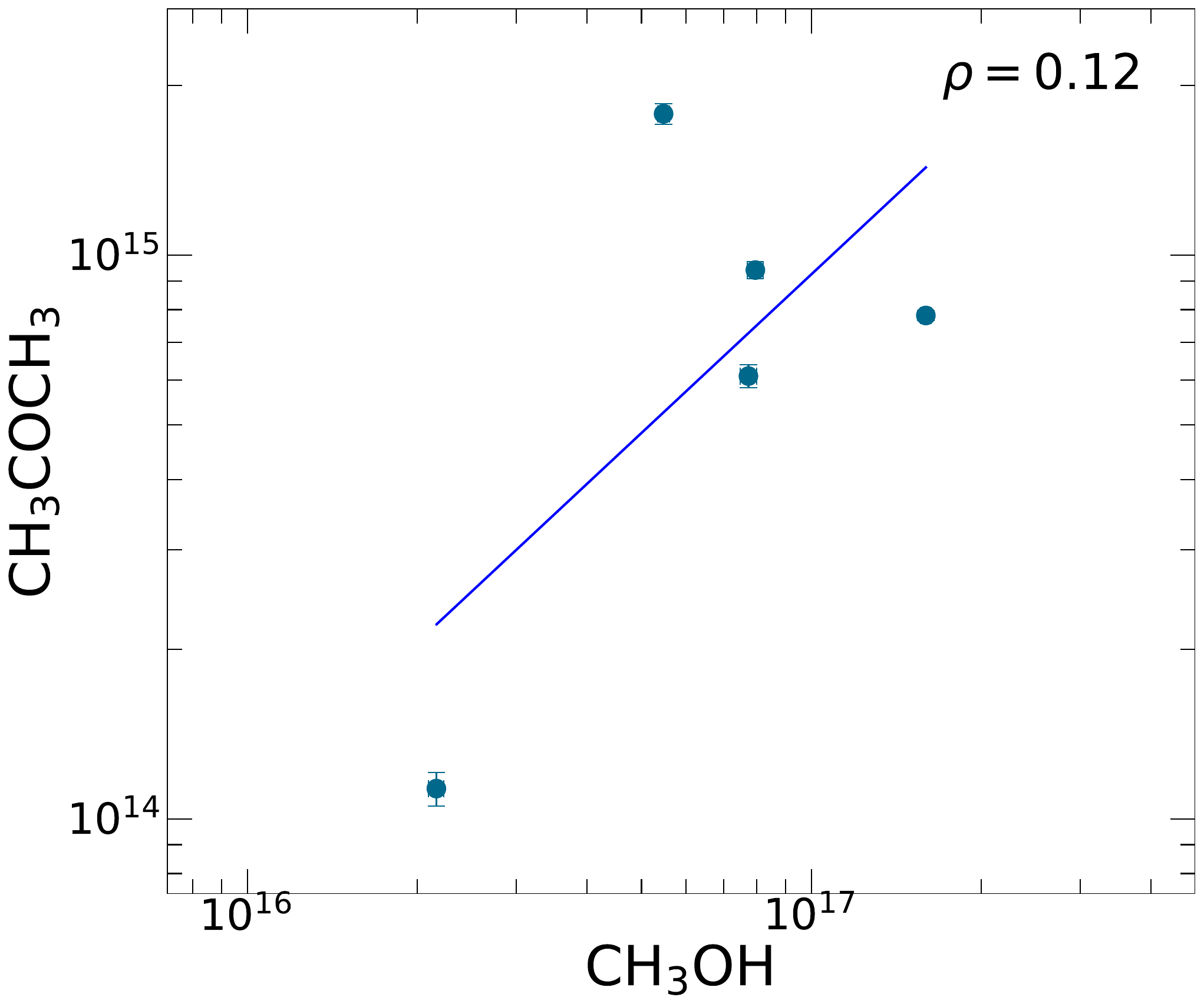}
                    \end{subfigure}
                    
                    \begin{subfigure}{0.247\textwidth}
                        \centering
                        \includegraphics[width=\textwidth]{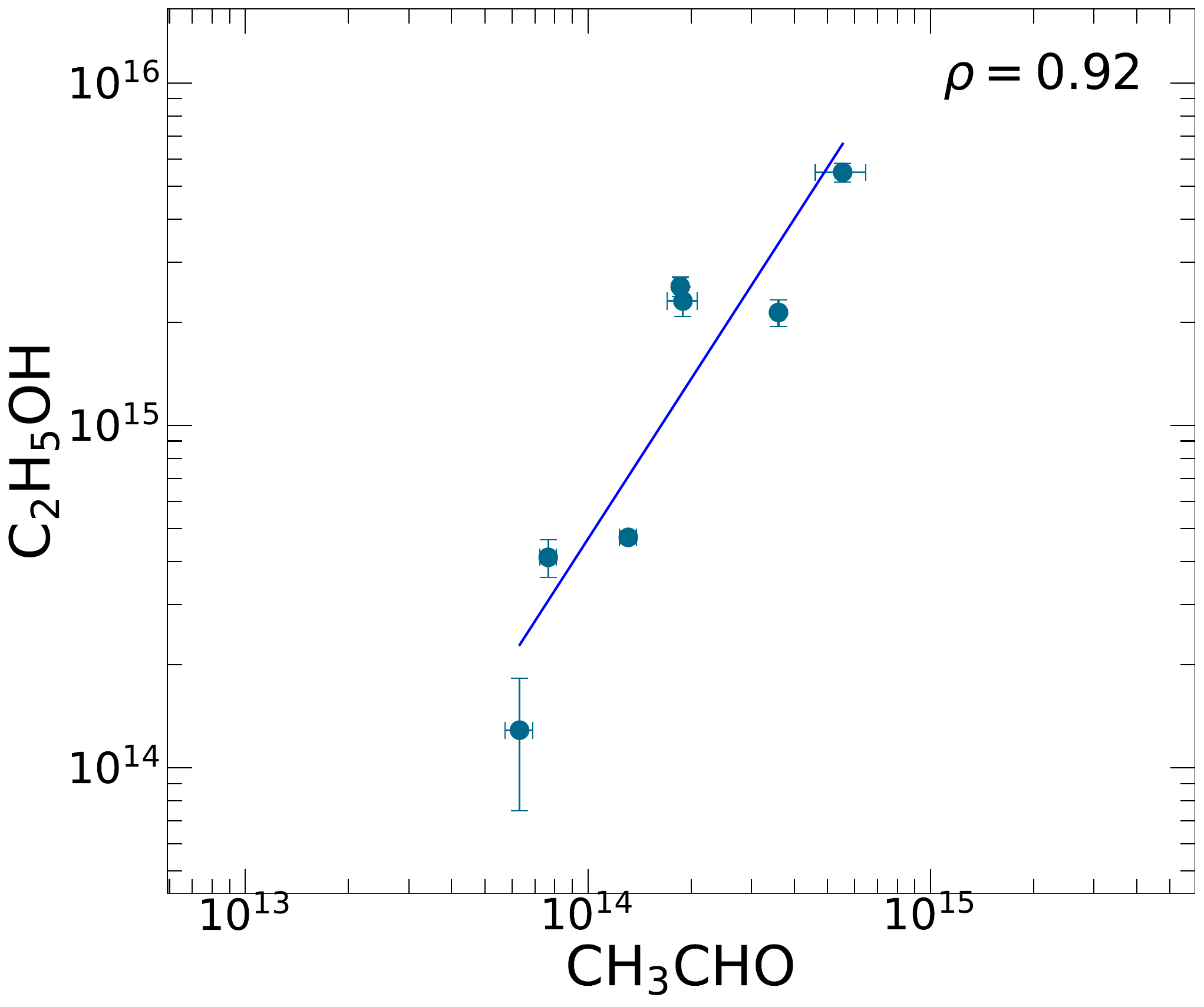}
                    \end{subfigure}
                    \begin{subfigure}{0.247\textwidth}
                        \centering
                        \includegraphics[width=\textwidth]{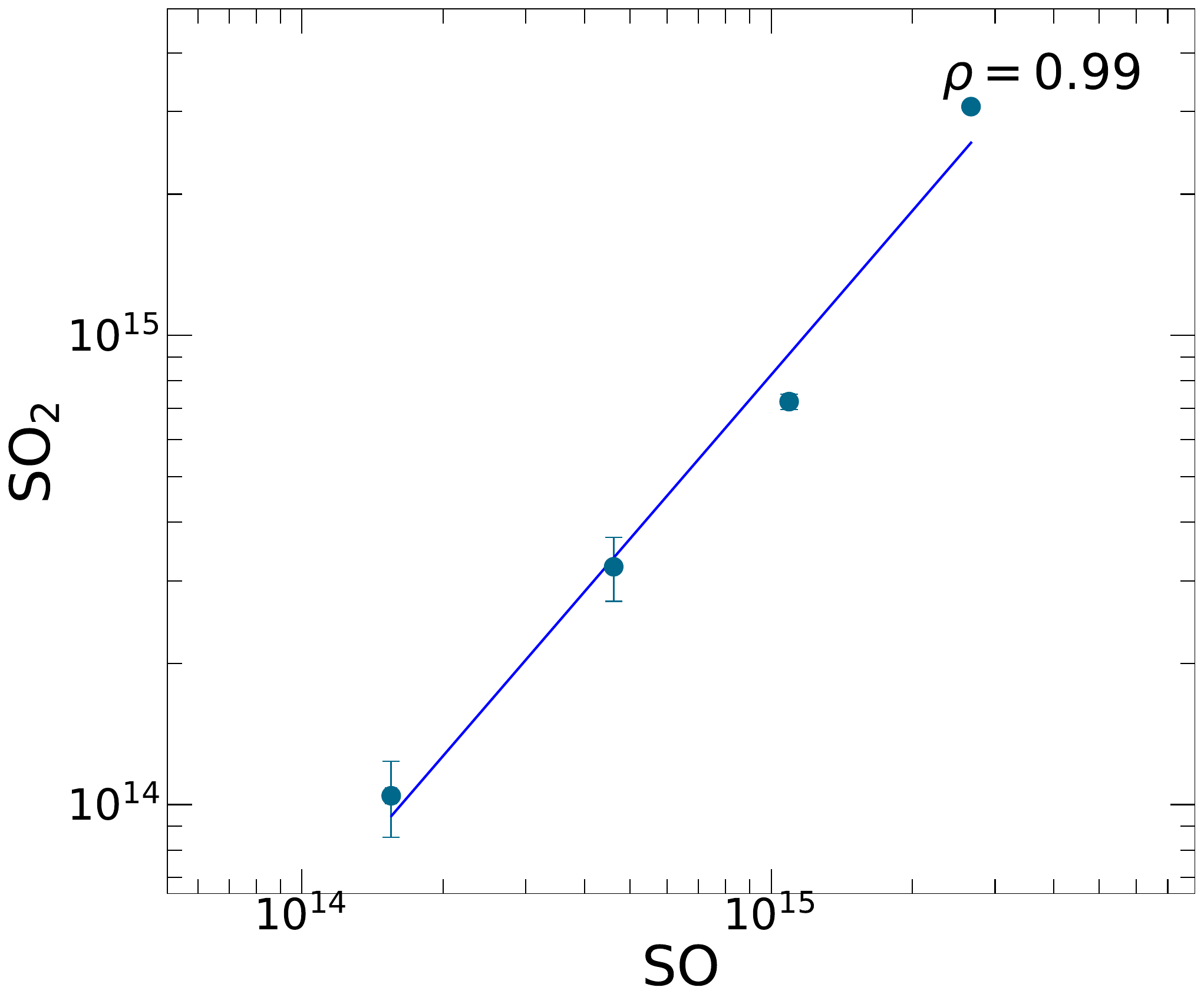}
                    \end{subfigure}
                    \begin{subfigure}{0.247\textwidth}
                        \centering
                        \includegraphics[width=\textwidth]{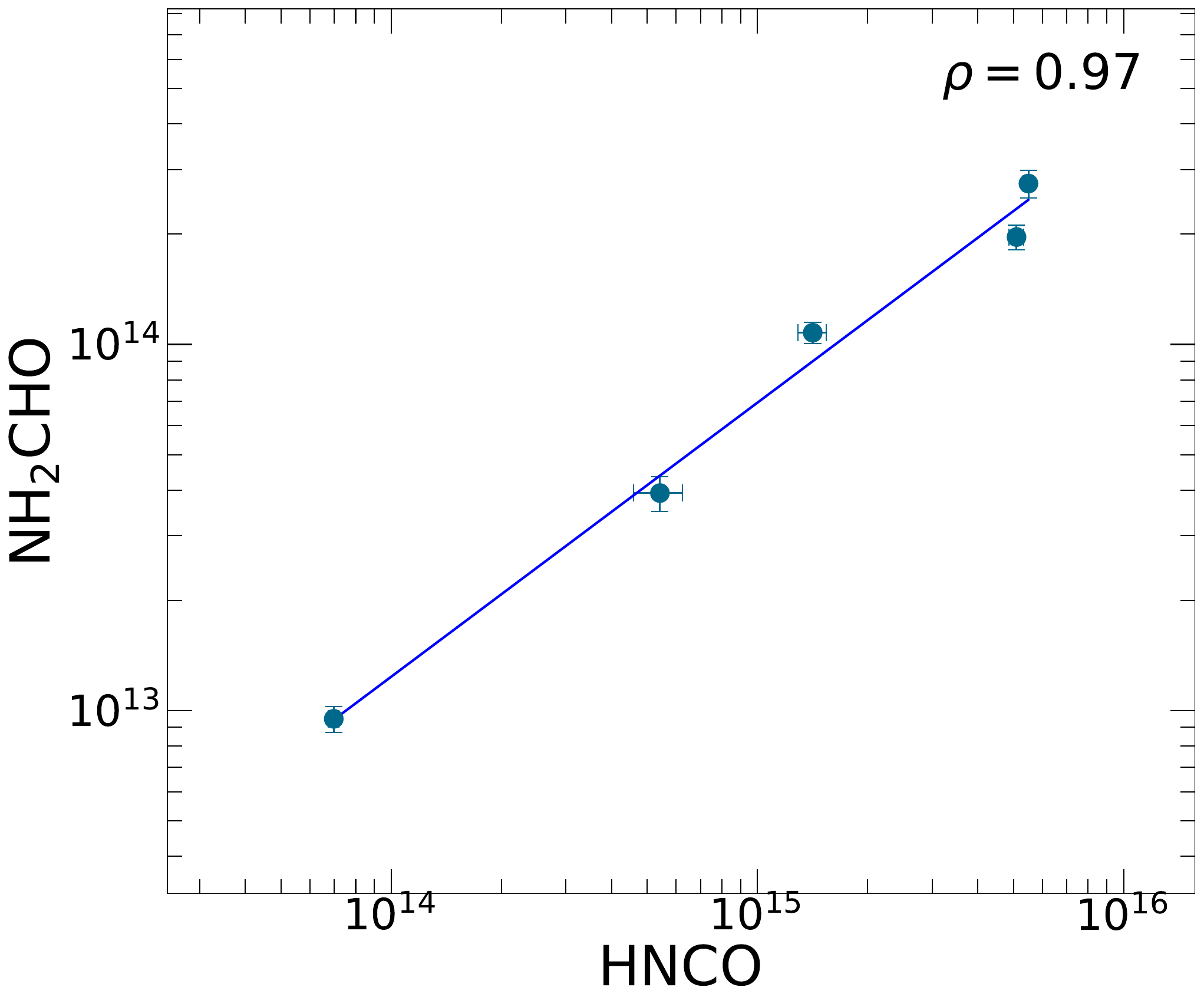}
                    \end{subfigure}
                    \begin{subfigure}{0.247\textwidth}
                        \centering
                        \includegraphics[width=\textwidth]{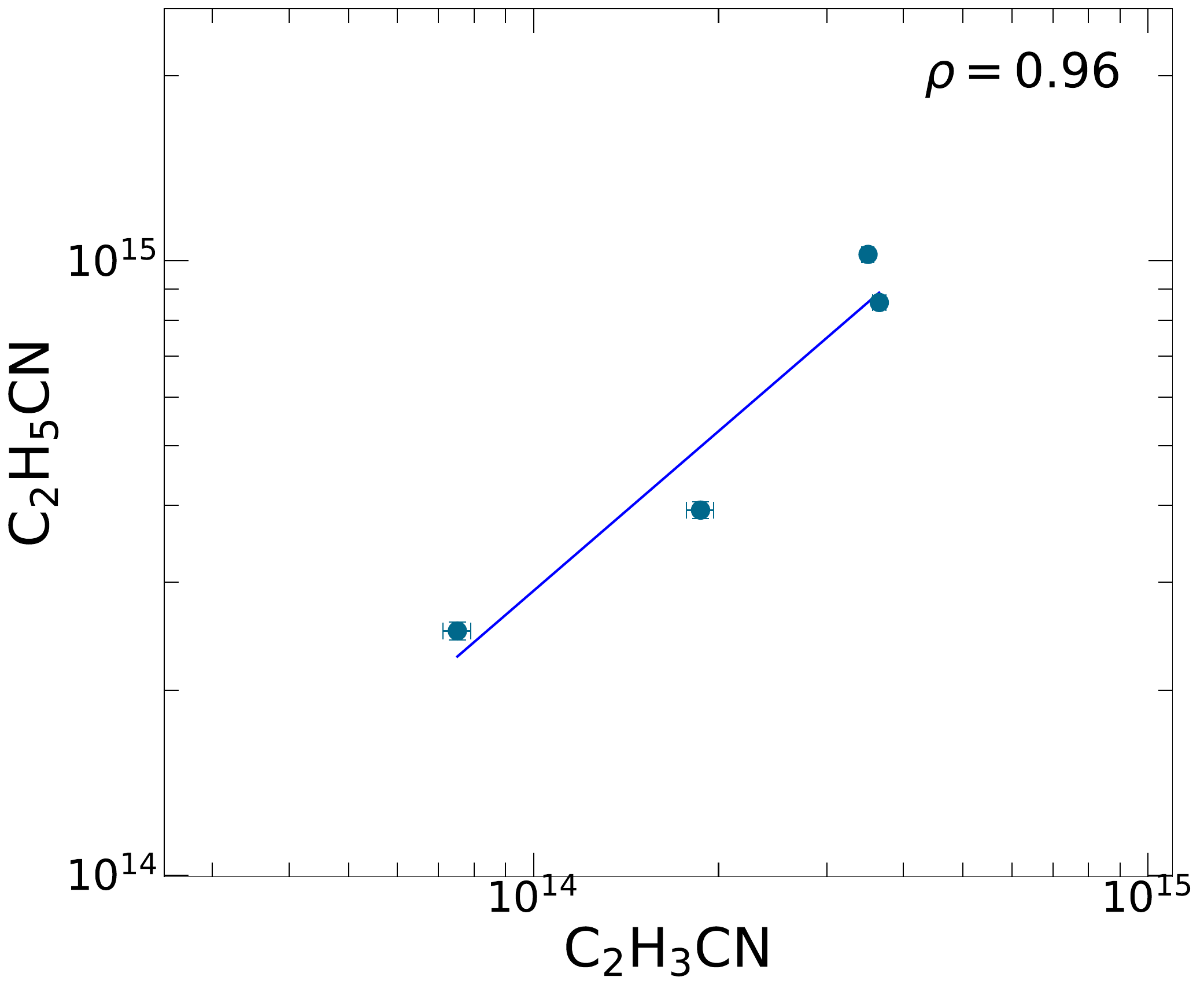}
                    \end{subfigure}
            
                    \begin{subfigure}{\textwidth}
                        \centering
                        \includegraphics[height=0.028\textwidth]{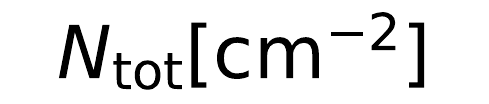}
                    \end{subfigure}
                \end{minipage}
                \caption{Correlation between the column densities ($N_\text{tot}$) of chemically linked pairs of molecules. A linear fit is drawn on the plot, and in the top-right corner, the Pearson correlation coefficient ($\rho$) for each dataset is displayed. Each data point represents a single targeted source.}
                \label{fig:denscorr}
            \end{figure*}

\begin{figure*}
                \centering
                \begin{minipage}{0.025\textwidth}
                    \centering
                    \includegraphics[height=0.12\textheight]{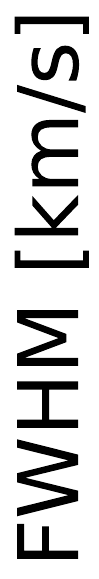}
                \end{minipage}%
                \begin{minipage}{0.97\textwidth}
                    \begin{subfigure}{0.247\textwidth}
                        \centering
                        \includegraphics[width=\textwidth]{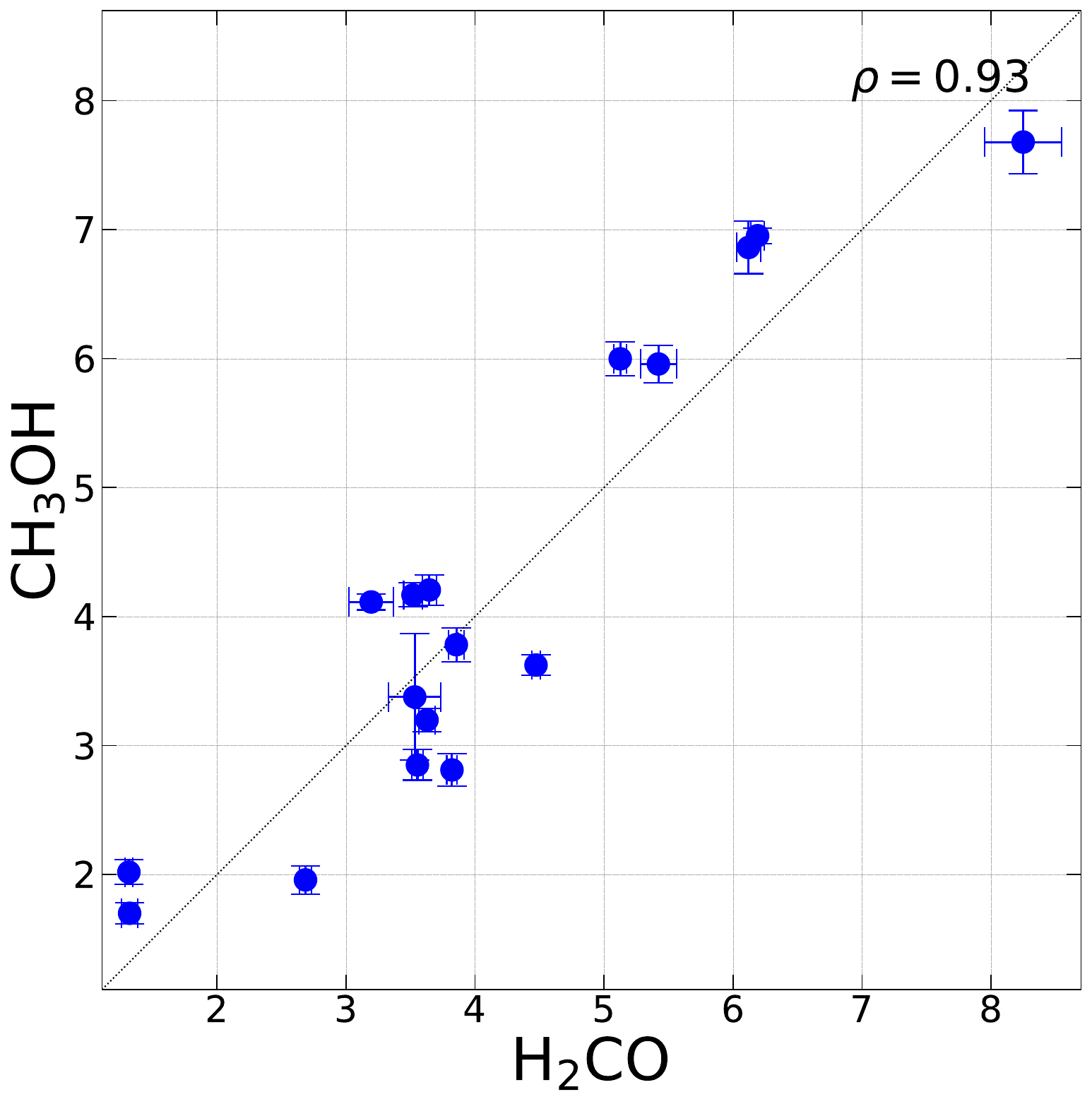}
                    \end{subfigure}
                    \begin{subfigure}{0.247\textwidth}
                        \centering
                        \includegraphics[width=\textwidth]{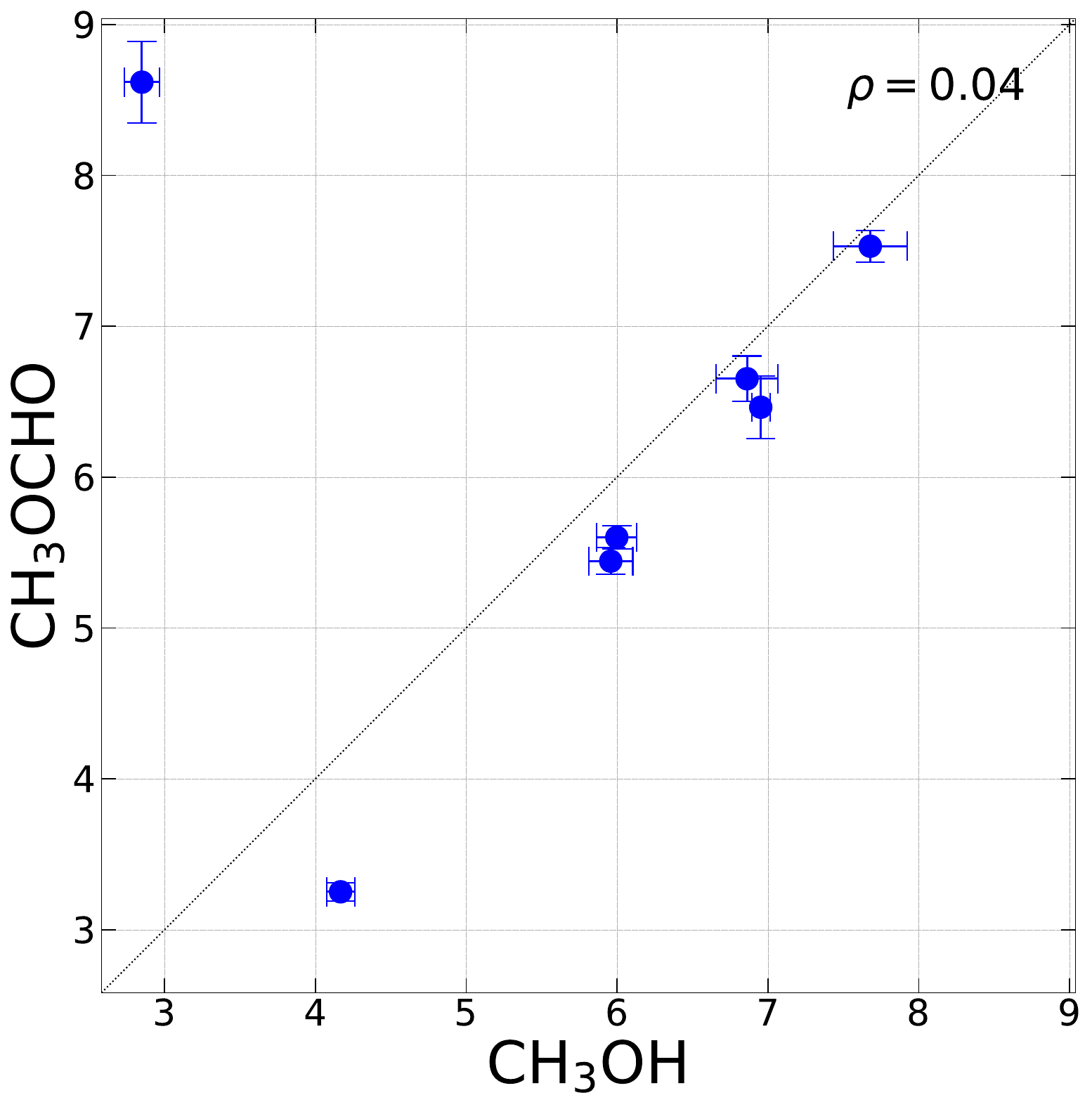}
                    \end{subfigure}
                    \begin{subfigure}{0.247\textwidth}
                        \centering
                        \includegraphics[width=\textwidth]{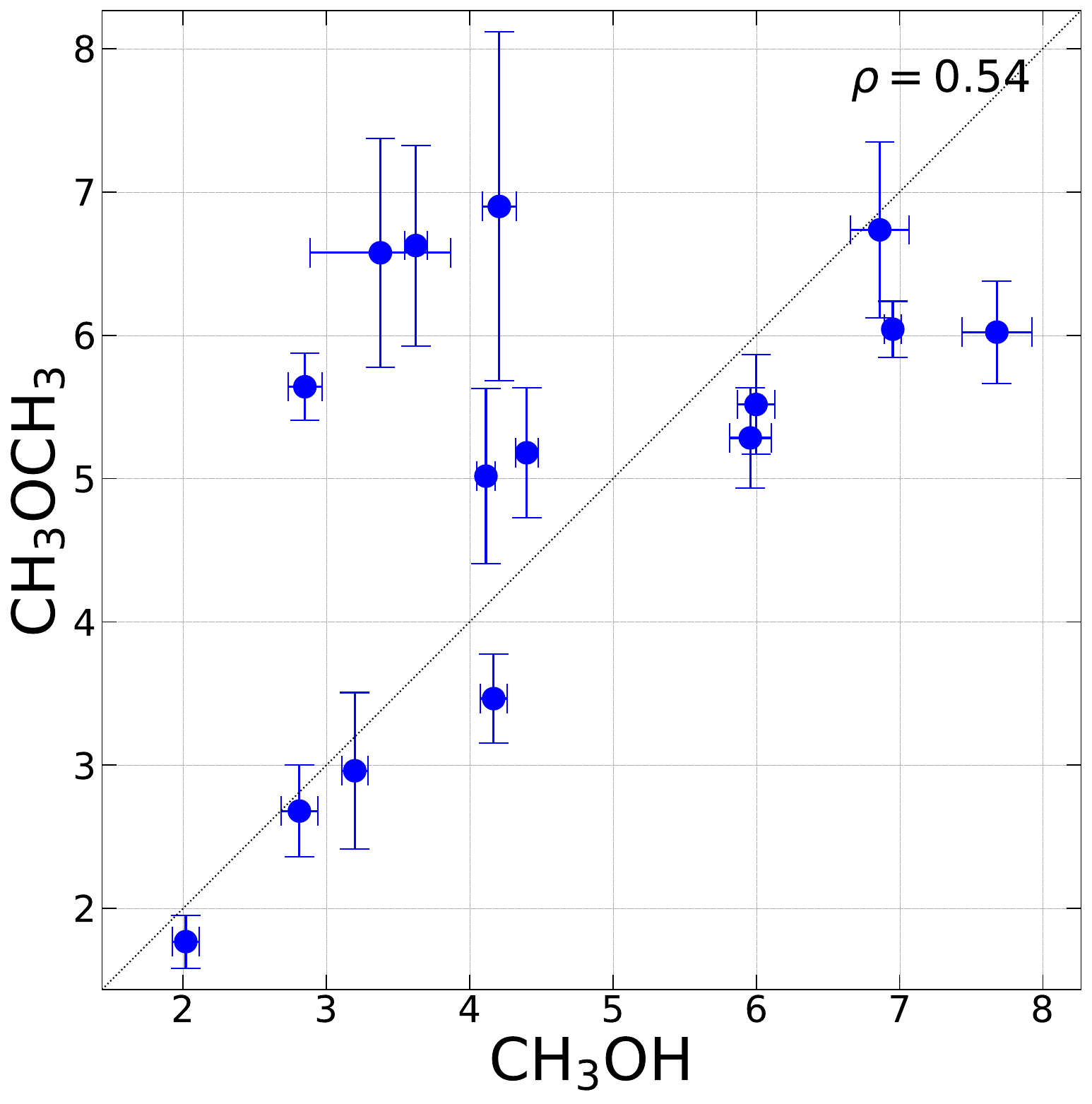}
                    \end{subfigure}
                    \begin{subfigure}{0.247\textwidth}
                        \centering
                        \includegraphics[width=\textwidth]{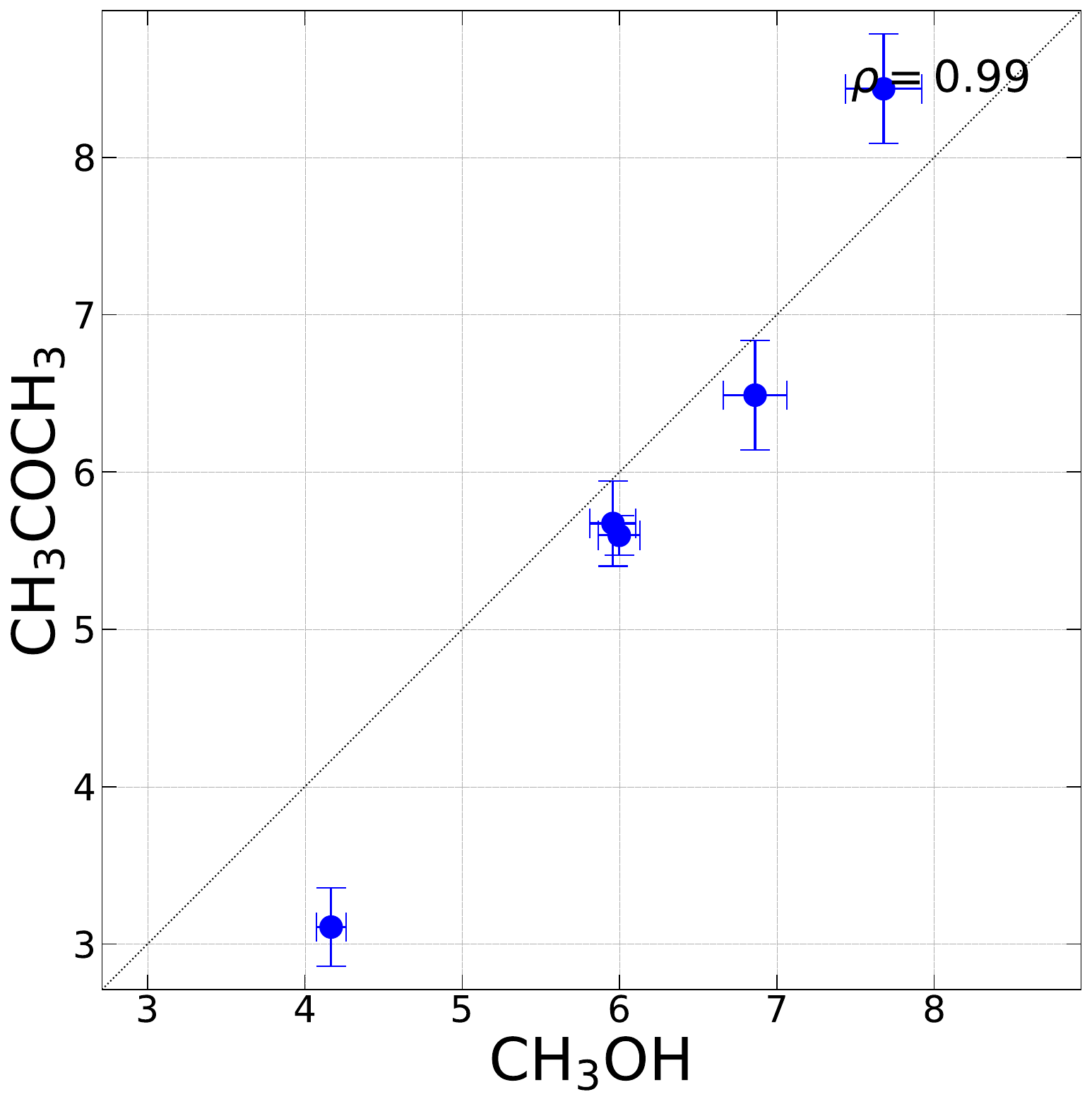}
                    \end{subfigure}
                    
                    \begin{subfigure}{0.247\textwidth}
                        \centering
                        \includegraphics[width=\textwidth]{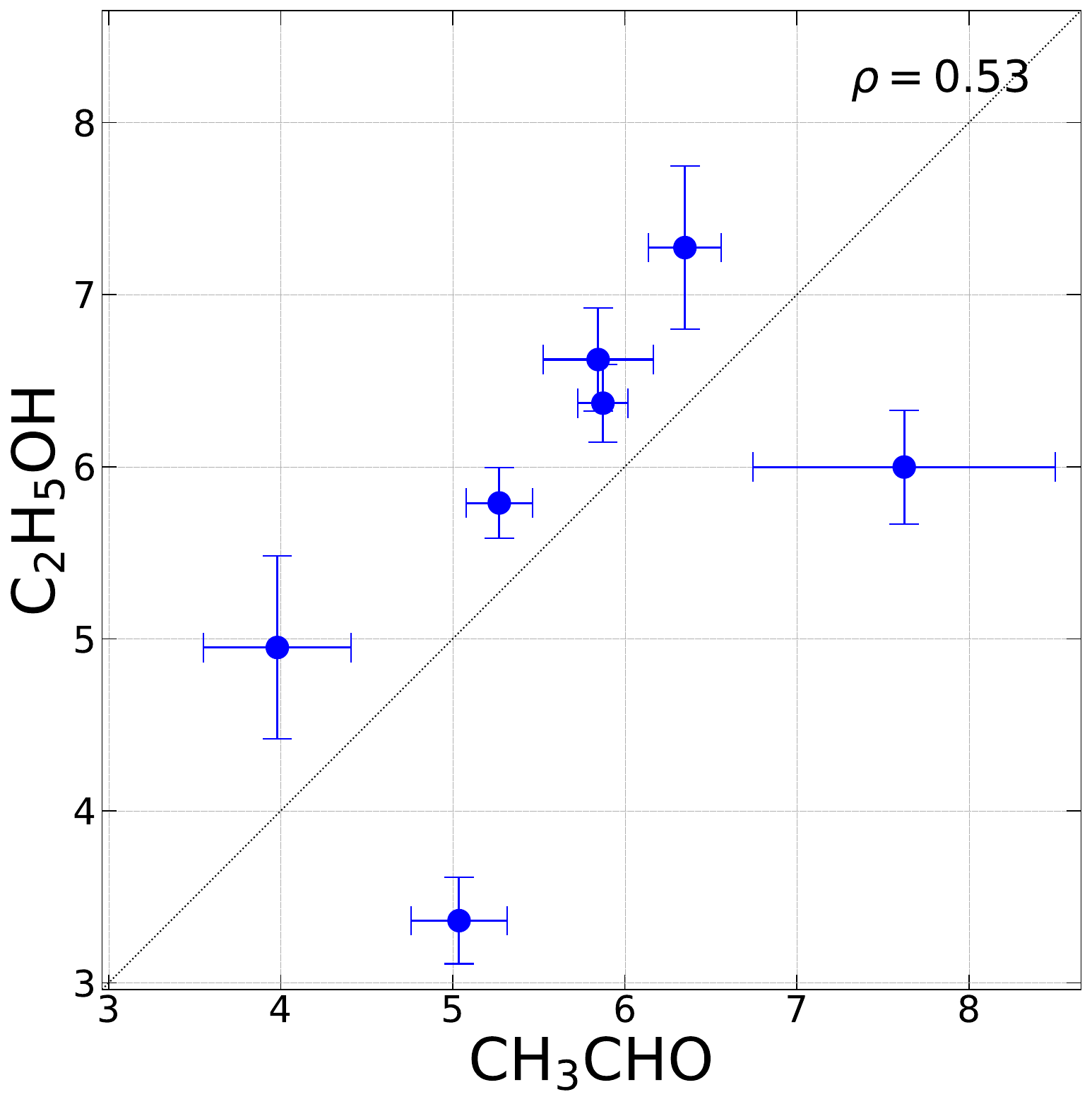}
                    \end{subfigure}
                    \begin{subfigure}{0.247\textwidth}
                        \centering
                        \includegraphics[width=\textwidth]{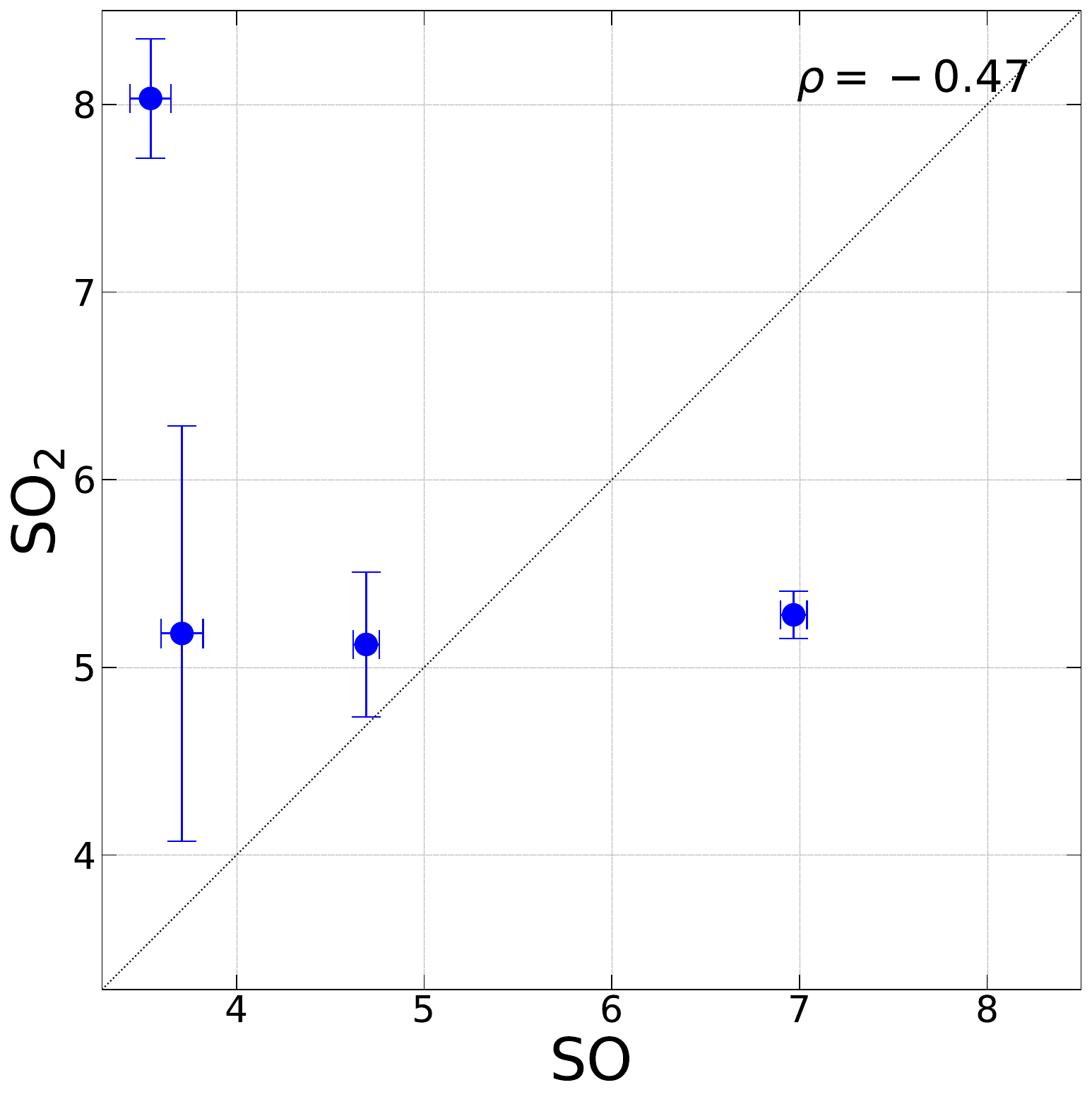}
                    \end{subfigure}
                    \begin{subfigure}{0.247\textwidth}
                        \centering
                        \includegraphics[width=\textwidth]{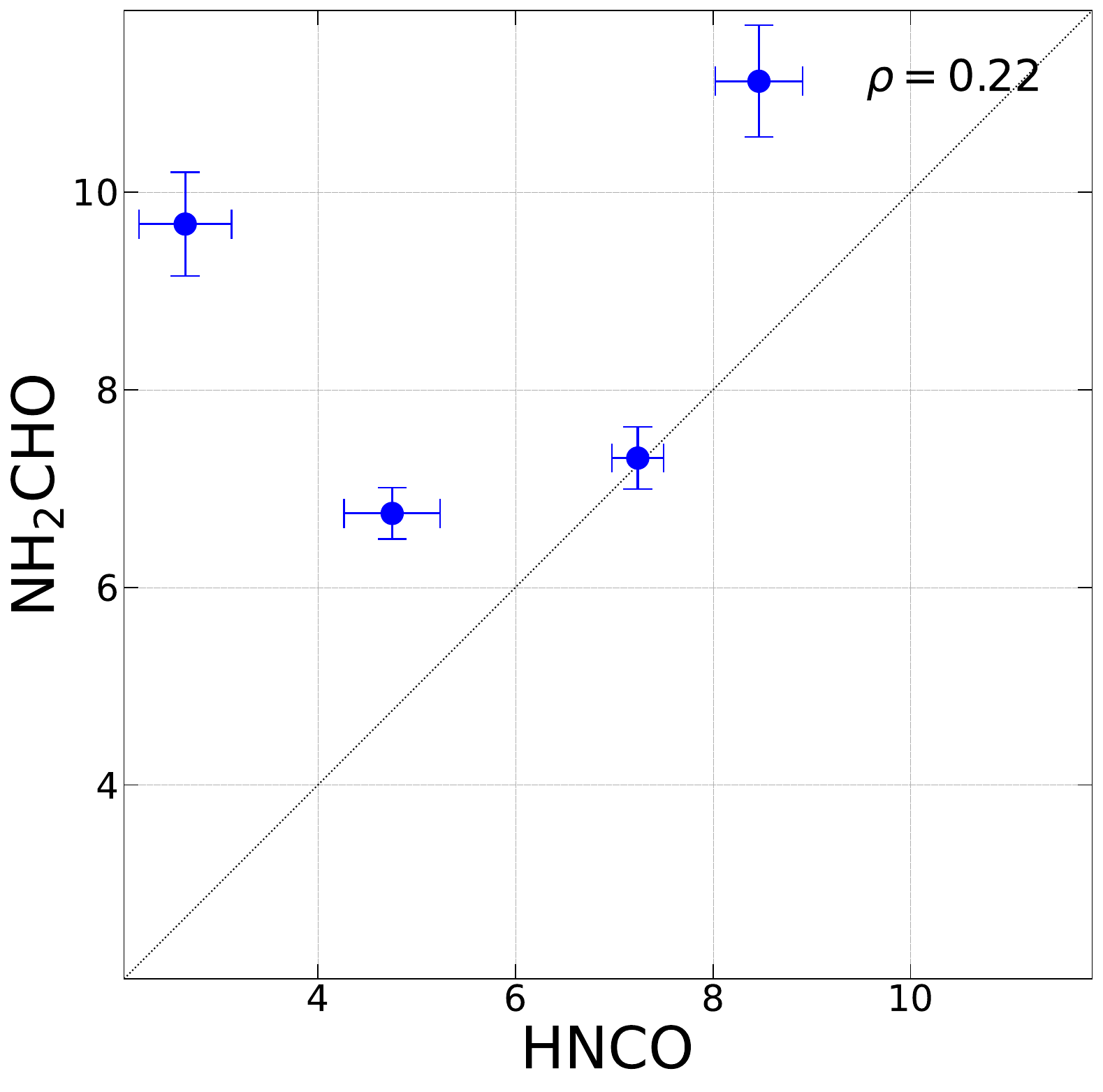}
                    \end{subfigure}
                    \begin{subfigure}{0.247\textwidth}
                        \centering
                        \includegraphics[width=\textwidth]{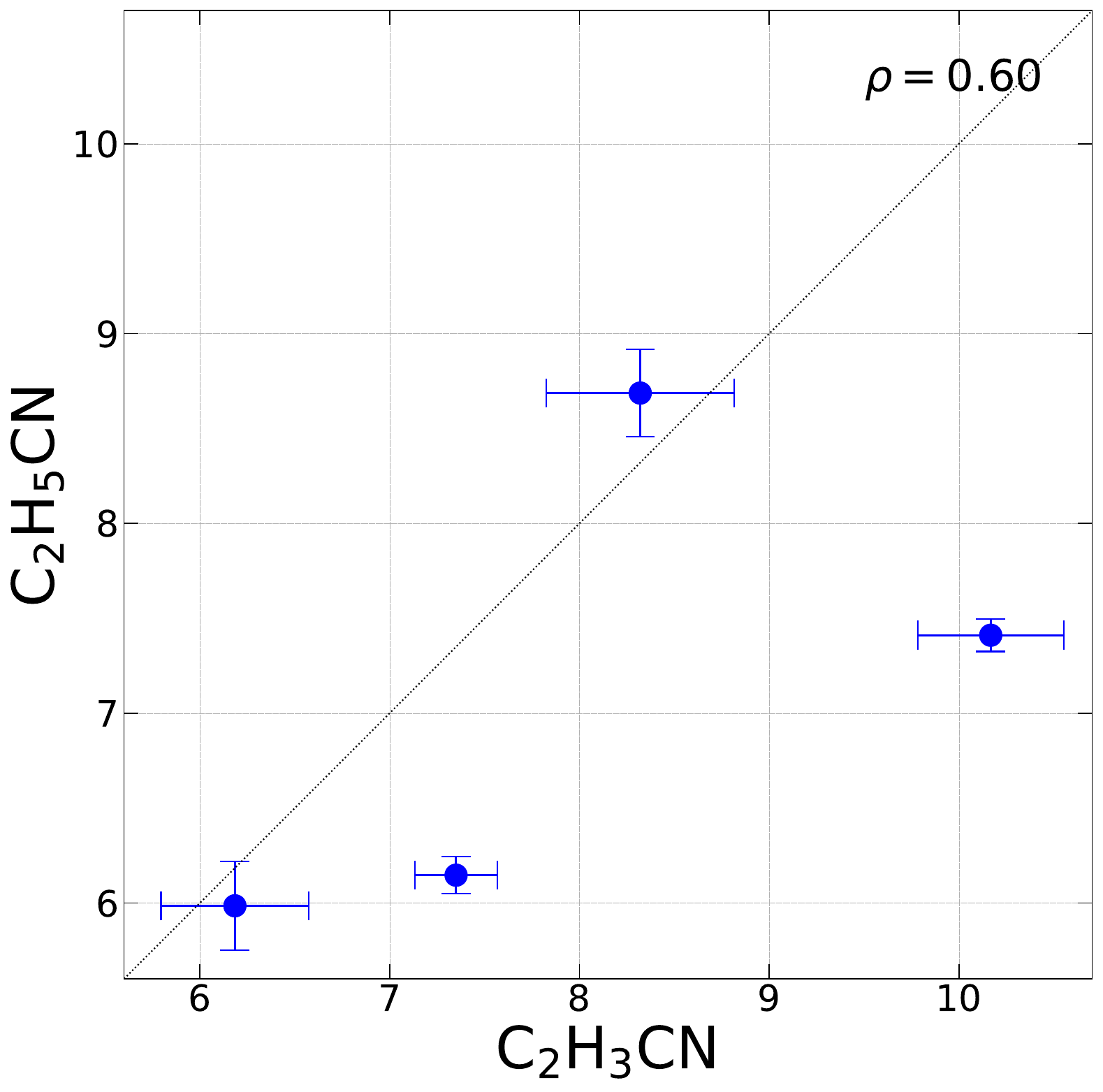}
                    \end{subfigure}
            
                    \begin{subfigure}{\textwidth}
                        \centering
                        \includegraphics[height=0\textwidth]{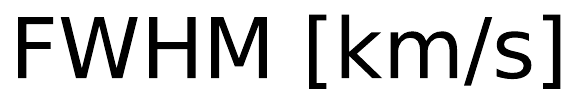}
                    \end{subfigure}
                    \begin{subfigure}{\textwidth}
                        \centering
                        \includegraphics[height=0.025\textwidth]{other/fwhm.pdf}
                    \end{subfigure}
                \end{minipage}
                \caption{Correlation between the line width (FWHM) of chemically linked pairs of molecules. Each data point represents a single targeted source. The Pearson correlation coefficient ($\rho$) for each dataset is displayed.}
                \label{fig:fwhmcorr}
            \end{figure*}

            
We focus on chemically linked species pairs, specifically:
            \begin{itemize}
                \item H$_2$CO and CH$_3$OH \citep[][]{watanabe2002efficient,fuchs2009hydrogenation,Mondal2021,Gorai2024,jimenez2025modelling};
                \item CH$_3$OH and CH$_3$OCHO \citep[][]{herbst2005chemistry,garrod2006formation,Gorai2024};
                \item CH$_3$OH and CH$_3$OCH$_3$ \citep[][]{peeters2006astrochemistry,skouteris2019interstellar,Gorai2024};
                \item CH$_3$OH and CH$_3$COCH$_3$  \citep{singh2022mechanistic,Gorai2024};
                \item CH$_3$CHO and C$_2$H$_5$OH \citep{molpeceres2025hydrogenation,Mondal2021,Gorai2024};
                \item SO and SO$_2$ \citep{prasad1982sulfur,maity2013electron};
                \item HNCO and NH$_2$CHO \citep{duvernay2005matrix,haupa2019hydrogen,chuang2022formation,Gorai2020,Taniguchi2023,Gorai2024};
                \item C$_2$H$_3$CN and C$_2$H$_5$CN \citep{garrod2008complex,belloche2009increased,Gorai2024}.
            \end{itemize}
Figure~\ref{fig:denscorr} shows the correlation plots for the column densities and the related Pearson correlation coefficient ($\rho$). Figure~\ref{fig:fwhmcorr} presents the analogous correlation regarding the line widths.
            
All the pairs of molecules, except CH$_3$COCH$_3$-CH$_3$OH, present a strong positive correlation between column densities (i.e., $\rho\geq0.7$). Comparing these correlations with the ones between the FWHM, we see how the strength of the correlation is maintained for fewer pairs of molecules. In particular, the chemical link for CH$_3$OH-H$_2$CO, and CH$_3$COCH$_3$-CH$_3$OH is notable given the correlation between their line widths. On the other hand, for SO$_2$-SO, NH$_2$CHO-HNCO, and C$_2$H$_5$CN-C$_2$H$_3$CN, the line widths are not well correlated, suggesting that the molecules trace different gases, characterized by different kinematics. Moreover, it seems that CH$_3$OCH$_3$ traces a more turbulent gas than methanol, having line widths higher than those of methanol.
            
The average values between the column densities ratios of these species are: 5.7 for CH$_3$OH/H$_2$CO; 0.20 for $\text{CH}_3\text{OCH}_3 / \text{CH}_3\text{OH}$; 0.012 for $\text{CH}_3\text{COCH}_3 /\text{CH}_3\text{OH}$; 0.15 for $\text{CH}_3\text{OCHO} / \text{CH}_3\text{OH}$; 7.5 for $\text{C}_2\text{H}_5\text{OH} / \text{CH}_3\text{CHO}$; 0.075 for $\text{NH}_2\text{CHO} / \text{HNCO}$; 2.7 for $\text{C}_2\text{H}_5\text{CN} / \text{C}_2\text{H}_3\text{CN}$; 0.80 for $\text{SO}_2 / \text{SO}$.

\subsection{Analysis of the excitation temperatures}\label{sect:temperature}

\begin{figure}
                \centering
                \includegraphics[width=\linewidth]{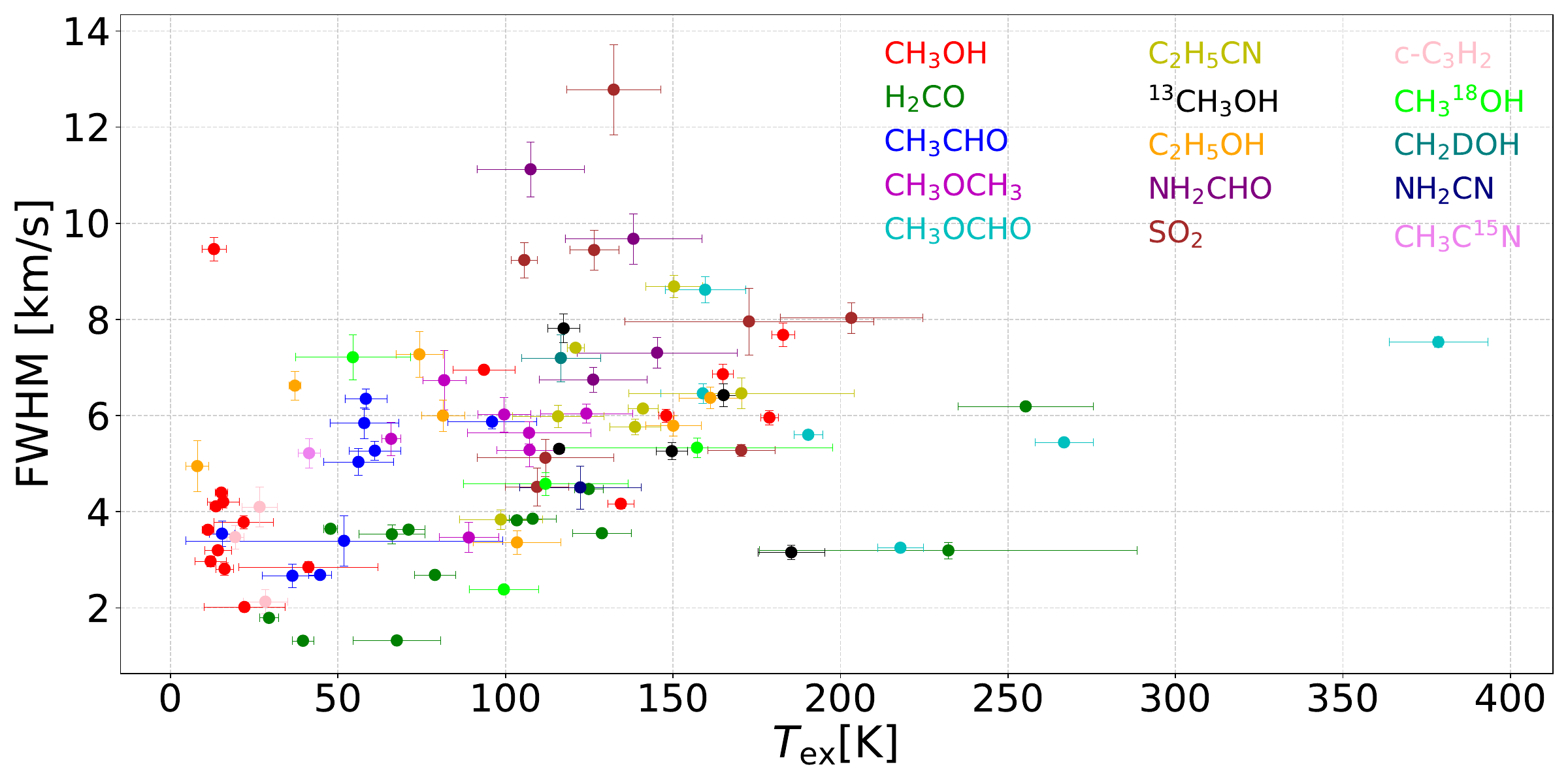} 
                \caption{Relationship between the excitation temperature and the line width for each line where these parameters were left free in the fit with \textsc{MADCUBA}. Each data point represents a single targeted source.}
                \label{fig:texfwhm}
            \end{figure}

Through the fitting procedure, we obtain values of the excitation temperature for each species for which we have at least two lines, and when the Boltzmann plot has no degeneracy. 
Figure~\ref{fig:texfwhm} presents the relationship between the derived excitation temperature and the line width, for each source and each molecular line where $T_\text{ex}$ was able to be fit. Considering all the molecules, we notice a general trend 
of an increasing line width with higher temperatures, although there is significant scatter in this relation.
Such a trend is generally expected in protostellar cores that have a temperature gradient set by internal heating. The warmer regions trace deeper parts of the gravitational potential where infall and/or orbital motions are expected to be larger.

            
In Fig.~\ref{fig:parameters}, a comparison between the excitation temperatures of the sources is presented. The line-rich sources (Sect.~\ref{sect:detections}) are characterized by $T_\text{ex}\gtrsim100~$K, considering the average over all the detected species (G45.47 is the only one characterized by a line-poor spectrum and by a high $T_\text{ex}$).
            
\subsection{Line widths and centroid velocities}

\begin{figure}
\centering
\includegraphics[width=\linewidth]{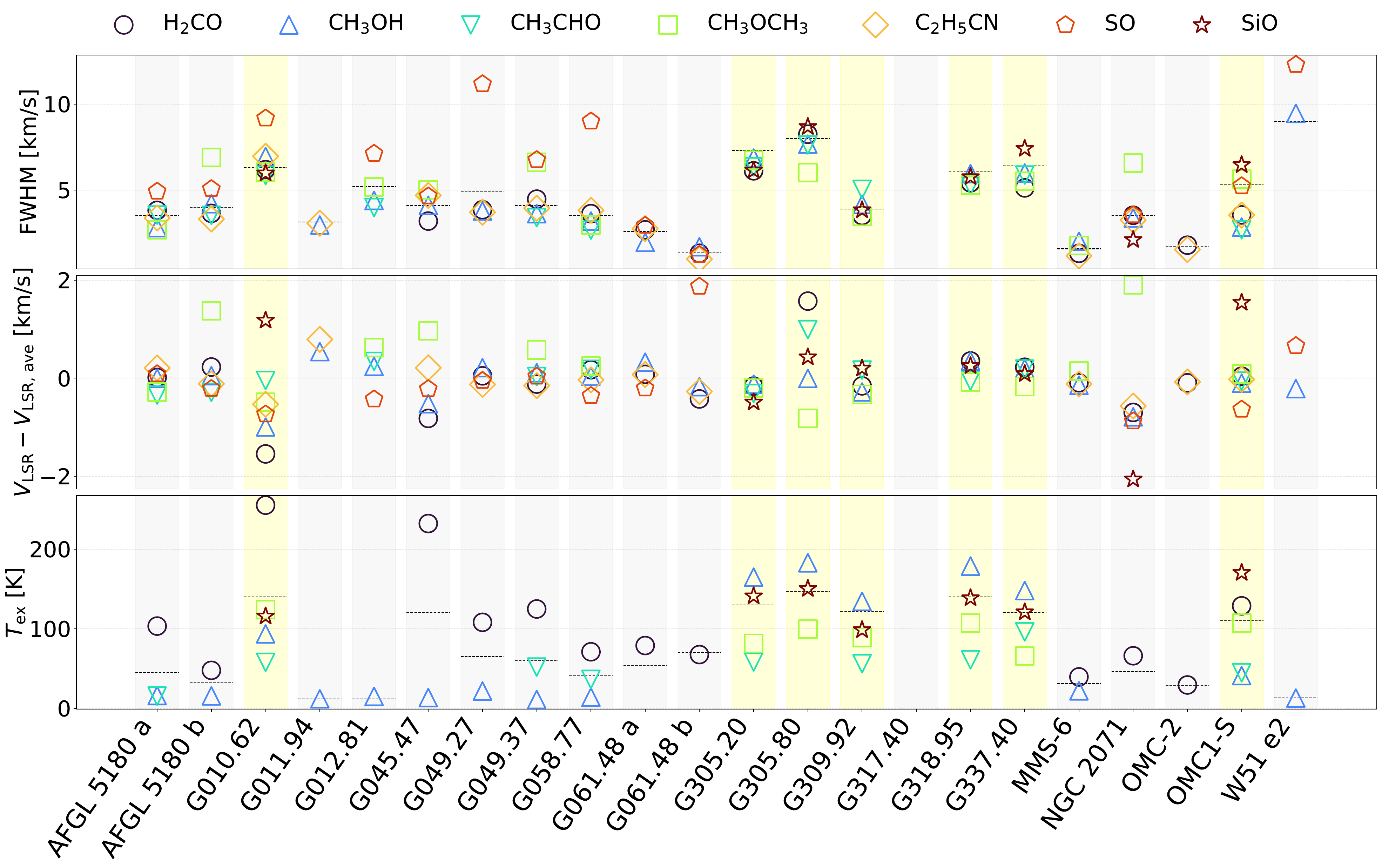}
\caption{Comparison between parameters obtained by the line analysis for each source. The average values and the ones for selected species (i.e., the most common species for each type of molecules; H$_2$CO, CH$_3$OH, CH$_3$CHO, CH$_3$OCH$_3$, C$_2$H$_5$CN, SO, and SiO) are shown. For the line width (\textit{Top}) and the excitation temperatures (\textit{Bottom}), a dashed line is shown for each source, presenting the average FWHM and $T_\text{ex}$, respectively, estimated over all the detected species. The centroid velocity (\textit{Middle}) is represented by subtracting the average value of each source from the absolute $v_\text{LSR}$ of each species. The bins highlighted in yellow show sources characterized by a line-rich spectrum ($\gtrsim100$ transitions).}
\label{fig:parameters}
\end{figure}

Figure~\ref{fig:parameters} shows a comparison of line width and centroid velocities obtained for each source in the sample from the analysis with \textsc{MADCUBA} (Sect.~\ref{sect:lineidentification}). The 8 sources characterized by higher $T_\text{ex}$ (Sect.~\ref{sect:temperature}) generally have higher line widths, perhaps indicative of deeper gravitational potentials and/or stronger feedback contributing to the observed line broadening. However, the source with the highest average FWHM is W51 e2, which is a relatively low-temperature ($T_\text{ex}<100~$K) and line-poor source. We note that this source and the surrounding clouds have been studied at higher spatial resolution by \citet{ginsburg2017thermal} and \citet{bonfand2024alma}, with several hot cores and COMs emission lines detected on scales smaller than our resolution. This highlights the possibility of source multiplicity contributing to our observed line widths and motivates the need for higher sensitivity and higher angular resolution follow-up of our sample.

To study the centroid velocities, we subtracted the average $v_\text{LSR}$ estimated for each source from the $v_\text{LSR}$ estimated for each species detected. No clear pattern emerges between the centroid velocities and the composition, the number of detected species, or the other parameters derived from the line analysis. However, we note that some sources show significant dispersion in the LSR velocities of the displayed species, which likely indicates a more complex internal kinematic structure.

\subsection{Correlation of spectroscopic parameters with SED properties}\label{sect:sed}

\begin{figure*}
  \centering
  \begin{minipage}{\textwidth}
    \centering
    \includegraphics[width=0.65\textwidth]{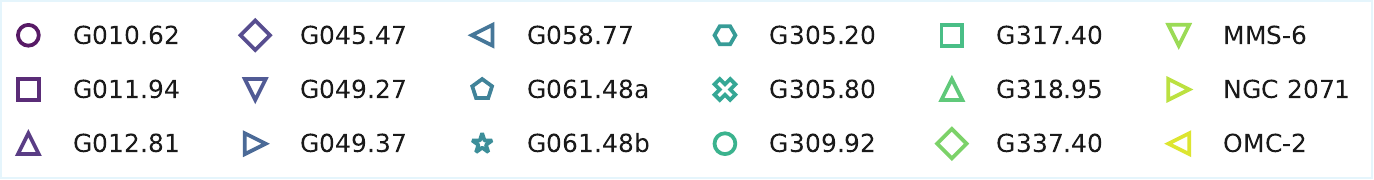}
  \end{minipage}
  \noindent
  \begin{minipage}{0.025\textwidth}
    \centering
    \includegraphics[width=\linewidth]{other/fwhm1.pdf}
  \end{minipage}\hfill
  \begin{minipage}{0.29\textwidth}
    \centering
    \includegraphics[width=\linewidth]{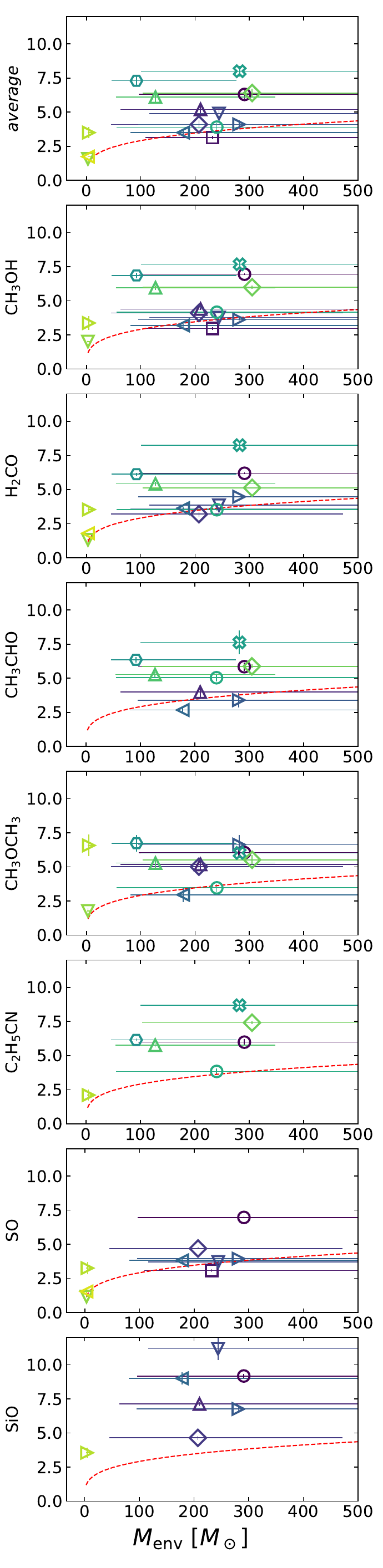}
  \end{minipage}\hfill
  \begin{minipage}{0.29\textwidth}
    \centering
    \includegraphics[width=\linewidth]{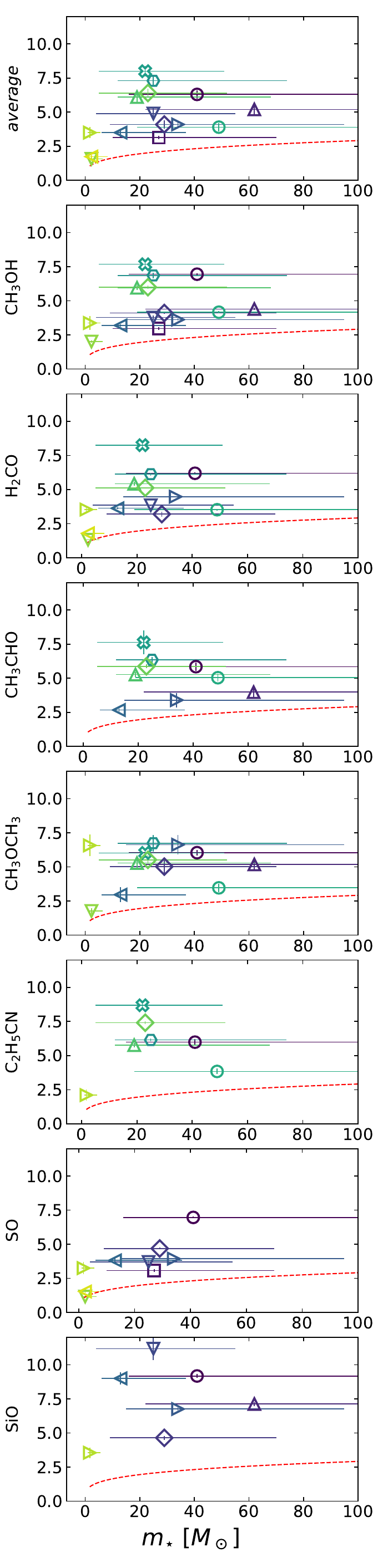}
  \end{minipage}\hfill
  \begin{minipage}{0.29\textwidth}
    \centering
    \includegraphics[width=\linewidth]{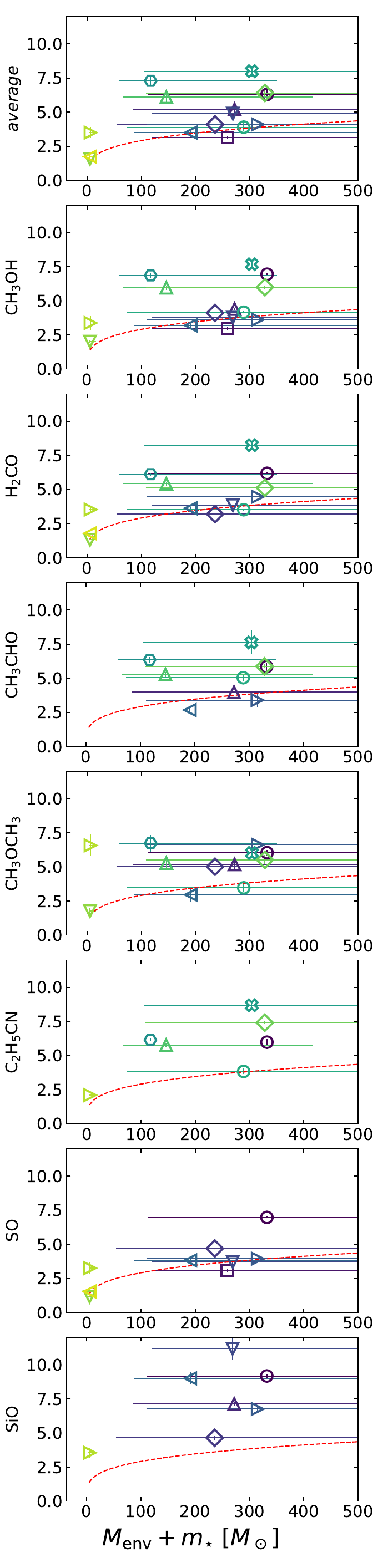}
  \end{minipage}
\vspace{-0.4cm}
  \caption{Correlation between envelope mass $M_\text{env}$ (\textit{Left}), stellar mass $m_\star$ (\textit{Center}), the sum of these two masses (\textit{Right}), and FWHM of the detected molecular lines. The species selected are CH$_3$OH, H$_2$CO, CH$_3$CHO, CH$_3$OCH$_3$, C$_2$H$_5$CN, SO, SiO, which are the most common species detected in the sample, for each type of molecule. The averaged values are also presented in the top row. The red dashed line shows the FWHM ($=\sqrt{8\:{\rm ln\:}2}\sigma_{\rm c,vir}$) of a fiducial pre-stellar core in the Turbulent Core Model for $\Sigma_\text{cl}= 1 \text{g}~\text{cm}^{-2}$ \citep[see eq.~4 of][]{tan2013dynamics}. We see that our various tracers have higher velocity dispersions, which is expected as the potential deepens during the protostellar phase.}
  \label{fig:sed_fwhm}
\end{figure*}

Figure~\ref{fig:sed_fwhm} examines the correlation between line widths (FWHM) and various mass estimates of the protostellar system, i.e., envelope mass, current protostellar mass, and total of these two masses. One generally expects that the line width should increase as the system mass increases. For example, for fixed $\Sigma_{\rm cl}=1\:{\rm g\:cm}^{-2}$, the red dashed lines in Figure~\ref{fig:sed_fwhm} show the line widths of pre-stellar cores in the Turbulent Core Model, i.e., assuming virialized conditions in the core. Protostellar cores are expected to have larger line widths, since the gravitational potential deepens during the growth of the star, and this expectation is born out in our data. We also see that our three lowest mass systems (MMS-6, NGC 2071, OMC-2) often have the smallest line widths. However, overall the correlations between line width and mass do not appear very strong and larger samples will be needed for more definitive conclusions to be drawn. Nevertheless, the data shown in Fig.~\ref{fig:sed_fwhm} are an important first observational guide for theoretical models of the chemodynamical evolution of massive protostellar systems.

\begin{figure*}
    \centering
    \begin{minipage}{\textwidth}
        \centering
        \includegraphics[width=0.65\textwidth]{other/legend.pdf}
    \end{minipage}
    \vspace{1em}
    \includegraphics[width=0.82\textwidth]{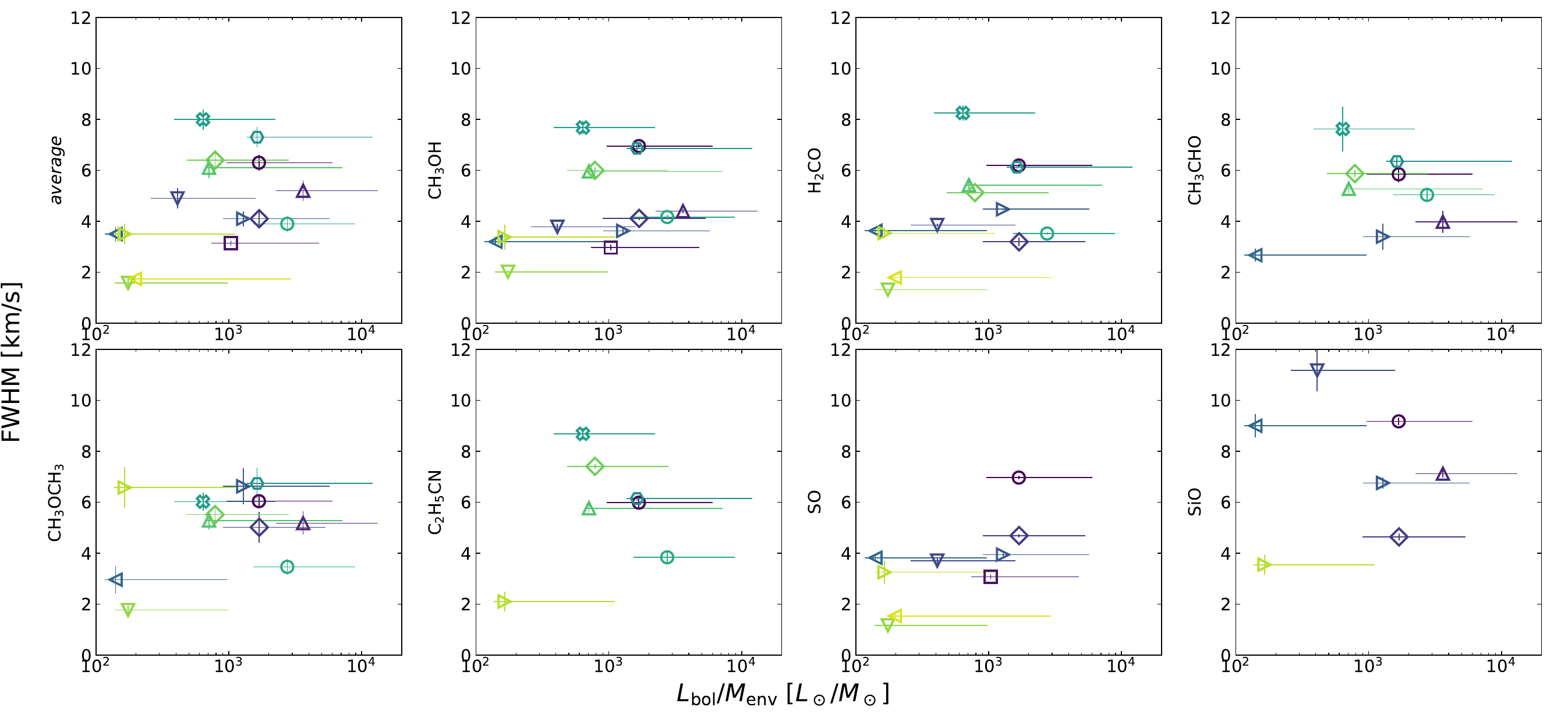}
    \includegraphics[width=0.82\textwidth]{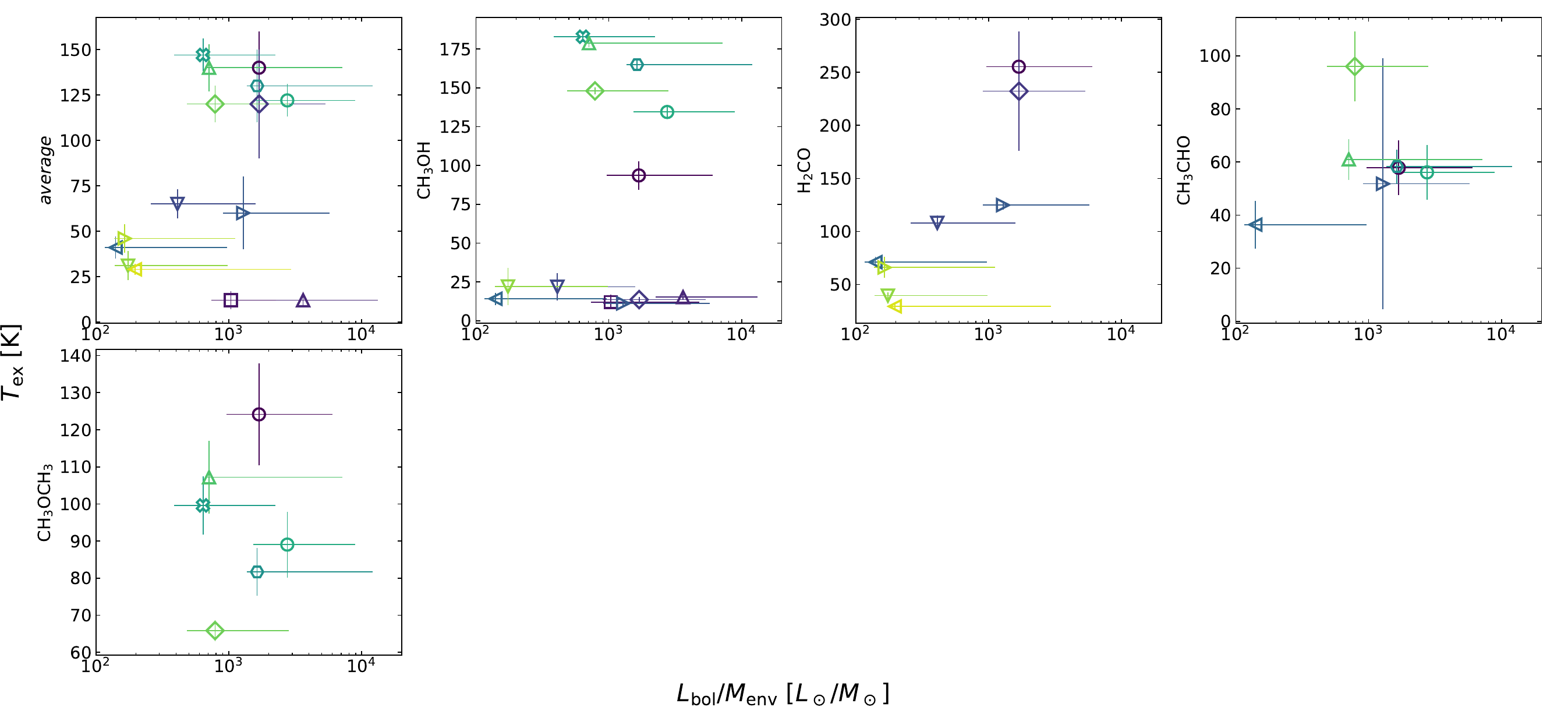}
    \includegraphics[width=0.82\textwidth]{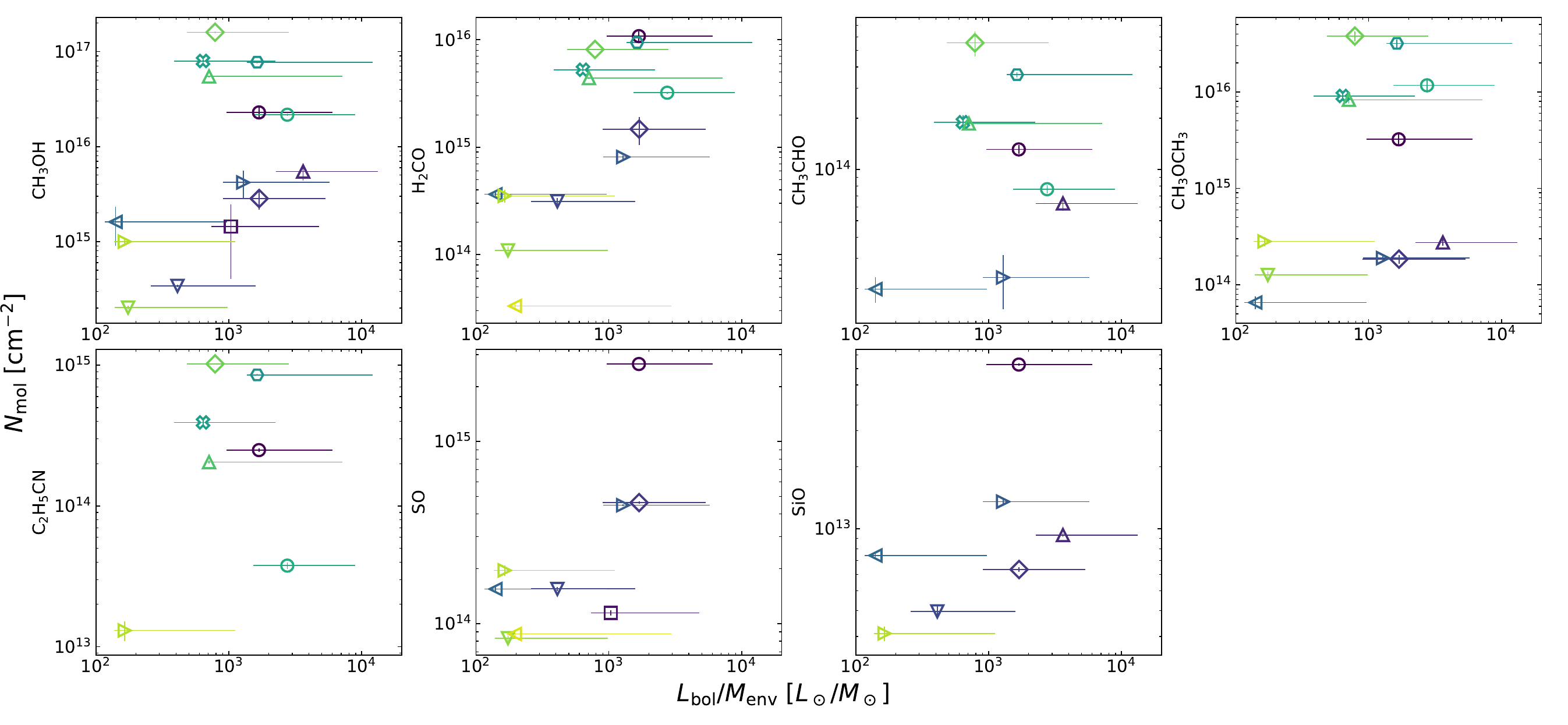}
    \caption{Correlation between luminosity-to-mass ratio $L_{\rm bol}/M_{\rm env}$ and line widths (\textit{Top}), excitation temperatures (\textit{Middle}), and column densities (\textit{Bottom}). The species selected are CH$_3$OH, H$_2$CO, CH$_3$CHO, CH$_3$OCH$_3$, C$_2$H$_5$CN, SO, SiO, which are the most common species detected in the sample, for each type of molecule. For FWHM and $T_\text{ex}$, the averaged values are also presented in the first subplot. The correlations for $T_\text{ex}$ are shown only for the species where an estimation of the excitation temperature was possible.}
    \label{fig:sed}
\end{figure*}

Figure~\ref{fig:sed} shows the correlations between the luminosity-to-mass ratio ($L_{\rm bol}/M_{\rm env}$) and line parameters such as line width, excitation temperature, and molecular column density. The bolometric luminosities and envelope masses, as reported in Table~\ref{tab:sources}, were derived from the SED fitting described in Sect.~\ref{sect:sed}. The $L_{\rm bol}/M_{\rm env}$ values are expected to increase with the evolution of the protostars. Figure~\ref{fig:sed} shows a slight increasing trend in the average line width, excitation temperature, and molecular column density with $L_{\rm bol}/M_{\rm env}$. However, the relatively limited $L_{\rm bol}/M_{\rm env}$ range covered by our sample makes it difficult to draw firm conclusions.
In particular, the low $L_{\rm bol}/M_{\rm env}$ value of MMS 6 strongly influences the correlations. Among the remaining sources, which all exhibit relatively high and comparable $L_{\rm bol}/M_{\rm env}$ values, the significant scatter in FWHM,  $T_\text{ex}$, and $N_\text{mol}$ hinders the identification of clear trends. 


\section{Conclusions}\label{sect:5}
        
We have analyzed ALMA ACA and TP Band 6 data for 20 massive star-forming regions selected from the SOMA survey. We have presented the results of the 1.33 mm dust continuum and spectral line analyses. The major findings of this work are as follows:

\begin{enumerate}
\item 
A total of 22 continuum sources were identified, and we measured the molecular hydrogen column density based on the 1.33~mm dust continuum emission.

\item
From the spectra extracted from these continuum sources, 35 molecular species were detected, including 14 rarer isotopologues (D, $^{13}$C, $^{18}$O, $^{15}$N, $^{34}$S) and, in 10 cores, hydrogen and helium recombination lines (H$_{30}\alpha$ and He$_{30}\alpha$). We report the line width, excitation temperature, systemic velocity, and molecular column density for the observed species across the 22 cores.

\item 
The SED fitting provided estimates of the physical properties of the sources, including bolometric luminosity and envelope mass, enabling the classification of the protostellar cores. While some sources (i.e., OMC1-S and W51 e2) lacked sufficient ancillary data for full SED characterization, the available results offer a representative overview of the sample's physical properties.

\item
Among the sample, we detect the presence of seven line-rich sources within our sample (G337.40, G318.95, G309.92, G305.80, G305.20, G10.62, OMC1-S), characterized by more than 100 transitions detected. 

\item 
We explored the correlations among chemically related species, finding strong associations in column densities and line widths (e.g., CH$_3$OH-H$_2$CO, CH$_3$OCHO-CH$_3$OH), supporting known formation pathways. We anticipate that these results will be important constraints on astrochemical models of massive protostellar cores.

\item 
Excitation temperatures, derived when possible through multi-line analysis, show an increasing trend with line width. This may indicate a trend of increasing temperature with deeper gravitational potential well and/or stronger kinetic feedback. The line-rich sources exhibit $T_\text{ex} \gtrsim 100$~K and broader line profiles, confirming their chemically and kinematically active nature.

\item 
Examining the correlation of spectroscopic parameters with $L_{\rm bol}/M_{\rm env}$, used as a proxy for evolutionary stage, reveals tentative trends of increasing line widths, excitation temperatures, and molecular column densities with increasing $L_{\rm bol}/M_{\rm env}$. However, the sample shows relatively large scatter: e.g., at a representative value of $L_{\rm bol}/M_{\rm env}\sim10^3\:L_\odot/M_\odot$ we see that a wide range of FWHM, $T_{\rm ex}$ and molecular column densities are present. Larger samples are needed to more reliably assess the significance of these potential correlations. 



\item
The detection of H${30}\alpha$ and He${30}\alpha$ across sources with both chemically rich and poor spectra indicates a variety of ionized environments, which do not necessarily directly influence the chemical complexity. This could indicate that in the final stages of evolution, when ionization is becoming stronger, molecular line emission can become relatively diminished, perhaps because the infall envelope and/or disk have mostly been accreted or destroyed by feedback.

\item 
This study presents a first examination of observed chemodynamical properties of massive protostars and their potential correlation with physical properties derived from SED fitting of Turbulent Core Accretion models. As such, the data provide important constraints for theoretical models of the chemical evolution of massive protostellar systems.


\end{enumerate}

\begin{acknowledgements}
This paper makes use of the following ALMA data: ADS/JAO.ALMA\#2021.2.00177.S ALMA is a partnership of ESO (representing its member states), NSF (USA), and NINS (Japan), together with NRC (Canada), MOST and ASIAA (Taiwan), and KASI (Republic of Korea), in cooperation with the Republic of Chile. The Joint ALMA Observatory is operated by ESO, AUI/NRAO and NAOJ. This research made use of sedcreator (Fedriani et al. 2023, ApJ, 942, 7). ADS url: https://ui.adsabs.harvard.edu/abs/2023ApJ...942....7F/abstract. This research made use of Photutils, an Astropy package for detection and photometry of astronomical sources (Bradley et al. 2020). 
D.G. acknowledges support from a CASSUM fellowship at Chalmers University of Technology. J.C.T. acknowledges support from ERC Advanced Grant MSTAR (788829) and NSF grant AST-2206450. K.T. is supported by JSPS KAKENHI grant Nos. 21H01142, 24K17096, and 24H00252. R.F. acknowledges support from the grants PID2023-146295NB-I00, and from the Severo Ochoa grant CEX2021-001131-S funded by MCIN/AEI/ 10.13039/501100011033 and by ``European Union NextGenerationEU/PRTR''.
\end{acknowledgements}

\bibliographystyle{aa} 
    \bibliography{bibliography} 
    \begin{appendix}
        \onecolumn
        \section{Observational parameters summary}
            In Table~\ref{tab:observations}, the setup properties of the observations carried out for the 22 cores detected in the \obsA\,SOMA regions are summarized.
            \begin{table*}[h]
                \centering
                \caption{Summary of the observational parameters of the ACA continuum maps, and the ACA+TP spectra cubes, including rms noise, beam size, and frequency range.}
                \label{tab:observations}
                \begin{tabular}{lccccccccc}
                    \hline
                    \hline
                    \noalign{\smallskip}
                    & &\multicolumn{2}{c}{Continuum maps} & & \multicolumn{3}{c}{Spectra} & & \\[-0.1ex]
                    \noalign{\smallskip}
                    \cline{2-4} \cline{6-8} 
                    \noalign{\smallskip}
                    Source & & rms & Synthesized beam & & rms\tablefootmark{a} & Synthesized beam \tablefootmark{b} & Frequency range \tablefootmark{c} & & Physical size \tablefootmark{d}\\[-0.1ex]
                    \noalign{\smallskip}
                    & & (Jy/beam) & ($''$ × $''$)& & (mK) & ($''$ × $''$) & (GHz) & & (AU)\\[-0.1ex]
                    \noalign{\smallskip}
                    \hline
                    \noalign{\smallskip}
                    AFGL 5180 a  & & 0.031 & 6.36 × 5.82 & & 43 & 6.81 × 5.86 & 216.58 - 234.91 & & 9180 \\[-0.1ex]
                    AFGL 5180 b  & & 0.031 & 6.36 × 5.82 & & 30 & 6.81 × 5.86 & 216.58 - 234.91 & & 9180 \\[-0.1ex]
                    G010.62 & & 0.291 & 9.32 × 4.38 & & 52 & 9.36 × 4.59 & 216.59 - 234.90 & & 24990 \\[-0.1ex]
                    G011.94 & & 0.058 & 9.35 × 4.14 & & 35 & 9.33 × 4.61 & 216.62 - 234.93 & & 19380 \\[-0.1ex]
                    G012.81 & & 0.346 & 9.36 × 4.09 & & 37 & 9.38 × 4.62 & 216.62 - 234.93 & & 12240 \\[-0.1ex]
                    G045.47 & & 0.027 & 7.09 × 5.08 & & 36 & 7.38 × 5.26 & 216.58 - 234.92 & & 42840 \\[-0.1ex]
                    G049.27 & & 0.020 & 6.97 × 5.30 & & 33 & 7.31 × 5.55 & 216.59 - 234.92 & & 28050 \\[-0.1ex]
                    G049.37 & & 0.044 & 6.97 × 5.26 & & 36 & 7.30 × 5.52 & 216.57 - 234.91 & & 27540 \\[-0.1ex]
                    G058.77 & & 0.011 & 8.31 × 5.17 & & 25 & 8.66 × 5.42 & 216.55 - 234.88 & & 16830 \\[-0.1ex]
                    G061.48 a & & 0.053 & 8.51 × 5.35 & & 19 & 8.78 × 5.65 & 216.54 - 234.87 & & 11220 \\[-0.1ex]
                    G061.48 b & & 0.053 & 8.51 × 5.35 & & 23 & 8.78 × 5.65 & 216.54 - 234.87 & & 11220\\[-0.1ex]
                    G305.20 & & 0.097 & 7.18 × 4.93 & & 93 & 7.32 × 5.16 & 216.49 - 234.83 & & 20910 \\[-0.1ex]
                    G305.80 & & 0.098 & 7.43 × 5.00 & & 79 & 7.21 × 5.21 & 216.50 - 234.83 & & 20400 \\[-0.1ex]
                    G309.92 & & 0.065 & 6.80 × 4.88 & & 31 & 7.28 × 5.18 & 216.48 - 234.81 & & 28050 \\[-0.1ex]
                    G317.43 & & 0.015 & 6.73 × 4.98 & & 37 & 6.98 × 5.31 & 216.54 - 234.88 & & 72420 \\[-0.1ex]
                    G318.95 & & 0.036 & 6.73 × 4.96 & & 51 & 6.94 × 5.15 & 216.50 - 234.83 & & 12240 \\[-0.1ex]
                    G337.40 & & 0.055 & 6.91 × 4.57 & & 52 & 7.09 × 4.85 & 216.50 - 234.83 & & 15810 \\[-0.1ex]
                    MMS 6  & & 0.065 & 7.20 × 4.39 & & 37 & 7.34 × 4.69 & 216.55 - 234.89 & & 2040 \\[-0.1ex]
                    NGC 2071 & & 0.050 & 6.94 × 4.53 & & 35 & 7.33 × 4.88 & 216.55 - 234.89 & & 2040 \\[-0.1ex]
                    OMC-2 & & 0.013 & 7.25 × 4.43 & & 34 & 7.39 × 4.71 & 216.55 - 234.89 & & 2040\\[-0.1ex]
                    OMC1-S & & 0.479 & 7.23 × 4.41 & & 49 & 7.41 × 4.74 & 216.55 - 234.89 & & 2040\\[-0.1ex]
                    W51 e2 & & 0.208 & 6.96 × 5.28 & & 55 & 7.29 × 5.54 & 216.58 - 234.92 & & 27540\\[-0.1ex]
                    \noalign{\smallskip}   
                    \hline
                \end{tabular}
                    \tablefoot{
                        \tablefoottext{a}{Averaged noise over all the spectral windows.}
                        \tablefoottext{b}{Average beam of the data cubes, from which the spectra were extracted with a circular aperture with a diameter of $\sim5.1''$.}
                        \tablefoottext{c}{These are the frequency limits covered by our observations. The coverage is not continuous across these ranges.}
                        \tablefoottext{d}{Physical size corresponding to $5.1''$.}
                    }
            \end{table*}
            
        \section{Best-fit molecular parameters}
            \begin{sidewaystable*}
    \centering
    \setlength{\tabcolsep}{3pt}
    \caption{Column density ($N_\text{tot}$) estimated for every detection with \textsc{MADCUBA}}
    \label{tab:dens1}
    \begin{tabular}{lcccccccccccc}
        \hline
        \hline
        \noalign{\smallskip}
        \multicolumn{2}{l}{} & \multicolumn{11}{c}{$N_\text{tot}$ (cm$^{-2}$)} \\
        \noalign{\smallskip}
        \cline{3-13}
        \noalign{\smallskip}
        Molecule & & AFGL 5180 a & AFGL 5180 b & G010.62 & G011.94 & G012.81 & G045.47 & G049.27 & G049.37 & G058.77 & G061.48 a & G061.48 b \\
        \noalign{\smallskip}
        \hline
        \noalign{\smallskip}
        $^{13}$CO & $(\times 10^{15})$ & ... & ... & ... & ... & ... & ... & ... & ... & 49(5) & 285(4) & 56(5) \\
        C$^{18}$O & $(\times 10^{16})$ & 2.21(0.03) & 2.35(0.05) & 36.8(0.4) & ... & ... & 4.76(0.09) & 3.76(0.13) & ... & 2.32(0.05) & 5.76(0.09) & 1.14(0.09) \\
        H$_{2}$CO & $(\times 10^{14})$ & 20.0(1.1) & 3.7(0.2) & 110(12) & ... & ... & 15(4) & 3.1(0.3) & 8.1(0.3) & 3.6(0.2) & 2.1(0.2) & 0.34(0.05) \\
        H$_{2}$$^{13}$CO & $(\times 10^{12})$ & 38.0(1.3) & 15.3(0.8) & 149(3) & ... & ... & 19.0(1.0) & 4.4(0.7) & 15.9(0.8) & 11.5(0.7) & 3.4(0.4) & ... \\
        D$_{2}$CO & $(\times 10^{13})$ & ... & ... & ... & ... & ... & ... & ... & ... & ... & ... & ... \\
        CH$_{3}$OH & $(\times 10^{15})$ & 7(2) & 1.4(0.5) & 23(3) & 1.4(1.0) & 5.5(1.1) & 2.8(0.7) & 0.342(0.011) & 4.2(1.4) & 1.6(0.7) & 0.43(0.02) & 0.144(0.006) \\
        $^{13}$CH$_{3}$OH & $(\times 10^{14})$ & ... & ... & 9.1(0.6) & ... & 1.5(0.3) & ... & ... & ... & ... & ... & ... \\
        CH$_{2}$DOH & $(\times 10^{15})$ & ... & ... & ... & ... & ... & ... & ... & ... & ... & ... & ... \\
        CH$_{3}$$^{18}$OH & $(\times 10^{14})$ & ... & ... & ... & ... & ... & ... & ... & ... & ... & ... & ... \\
        CH$_{3}$CHO & $(\times 10^{13})$ & 20(14) & 4.9(0.6) & 13.1(0.8) & ... & 6.3(0.6) & ... & ... & 2.3(0.8) & 2.0(0.3) & ... & ... \\
        CH$_{3}$OCHO & $(\times 10^{15})$ & ... & ... & 2.1(0.2) & ... & ... & ... & ... & ... & ... & ... & ... \\
        CH$_{3}$OCH$_{3}$ & $(\times 10^{14})$ & 1.6(0.2) & 1.1(0.2) & 32(4) & ... & 2.8(0.2) & 1.8(0.2) & ... & 1.9(0.2) & 0.7(0.1) & ... & ... \\
        CH$_{3}$COCH$_{3}$ & $(\times 10^{14})$ & ... & ... & ... & ... & ... & ... & ... & ... & ... & ... & ... \\
        C$_{2}$H$_{5}$OH & $(\times 10^{14})$ & ... & ... & 4.7(0.2) & ... & 1.3(0.5) & ... & ... & ... & ... & ... & ... \\
        $^{13}$CN & $(\times 10^{16})$ & ... & ... & ... & 8.0(0.2) & 128(2) & ... & ... & ... & ... & ... & ... \\
        C$^{15}$N & $(\times 10^{12})$ & ... & ... & ... & ... & ... & ... & ... & 7.7(0.8) & 1.8(0.5) & ... & ... \\
        HNCO & $(\times 10^{14})$ & ... & ... & ... & ... & ... & ... & ... & ... & ... & ... & ... \\
        CH$_{3}$C$^{15}$N & $(\times 10^{13})$ & ... & ... & ... & ... & ... & ... & ... & ... & ... & ... & ... \\
        C$_{2}$H$_{3}$CN & $(\times 10^{12})$ & ... & ... & 75(4) & ... & ... & ... & ... & ... & ... & ... & ... \\
        C$_{2}$H$_{5}$CN & $(\times 10^{13})$ & ... & ... & 25.0(0.8) & ... & ... & ... & ... & ... & ... & ... & ... \\
        CH$_{3}$NCO & $(\times 10^{13})$ & ... & ... & ... & ... & ... & ... & ... & ... & ... & ... & ... \\
        NH$_{2}$CHO & $(\times 10^{13})$ & ... & ... & 7(2) & ... & ... & ... & ... & ... & ... & ... & ... \\
        NH$_{2}$CN & $(\times 10^{13})$ & ... & ... & ... & ... & ... & ... & ... & ... & ... & ... & ... \\
        NH$_{2}$D & $(\times 10^{14})$ & ... & ... & ... & ... & ... & ... & ... & ... & ... & ... & ... \\
        HC$_{3}$N & $(\times 10^{12})$ & ... & ... & ... & ... & ... & ... & ... & ... & ... & ... & ... \\
        SO & $(\times 10^{14})$ & 6.98(0.09) & 4.25(0.11) & 26.7(0.2) & 1.15(0.04) & ... & 4.62(0.06) & 1.55(0.04) & 4.47(0.05) & 1.55(0.05) & 2.28(0.04) & 0.113(0.009) \\
        SO$_{2}$ & $(\times 10^{14})$ & ... & ... & 30.7(0.7) & ... & ... & 3.2(0.5) & 1.0(0.2) & ... & ... & ... & ... \\
        $^{34}$SO$_{2}$ & $(\times 10^{14})$ & ... & ... & 5.9(0.8) & ... & ... & ... & ... & ... & ... & ... & ... \\
        CCS & $(\times 10^{14})$ & ... & ... & 3.6(0.3) & ... & ... & ... & ... & ... & ... & ... & ... \\
        O$^{13}$CS & $(\times 10^{14})$ & ... & ... & 1.5(0.1) & ... & ... & ... & ... & ... & ... & ... & ... \\
        HDCS & $(\times 10^{12})$ & 36(4) & 14(2) & ... & ... & ... & 13(3) & ... & 11(2) & ... & ... & ... \\
        H$_{2}$$^{13}$CS & $(\times 10^{13})$ & ... & ... & 4.8(0.4) & ... & ... & ... & ... & ... & ... & ... & ... \\
        SiO & $(\times 10^{11})$ & 155(5) & 97(3) & 628(9) & ... & 93(2) & 63(2) & 40(3) & 135(4) & 74(3) & 4.9(0.7) & 3.5(0.5) \\
        c-C$_{3}$H$_{2}$ & $(\times 10^{12})$ & ... & ... & ... & ... & ... & ... & ... & 8.4(1.1) & ... & 5.8(0.6) & ... \\
        \noalign{\smallskip}
        \hline
        \noalign{\smallskip}
    \end{tabular}
\end{sidewaystable*}

            \addtocounter{table}{-1}
\begin{sidewaystable*}
    \centering
    \setlength{\tabcolsep}{3pt}
    \caption{continued.}
    \label{tab:dens2}
    \begin{tabular}{lcccccccccccc}
        \hline
        \hline
        \noalign{\smallskip}
        \multicolumn{2}{l}{} & \multicolumn{11}{c}{$N_\text{tot}$ (cm$^{-2}$)} \\
        \noalign{\smallskip}
        \cline{3-13}
        \noalign{\smallskip}
        Molecule & & G305.20 & G305.80 & G309.92 & G317.40 & G318.95 & G337.40 & MMS-6 & NGC 2071 & OMC-2 & OMC1-S & W51 e2 \\
        \noalign{\smallskip}
        \hline
        \noalign{\smallskip}
        $^{13}$CO & $(\times 10^{15})$ & ... & ... & ... & ... & ... & ... & 48.0(1.0) & ... & 56(2) & ... & ... \\
        C$^{18}$O & $(\times 10^{16})$ & ... & ... & ... & ... & ... & ... & 2.36(0.03) & 1.40(0.09) & 1.81(0.04) & ... & ... \\
        H$_{2}$CO & $(\times 10^{14})$ & 94.0(1.2) & 52(2) & 32.1(0.6) & ... & 44.0(1.0) & 81.2(0.7) & 1.1(0.1) & 3.5(0.5) & 0.33(0.02) & 28(2) & ... \\
        H$_{2}$$^{13}$CO & $(\times 10^{12})$ & ... & 118(4) & ... & ... & ... & 341(8) & 6.0(0.4) & 13.0(1.0) & 3.2(0.6) & 47.0(1.5) & 21(2) \\
        D$_{2}$CO & $(\times 10^{13})$ & ... & ... & ... & ... & ... & ... & 1.00(0.14) & ... & ... & ... & ... \\
        CH$_{3}$OH & $(\times 10^{15})$ & 77(3) & 80(3) & 21.6(0.7) & ... & 55.0(1.3) & 160(3) & 0.202(0.008) & 1.00(0.06) & ... & 7(3) & 2.2(1.0) \\
        $^{13}$CH$_{3}$OH & $(\times 10^{14})$ & 42(2) & 56(3) & 25(2) & ... & 34.0(1.2) & 68.0(1.1) & 1.0(0.2) & ... & ... & 7.1(0.7) & ... \\
        CH$_{2}$DOH & $(\times 10^{15})$ & ... & ... & ... & ... & 1.8(0.2) & ... & ... & ... & ... & ... & ... \\
        CH$_{3}$$^{18}$OH & $(\times 10^{14})$ & 5.1(0.6) & 3.2(1.0) & 2.5(0.3) & ... & 4.0(1.1) & 17(8) & ... & ... & ... & ... & ... \\
        CH$_{3}$CHO & $(\times 10^{13})$ & 36.0(1.3) & 19(2) & 7.7(0.4) & ... & 18.6(0.8) & 55(9) & ... & ... & ... & 8.6(0.2) & ... \\
        CH$_{3}$OCHO & $(\times 10^{15})$ & 6.3(0.1) & 16.0(0.6) & 2.56(0.08) & ... & 5.8(0.2) & 11.1(0.3) & ... & ... & ... & 2.5(0.2) & ... \\
        CH$_{3}$OCH$_{3}$ & $(\times 10^{14})$ & 320(30) & 91(8) & 120(20) & ... & 83(8) & 380(50) & 1.27(0.11) & 2.8(0.3) & ... & 12(2) & ... \\
        CH$_{3}$COCH$_{3}$ & $(\times 10^{14})$ & 6.1(0.3) & 9.4(0.3) & 1.13(0.08) & ... & 17.8(0.7) & 7.8(0.2) & ... & ... & ... & ... & ... \\
        C$_{2}$H$_{5}$OH & $(\times 10^{14})$ & 21(2) & 23(2) & 4.1(0.5) & ... & 25(2) & 55(3) & ... & ... & ... & ... & ... \\
        $^{13}$CN & $(\times 10^{16})$ & ... & ... & ... & ... & ... & ... & ... & ... & ... & ... & 0.0175(0.0020) \\
        C$^{15}$N & $(\times 10^{12})$ & ... & ... & ... & ... & ... & ... & ... & ... & ... & ... & ... \\
        HNCO & $(\times 10^{14})$ & ... & 51(2) & 0.70(0.02) & ... & 14.0(1.3) & 55(2) & ... & ... & ... & 5.4(0.8) & ... \\
        CH$_{3}$C$^{15}$N & $(\times 10^{13})$ & ... & ... & ... & ... & ... & 1.49(0.11) & ... & ... & ... & ... & ... \\
        C$_{2}$H$_{3}$CN & $(\times 10^{12})$ & 365(9) & 190(10) & ... & ... & ... & 350(8) & ... & ... & ... & ... & ... \\
        C$_{2}$H$_{5}$CN & $(\times 10^{13})$ & 86.0(1.1) & 39.0(1.2) & 3.8(0.2) & ... & 20.5(0.5) & 100.0(1.0) & ... & 1.3(0.2) & ... & 7.8(1.0) & ... \\
        CH$_{3}$NCO & $(\times 10^{13})$ & ... & ... & 8(2) & ... & ... & 31(2) & ... & ... & ... & ... & ... \\
        NH$_{2}$CHO & $(\times 10^{13})$ & 30(14) & 20.0(1.5) & 0.95(0.08) & ... & 10.8(0.7) & 27(2) & ... & ... & ... & 3.9(0.4) & ... \\
        NH$_{2}$CN & $(\times 10^{13})$ & ... & ... & ... & ... & 1.5(0.2) & ... & ... & ... & ... & ... & ... \\
        NH$_{2}$D & $(\times 10^{14})$ & ... & ... & ... & ... & 2.9(0.2) & 1.9(0.2) & ... & ... & ... & ... & ... \\
        HC$_{3}$N & $(\times 10^{12})$ & 286(2) & 124(2) & 53(2) & ... & 64.0(1.1) & ... & ... & ... & ... & ... & ... \\
        SO & $(\times 10^{14})$ & ... & ... & ... & ... & ... & ... & 0.83(0.03) & 1.96(0.13) & 0.88(0.05) & 10.9(0.3) & ... \\
        SO$_{2}$ & $(\times 10^{14})$ & 35(6) & 125(9) & 5.8(1.0) & ... & 9.8(1.5) & 65(6) & ... & ... & ... & 7.2(0.3) & ... \\
        $^{34}$SO$_{2}$ & $(\times 10^{14})$ & ... & ... & ... & ... & ... & ... & ... & ... & ... & ... & ... \\
        CCS & $(\times 10^{14})$ & ... & ... & ... & ... & ... & ... & ... & ... & ... & 0.0551(0.0060) & ... \\
        O$^{13}$CS & $(\times 10^{14})$ & ... & ... & 1.99(0.07) & ... & ... & 7.62(0.12) & ... & ... & ... & 0.31(0.05) & ... \\
        HDCS & $(\times 10^{12})$ & ... & ... & ... & ... & ... & ... & ... & 9(2) & ... & 25(2) & ... \\
        H$_{2}$$^{13}$CS & $(\times 10^{13})$ & 6.9(1.1) & 7(2) & 2.0(0.2) & ... & 6.1(1.1) & 11(2) & ... & ... & ... & 4.1(0.7) & ... \\
        SiO & $(\times 10^{11})$ & ... & ... & ... & ... & ... & ... & ... & 31(3) & ... & 370(20) & 125(4) \\
        c-C$_{3}$H$_{2}$ & $(\times 10^{12})$ & 180(20) & ... & ... & ... & ... & ... & 12.2(0.8) & 17.0(1.5) & ... & 31(2) & ... \\
        \noalign{\smallskip}
        \hline
        \noalign{\smallskip}
    \end{tabular}
\end{sidewaystable*}

            \begin{sidewaystable*}
    \centering
    \setlength{\tabcolsep}{3pt}
    \caption{Line width (FWHM) estimated for every detection with \textsc{MADCUBA}}
    \label{tab:fwhm1}
    \begin{tabular}{lccccccccccc}
        \hline
        \hline
        \noalign{\smallskip}
        \multicolumn{1}{l}{} & \multicolumn{11}{c}{FWHM (km/s)} \\
        \noalign{\smallskip}
        \cline{2-12}
        \noalign{\smallskip}
        Molecule & AFGL 5180 a & AFGL 5180 b & G010.62 & G011.94 & G012.81 & G045.47 & G049.27 & G049.37 & G058.77 & G061.48 a & G061.48 b \\
        \noalign{\smallskip}
        \hline
        \noalign{\smallskip}
        $^{13}$CO & ... & ... & ... & ... & ... & ... & ... & ... & 2.8(0.3) & 4.24(0.06) & 1.59(0.11) \\
        C$^{18}$O & 3.09(0.04) & 2.96(0.08) & 6.59(0.07) & ... & ... & 3.94(0.09) & 3.61(0.14) & ... & 3.14(0.07) & 2.63(0.04) & 1.30(0.08) \\
        H$_{2}$CO & 3.82(0.04) & 3.65(0.06) & 6.19(0.05) & ... & ... & 3.2(0.2) & 3.86(0.06) & 4.47(0.03) & 3.63(0.06) & 2.69(0.05) & 1.32(0.06) \\
        H$_{2}$$^{13}$CO & 3.37(0.13) & 3.2(0.2) & 6.19(0.12) & ... & ... & 3.9(0.2) & 2.7(0.5) & 3.8(0.2) & 3.0(0.2) & 1.7(0.2) & ... \\
        D$_{2}$CO & ... & ... & ... & ... & ... & ... & ... & ... & ... & ... & ... \\
        CH$_{3}$OH & 2.81(0.13) & 4.21(0.12) & 6.95(0.06) & 2.97(0.11) & 4.40(0.08) & 4.11(0.06) & 3.78(0.13) & 3.62(0.08) & 3.20(0.09) & 1.96(0.11) & 1.70(0.08) \\
        $^{13}$CH$_{3}$OH & ... & ... & 5.9(0.4) & ... & 5.3(1.1) & ... & ... & ... & ... & ... & ... \\
        CH$_{2}$DOH & ... & ... & ... & ... & ... & ... & ... & ... & ... & ... & ... \\
        CH$_{3}$$^{18}$OH & ... & ... & ... & ... & ... & ... & ... & ... & ... & ... & ... \\
        CH$_{3}$CHO & 3.5(0.3) & 3.5(0.5) & 5.8(0.3) & ... & 4.0(0.4) & ... & ... & 3.4(0.5) & 2.7(0.2) & ... & ... \\
        CH$_{3}$OCHO & ... & ... & 6.5(0.2) & ... & ... & ... & ... & ... & ... & ... & ... \\
        CH$_{3}$OCH$_{3}$ & 2.7(0.3) & 6.9(1.2) & 6.0(0.2) & ... & 5.2(0.5) & 5.0(0.6) & ... & 6.6(0.7) & 3.0(0.5) & ... & ... \\
        CH$_{3}$COCH$_{3}$ & ... & ... & ... & ... & ... & ... & ... & ... & ... & ... & ... \\
        C$_{2}$H$_{5}$OH & ... & ... & 6.6(0.3) & ... & 4.9(0.5) & ... & ... & ... & ... & ... & ... \\
        $^{13}$CN & ... & ... & ... & 3.4(0.1) & 5.26(0.09) & ... & ... & ... & ... & ... & ... \\
        C$^{15}$N & ... & ... & ... & ... & ... & ... & ... & 3.8(0.5) & 1.2(0.7) & ... & ... \\
        HNCO & ... & ... & ... & ... & ... & ... & ... & ... & ... & ... & ... \\
        CH$_{3}$C$^{15}$N & ... & ... & ... & ... & ... & ... & ... & ... & ... & ... & ... \\
        C$_{2}$H$_{3}$CN & ... & ... & 6.2(0.4) & ... & ... & ... & ... & ... & ... & ... & ... \\
        C$_{2}$H$_{5}$CN & ... & ... & 6.0(0.2) & ... & ... & ... & ... & ... & ... & ... & ... \\
        CH$_{3}$NCO & ... & ... & ... & ... & ... & ... & ... & ... & ... & ... & ... \\
        NH$_{2}$CHO & ... & ... & 6.5$^{\tablefootmark{a}}$ & ... & ... & ... & ... & ... & ... & ... & ... \\
        NH$_{2}$CN & ... & ... & ... & ... & ... & ... & ... & ... & ... & ... & ... \\
        NH$_{2}$D & ... & ... & ... & ... & ... & ... & ... & ... & ... & ... & ... \\
        HC$_{3}$N & ... & ... & ... & ... & ... & ... & ... & ... & ... & ... & ... \\
        SO & 3.38(0.05) & 3.3(0.1) & 6.97(0.07) & 3.07(0.11) & ... & 4.69(0.07) & 3.71(0.11) & 3.94(0.06) & 3.82(0.14) & 2.76(0.05) & 0.98(0.11) \\
        SO$_{2}$ & ... & ... & 5.28(0.13) & ... & ... & 5.1(0.4) & 5.2(1.1) & ... & ... & ... & ... \\
        $^{34}$SO$_{2}$ & ... & ... & 4.6(0.8) & ... & ... & ... & ... & ... & ... & ... & ... \\
        CCS & ... & ... & 6.5(0.7) & ... & ... & ... & ... & ... & ... & ... & ... \\
        O$^{13}$CS & ... & ... & 5.4(0.4) & ... & ... & ... & ... & ... & ... & ... & ... \\
        HDCS & 3.7(0.5) & 3.2(0.6) & ... & ... & ... & 2.7(0.6) & ... & 2.2(0.5) & ... & ... & ... \\
        H$_{2}$$^{13}$CS & ... & ... & 6.8(0.6) & ... & ... & ... & ... & ... & ... & ... & ... \\
        SiO & 4.9(0.2) & 5.1(0.2) & 9.2(0.2) & ... & 7.12(0.15) & 4.64(0.14) & 11.2(0.8) & 6.8(0.2) & 9.0(0.5) & 2.9(0.5) & 1.2(0.2) \\
        c-C$_{3}$H$_{2}$ & ... & ... & ... & ... & ... & ... & ... & 2.7(0.4) & ... & 2.1(0.2) & ... \\
        \noalign{\smallskip}
        \hline
        \noalign{\smallskip}
        average & $3.5(0.2)$ & $4.0(0.3)$ & $6.3(0.3)$ & $3.14(0.11)$ & $5.2(0.4)$ & $4.1(0.3)$ & $4.9(0.4)$ & $4.1(0.3)$ & $3.5(0.3)$ & $2.6(0.2)$ & $1.35(0.11)$ \\
        \noalign{\smallskip}
        \hline
        \noalign{\smallskip}
    \end{tabular}
    \tablefoot{
        \tablefoottext{a}{Fixed parameter in the line fit.}
    }
\end{sidewaystable*}

            \addtocounter{table}{-1}
\begin{sidewaystable*}
    \centering
    \setlength{\tabcolsep}{3pt}
    \caption{continued.}
    \label{tab:fwhm2}
    \begin{tabular}{lccccccccccc}
        \hline
        \hline
        \noalign{\smallskip}
        \multicolumn{1}{l}{} & \multicolumn{11}{c}{FWHM (km/s)} \\
        \noalign{\smallskip}
        \cline{2-12}
        \noalign{\smallskip}
        Molecule & G305.20 & G305.80 & G309.92 & G317.40 & G318.95 & G337.40 & MMS-6 & NGC 2071 & OMC-2 & OMC1-S & W51 e2 \\
        \noalign{\smallskip}
        \hline
        \noalign{\smallskip}
        $^{13}$CO & ... & ... & ... & ... & ... & ... & 1.95(0.05) & ... & 2.61(0.09) & ... & ... \\
        C$^{18}$O & ... & ... & ... & ... & ... & ... & 1.23(0.02) & 3.5(0.5) & 1.45(0.04) & ... & ... \\
        H$_{2}$CO & 6.12(0.09) & 8.2(0.3) & 3.52(0.07) & ... & 5.42(0.14) & 5.13(0.05) & 1.32(0.03) & 3.5(0.2) & 1.79(0.04) & 3.55(0.04) & ... \\
        H$_{2}$$^{13}$CO & ... & 7.9(0.3) & ... & ... & ... & 6.4(0.2) & 1.02(0.07) & 3.4(0.4) & 1.3(0.3) & 2.8(0.1) & 9.7(1.1) \\
        D$_{2}$CO & ... & ... & ... & ... & ... & ... & 1.9(0.4) & ... & ... & ... & ... \\
        CH$_{3}$OH & 6.9(0.2) & 7.7(0.2) & 4.17(0.09) & ... & 5.96(0.15) & 6.00(0.13) & 2.02(0.09) & 3.4(0.5) & ... & 2.85(0.12) & 9.5(0.2) \\
        $^{13}$CH$_{3}$OH & 7.8(0.3) & 6.4(0.2) & 3.2(0.2) & ... & 5.3(0.2) & 5.31(0.06) & 2.1(0.4) & ... & ... & 8.7(1.0) & ... \\
        CH$_{2}$DOH & ... & ... & ... & ... & 7.2(0.5) & ... & ... & ... & ... & ... & ... \\
        CH$_{3}$$^{18}$OH & 6.4(0.9) & 7.2(0.5) & 2.38(0.07) & ... & 4.6(0.2) & 5.3(0.2) & ... & ... & ... & ... & ... \\
        CH$_{3}$CHO & 6.3(0.2) & 7.6(0.9) & 5.0(0.3) & ... & 5.3(0.2) & 5.87(0.15) & ... & ... & ... & 2.69(0.08) & ... \\
        CH$_{3}$OCHO & 6.7(0.2) & 7.53(0.11) & 3.25(0.06) & ... & 5.44(0.09) & 5.60(0.08) & ... & ... & ... & 8.6(0.3) & ... \\
        CH$_{3}$OCH$_{3}$ & 6.7(0.6) & 6.0(0.4) & 3.5(0.3) & ... & 5.3(0.3) & 5.5(0.3) & 1.8(0.2) & 6.6(0.8) & ... & 5.6(0.2) & ... \\
        CH$_{3}$COCH$_{3}$ & 6.5(0.3) & 8.4(0.3) & 3.1(0.2) & ... & 5.7(0.3) & 5.60(0.13) & ... & ... & ... & ... & ... \\
        C$_{2}$H$_{5}$OH & 7.3(0.5) & 6.0(0.3) & 3.4(0.3) & ... & 5.8(0.2) & 6.4(0.2) & ... & ... & ... & ... & ... \\
        $^{13}$CN & ... & ... & ... & ... & ... & ... & ... & ... & ... & ... & 4.5(0.8) \\
        C$^{15}$N & ... & ... & ... & ... & ... & ... & ... & ... & ... & ... & ... \\
        HNCO & ... & 8.5(0.4) & 4.2(0.2) & ... & 4.8(0.5) & 7.2(0.3) & ... & ... & ... & 2.7(0.5) & ... \\
        CH$_{3}$C$^{15}$N & ... & ... & ... & ... & ... & 5.2(0.3) & ... & ... & ... & ... & ... \\
        C$_{2}$H$_{3}$CN & 7.3(0.2) & 8.3(0.5) & ... & ... & ... & 10.2(0.4) & ... & ... & ... & ... & ... \\
        C$_{2}$H$_{5}$CN & 6.1(0.1) & 8.7(0.2) & 3.8(0.2) & ... & 5.8(0.2) & 7.41(0.09) & ... & 2.1(0.4) & ... & 6.5(0.3) & ... \\
        CH$_{3}$NCO & ... & ... & 7(2) & ... & ... & 6.6(0.6) & ... & ... & ... & ... & ... \\
        NH$_{2}$CHO & 7.0$^{\tablefootmark{a}}$ & 11.1(0.6) & 4.5$^{\tablefootmark{a}}$ & ... & 6.7(0.3) & 7.3(0.3) & ... & ... & ... & 9.7(0.5) & ... \\
        NH$_{2}$CN & ... & ... & ... & ... & 4.5(0.4) & ... & ... & ... & ... & ... & ... \\
        NH$_{2}$D & ... & ... & ... & ... & 4.5(0.3) & 5.4(0.6) & ... & ... & ... & ... & ... \\
        HC$_{3}$N & 7.81(0.07) & 8.61(0.13) & 4.1(0.2) & ... & 5.63(0.11) & ... & ... & ... & ... & ... & ... \\
        SO & ... & ... & ... & ... & ... & ... & 1.17(0.05) & 3.3(0.5) & 1.5(0.1) & 3.54(0.11) & ... \\
        SO$_{2}$ & 12.8(0.9) & 9.2(0.4) & 4.5(0.4) & ... & 8.0(0.7) & 9.4(0.4) & ... & ... & ... & 8.0(0.3) & ... \\
        $^{34}$SO$_{2}$ & ... & ... & ... & ... & ... & ... & ... & ... & ... & ... & ... \\
        CCS & ... & ... & ... & ... & ... & ... & ... & ... & ... & 3.3(0.4) & ... \\
        O$^{13}$CS & ... & ... & 4.2(0.2) & ... & ... & 4.80(0.09) & ... & ... & ... & 3.1(0.6) & ... \\
        HDCS & ... & ... & ... & ... & ... & ... & ... & 1.3(0.3) & ... & 1.6(0.2) & ... \\
        H$_{2}$$^{13}$CS & 5.0$^{\tablefootmark{a}}$ & 6.0$^{\tablefootmark{a}}$ & 3.1(0.3) & ... & 15(3) & 6.6(1.5) & ... & ... & ... & 13(2) & ... \\
        SiO & ... & ... & ... & ... & ... & ... & ... & 3.5(0.4) & ... & 5.2(0.2) & 12.3(0.5) \\
        c-C$_{3}$H$_{2}$ & 6.8(0.8) & ... & ... & ... & ... & ... & 1.3(0.1) & 4.1(0.4) & ... & 3.5(0.2) & ... \\
        \noalign{\smallskip}
        \hline
        \noalign{\smallskip}
        average & $7.3(0.4)$ & $8.0(0.4)$ & $3.9(0.3)$ & ... & $6.1(0.4)$ & $6.4(0.3)$ & $1.58(0.14)$ & $3.5(0.4)$ & $1.74(0.12)$ & $5.3(0.4)$ & $9.0(0.6)$ \\
        \noalign{\smallskip}
        \hline
        \noalign{\smallskip}
    \end{tabular}
    \tablefoot{
        \tablefoottext{a}{Fixed parameter in the line fit.}
    }
\end{sidewaystable*}

            \begin{sidewaystable*}
    \centering
    \setlength{\tabcolsep}{3pt}
    \renewcommand{\arraystretch}{0.8}
    \caption{Excitation temperature ($T_\text{ex}$) estimated for every detection with \textsc{MADCUBA}}
    \label{tab:tex1}
    \begin{tabular}{lccccccccccc}
        \hline
        \hline
        \noalign{\smallskip}
        \multicolumn{1}{l}{} & \multicolumn{11}{c}{$T_\text{ex}$ (K)} \\
        \noalign{\smallskip}
        \cline{2-12}
        \noalign{\smallskip}
        Molecule & AFGL 5180 a & AFGL 5180 b & G010.62 & G011.94 & G012.81 & G045.47 & G049.27 & G049.37 & G058.77 & G061.48 a & G061.48 b \\
        \noalign{\smallskip}
        \hline
        \noalign{\smallskip}
        $^{13}$CO & ... & ... & ... & ... & ... & ... & ... & ... & 100.0$^{\tablefootmark{a}}$ & 100.0$^{\tablefootmark{a}}$ & 100.0$^{\tablefootmark{a}}$ \\
        C$^{18}$O & 100.0$^{\tablefootmark{a}}$ & 100.0$^{\tablefootmark{a}}$ & 100.0$^{\tablefootmark{a}}$ & ... & ... & 100.0$^{\tablefootmark{a}}$ & 100.0$^{\tablefootmark{a}}$ & ... & 100.0$^{\tablefootmark{a}}$ & 100.0$^{\tablefootmark{a}}$ & 100.0$^{\tablefootmark{a}}$ \\
        H$_{2}$CO & 103(5) & 48(2) & 260(20) & ... & ... & 230(60) & 108(7) & 125(4) & 71(5) & 79(6) & 70(13) \\
        H$_{2}$$^{13}$CO & 100.0$^{\tablefootmark{a}}$ & 100.0$^{\tablefootmark{a}}$ & 100.0$^{\tablefootmark{a}}$ & ... & ... & 100.0$^{\tablefootmark{a}}$ & 100.0$^{\tablefootmark{a}}$ & 100.0$^{\tablefootmark{a}}$ & 100.0$^{\tablefootmark{a}}$ & 100.0$^{\tablefootmark{a}}$ & ... \\
        D$_{2}$CO & ... & ... & ... & ... & ... & ... & ... & ... & ... & ... & ... \\
        CH$_{3}$OH & 16(3) & 16(5) & 94(9) & 12(5) & 15(2) & 14(2) & 22(9) & 11(2) & 14(4) & 9.7$^{\tablefootmark{a}}$ & 14.1$^{\tablefootmark{a}}$ \\
        $^{13}$CH$_{3}$OH & ... & ... & 100.0$^{\tablefootmark{a}}$ & ... & 100.0$^{\tablefootmark{a}}$ & ... & ... & ... & ... & ... & ... \\
        CH$_{2}$DOH & ... & ... & ... & ... & ... & ... & ... & ... & ... & ... & ... \\
        CH$_{3}$$^{18}$OH & ... & ... & ... & ... & ... & ... & ... & ... & ... & ... & ... \\
        CH$_{3}$CHO & 15(4) & 18.5$^{\tablefootmark{a}}$ & 60(10) & ... & 100.0$^{\tablefootmark{a}}$ & ... & ... & 50(50) & 36(9) & ... & ... \\
        CH$_{3}$OCHO & ... & ... & 160(13) & ... & ... & ... & ... & ... & ... & ... & ... \\
        CH$_{3}$OCH$_{3}$ & 100.0$^{\tablefootmark{a}}$ & 100.0$^{\tablefootmark{a}}$ & 120(14) & ... & 100.0$^{\tablefootmark{a}}$ & 100.0$^{\tablefootmark{a}}$ & ... & 100.0$^{\tablefootmark{a}}$ & 100.0$^{\tablefootmark{a}}$ & ... & ... \\
        CH$_{3}$COCH$_{3}$ & ... & ... & ... & ... & ... & ... & ... & ... & ... & ... & ... \\
        C$_{2}$H$_{5}$OH & ... & ... & 37(2) & ... & 8(3) & ... & ... & ... & ... & ... & ... \\
        $^{13}$CN & ... & ... & ... & 100.0$^{\tablefootmark{a}}$ & 100.0$^{\tablefootmark{a}}$ & ... & ... & ... & ... & ... & ... \\
        C$^{15}$N & ... & ... & ... & ... & ... & ... & ... & 100.0$^{\tablefootmark{a}}$ & 100.0$^{\tablefootmark{a}}$ & ... & ... \\
        HNCO & ... & ... & ... & ... & ... & ... & ... & ... & ... & ... & ... \\
        CH$_{3}$C$^{15}$N & ... & ... & ... & ... & ... & ... & ... & ... & ... & ... & ... \\
        C$_{2}$H$_{3}$CN & ... & ... & 100.0$^{\tablefootmark{a}}$ & ... & ... & ... & ... & ... & ... & ... & ... \\
        C$_{2}$H$_{5}$CN & ... & ... & 120(14) & ... & ... & ... & ... & ... & ... & ... & ... \\
        CH$_{3}$NCO & ... & ... & ... & ... & ... & ... & ... & ... & ... & ... & ... \\
        NH$_{2}$CHO & ... & ... & 260(70) & ... & ... & ... & ... & ... & ... & ... & ... \\
        NH$_{2}$CN & ... & ... & ... & ... & ... & ... & ... & ... & ... & ... & ... \\
        NH$_{2}$D & ... & ... & ... & ... & ... & ... & ... & ... & ... & ... & ... \\
        HC$_{3}$N & ... & ... & ... & ... & ... & ... & ... & ... & ... & ... & ... \\
        SO & 100.0$^{\tablefootmark{a}}$ & 100.0$^{\tablefootmark{a}}$ & 100.0$^{\tablefootmark{a}}$ & 100.0$^{\tablefootmark{a}}$ & ... & 100.0$^{\tablefootmark{a}}$ & 100.0$^{\tablefootmark{a}}$ & 100.0$^{\tablefootmark{a}}$ & 100.0$^{\tablefootmark{a}}$ & 100.0$^{\tablefootmark{a}}$ & 100.0$^{\tablefootmark{a}}$ \\
        SO$_{2}$ & ... & ... & 170(10) & ... & ... & 110(20) & 308.0$^{\tablefootmark{a}}$ & ... & ... & ... & ... \\
        $^{34}$SO$_{2}$ & ... & ... & 100.0$^{\tablefootmark{a}}$ & ... & ... & ... & ... & ... & ... & ... & ... \\
        CCS & ... & ... & 19.0$^{\tablefootmark{a}}$ & ... & ... & ... & ... & ... & ... & ... & ... \\
        O$^{13}$CS & ... & ... & 100.0$^{\tablefootmark{a}}$ & ... & ... & ... & ... & ... & ... & ... & ... \\
        HDCS & 100.0$^{\tablefootmark{a}}$ & 100.0$^{\tablefootmark{a}}$ & ... & ... & ... & 100.0$^{\tablefootmark{a}}$ & ... & 100.0$^{\tablefootmark{a}}$ & ... & ... & ... \\
        H$_{2}$$^{13}$CS & ... & ... & 100.0$^{\tablefootmark{a}}$ & ... & ... & ... & ... & ... & ... & ... & ... \\
        SiO & 100.0$^{\tablefootmark{a}}$ & 100.0$^{\tablefootmark{a}}$ & 100.0$^{\tablefootmark{a}}$ & ... & 100.0$^{\tablefootmark{a}}$ & 100.0$^{\tablefootmark{a}}$ & 100.0$^{\tablefootmark{a}}$ & 100.0$^{\tablefootmark{a}}$ & 100.0$^{\tablefootmark{a}}$ & 100.0$^{\tablefootmark{a}}$ & 100.0$^{\tablefootmark{a}}$ \\
        c-C$_{3}$H$_{2}$ & ... & ... & ... & ... & ... & ... & ... & 38.5$^{\tablefootmark{a}}$ & ... & 28(7) & ... \\
        \noalign{\smallskip}
        \hline
        \noalign{\smallskip}
        average & $45(4)$ & $32(3)$ & $140(20)$ & $12(5)$ & $12(3)$ & $120(30)$ & $65(8)$ & $60(20)$ & $41(6)$ & $54(6)$ & $70(13)$ \\
        \noalign{\smallskip}
        \hline
        \noalign{\smallskip}
        literature & \multicolumn{2}{c}{$50-150$ K$^{(1)}$} & $96-178$~K$^{(2)}$ & ... & ... & $20-81~$K$^{(3,4)}$; & ... & ... & $9.4~$K$^{(5)}$ & ... & ... \\
        \noalign{\smallskip}
        \hline
    \end{tabular}
    \tablefoot{
        \tablefoottext{a}{Fixed parameter in the line fit.}
    }
    \tablebib{
        (1)~\citet{minier2005star}; (2)~\citet{wong2018sma}; (3)~\citet{ge2014early};
        (4)~\citet{remijan2004survey};
        (5)~\citet{chen2025chemical}.
    }
\end{sidewaystable*}

            \addtocounter{table}{-1}
\begin{sidewaystable*}
    \centering
    \setlength{\tabcolsep}{3pt}
    \renewcommand{\arraystretch}{0.7}
    \caption{continued.}
    \label{tab:tex2}
    \begin{tabular}{lccccccccccc}
        \hline
        \hline
        \noalign{\smallskip}
        \multicolumn{1}{l}{} & \multicolumn{11}{c}{$T_\text{ex}$ (K)} \\
        \noalign{\smallskip}
        \cline{2-12}
        \noalign{\smallskip}
        Molecule & G305.20 & G305.80 & G309.92 & G317.40 & G318.95 & G337.40 & MMS-6 & NGC 2071 & OMC-2 & OMC1-S & W51 e2 \\
        \noalign{\smallskip}
        \hline
        \noalign{\smallskip}
        $^{13}$CO & ... & ... & ... & ... & ... & ... & 100.0$^{\tablefootmark{a}}$ & ... & 100.0$^{\tablefootmark{a}}$ & ... & ... \\
        C$^{18}$O & ... & ... & ... & ... & ... & ... & 100.0$^{\tablefootmark{a}}$ & 100.0$^{\tablefootmark{a}}$ & 100.0$^{\tablefootmark{a}}$ & ... & ... \\
        H$_{2}$CO & 100.0$^{\tablefootmark{a}}$ & 100.0$^{\tablefootmark{a}}$ & 100.0$^{\tablefootmark{a}}$ & ... & 100.0$^{\tablefootmark{a}}$ & 100.0$^{\tablefootmark{a}}$ & 40(3) & 70(10) & 29(3) & 129(9) & ... \\
        H$_{2}$$^{13}$CO & ... & 100.0$^{\tablefootmark{a}}$ & ... & ... & ... & 100.0$^{\tablefootmark{a}}$ & 100.0$^{\tablefootmark{a}}$ & 100.0$^{\tablefootmark{a}}$ & 100.0$^{\tablefootmark{a}}$ & 100.0$^{\tablefootmark{a}}$ & 100.0$^{\tablefootmark{a}}$ \\
        D$_{2}$CO & ... & ... & ... & ... & ... & ... & 36.8$^{\tablefootmark{a}}$ & ... & ... & ... & ... \\
        CH$_{3}$OH & 165(3) & 183(3) & 134(4) & ... & 179(3) & 148(2) & 20(12) & 20$^{\tablefootmark{a}}$ & ... & 40(20) & 13(4) \\
        $^{13}$CH$_{3}$OH & 117(5) & 165(5) & 190(10) & ... & 150(5) & 120.0(1.5) & 100.0$^{\tablefootmark{a}}$ & ... & ... & 100.0$^{\tablefootmark{a}}$ & ... \\
        CH$_{2}$DOH & ... & ... & ... & ... & 120(12) & ... & ... & ... & ... & ... & ... \\
        CH$_{3}$$^{18}$OH & 100.0$^{\tablefootmark{a}}$ & 50(20) & 100(10) & ... & 110(20) & 160(40) & ... & ... & ... & ... & ... \\
        CH$_{3}$CHO & 58(6) & 100.0$^{\tablefootmark{a}}$ & 60(10) & ... & 61(8) & 100(13) & ... & ... & ... & 45(3) & ... \\
        CH$_{3}$OCHO & 188.3$^{\tablefootmark{a}}$ & 380(15) & 218(7) & ... & 267(9) & 190(4) & ... & ... & ... & 160(12) & ... \\
        CH$_{3}$OCH$_{3}$ & 82(6) & 100(8) & 89(9) & ... & 110(10) & 66(3) & 100.0$^{\tablefootmark{a}}$ & 100.0$^{\tablefootmark{a}}$ & ... & 110(20) & ... \\
        CH$_{3}$COCH$_{3}$ & 100.0$^{\tablefootmark{a}}$ & 100.0$^{\tablefootmark{a}}$ & 100.0$^{\tablefootmark{a}}$ & ... & 176.7$^{\tablefootmark{a}}$ & 89.9$^{\tablefootmark{a}}$ & ... & ... & ... & ... & ... \\
        C$_{2}$H$_{5}$OH & 74(7) & 81(6) & 100(13) & ... & 150(8) & 161(9) & ... & ... & ... & ... & ... \\
        $^{13}$CN & ... & ... & ... & ... & ... & ... & ... & ... & ... & ... & 100.0$^{\tablefootmark{a}}$ \\
        C$^{15}$N & ... & ... & ... & ... & ... & ... & ... & ... & ... & ... & ... \\
        HNCO & ... & 100.0$^{\tablefootmark{a}}$ & 100.0$^{\tablefootmark{a}}$ & ... & 100.0$^{\tablefootmark{a}}$ & 100.0$^{\tablefootmark{a}}$ & ... & ... & ... & 100.0$^{\tablefootmark{a}}$ & ... \\
        CH$_{3}$C$^{15}$N & ... & ... & ... & ... & ... & 41(3) & ... & ... & ... & ... & ... \\
        C$_{2}$H$_{3}$CN & 100.0$^{\tablefootmark{a}}$ & 100.0$^{\tablefootmark{a}}$ & ... & ... & ... & 100.0$^{\tablefootmark{a}}$ & ... & ... & ... & ... & ... \\
        C$_{2}$H$_{5}$CN & 141(4) & 150(9) & 100(12) & ... & 139(8) & 121(3) & ... & 100.0$^{\tablefootmark{a}}$ & ... & 170(30) & ... \\
        CH$_{3}$NCO & ... & ... & 100.0$^{\tablefootmark{a}}$ & ... & ... & 100.0$^{\tablefootmark{a}}$ & ... & ... & ... & ... & ... \\
        NH$_{2}$CHO & 200(120) & 110(20) & 100.0$^{\tablefootmark{a}}$ & ... & 130(20) & 150(20) & ... & ... & ... & 140(20) & ... \\
        NH$_{2}$CN & ... & ... & ... & ... & 120(20) & ... & ... & ... & ... & ... & ... \\
        NH$_{2}$D & ... & ... & ... & ... & 100.0$^{\tablefootmark{a}}$ & 100.0$^{\tablefootmark{a}}$ & ... & ... & ... & ... & ... \\
        HC$_{3}$N & 100.0$^{\tablefootmark{a}}$ & 100.0$^{\tablefootmark{a}}$ & 100.0$^{\tablefootmark{a}}$ & ... & 100.0$^{\tablefootmark{a}}$ & ... & ... & ... & ... & ... & ... \\
        SO & ... & ... & ... & ... & ... & ... & 100.0$^{\tablefootmark{a}}$ & 100.0$^{\tablefootmark{a}}$ & 100.0$^{\tablefootmark{a}}$ & 100.0$^{\tablefootmark{a}}$ & ... \\
        SO$_{2}$ & 130(14) & 106(4) & 110(10) & ... & 170(40) & 126(7) & ... & ... & ... & 200(20) & ... \\
        $^{34}$SO$_{2}$ & ... & ... & ... & ... & ... & ... & ... & ... & ... & ... & ... \\
        CCS & ... & ... & ... & ... & ... & ... & ... & ... & ... & 100.0$^{\tablefootmark{a}}$ & ... \\
        O$^{13}$CS & ... & ... & 100.0$^{\tablefootmark{a}}$ & ... & ... & 100.0$^{\tablefootmark{a}}$ & ... & ... & ... & 100.0$^{\tablefootmark{a}}$ & ... \\
        HDCS & ... & ... & ... & ... & ... & ... & ... & 100.0$^{\tablefootmark{a}}$ & ... & 100.0$^{\tablefootmark{a}}$ & ... \\
        H$_{2}$$^{13}$CS & 100.0$^{\tablefootmark{a}}$ & 100.0$^{\tablefootmark{a}}$ & 100.0$^{\tablefootmark{a}}$ & ... & 100.0$^{\tablefootmark{a}}$ & 100.0$^{\tablefootmark{a}}$ & ... & ... & ... & 100.0$^{\tablefootmark{a}}$ & ... \\
        SiO & ... & ... & ... & ... & ... & ... & ... & 100.0$^{\tablefootmark{a}}$ & ... & 100.0$^{\tablefootmark{a}}$ & 100.0$^{\tablefootmark{a}}$ \\
        c-C$_{3}$H$_{2}$ & 100.0$^{\tablefootmark{a}}$ & ... & ... & ... & ... & ... & 18.2$^{\tablefootmark{a}}$ & 27(5) & ... & 19(3) & ... \\
        \noalign{\smallskip}
        \hline
        \noalign{\smallskip}
        average & $130(20)$ & $147(9)$ & $122(9)$ & ... & $140(13)$ & $120(10)$ & $31(8)$ & $46(8)$ & $29(3)$ & $110(20)$ & $13(4)$ \\
        \noalign{\smallskip}
        \hline
        \noalign{\smallskip}
        literature & $26-253$~K$^{(1, 2, 3)}$ & $139-335$~K$^{(1, 2)}$  & ... & ... & ... & $6-150$~K$^{(2, 4)}$ & $20-400$~K$^{(5)}$ & $100-215$~K$^{(6)}$ & $14-225$~K$^{(7, 8)}$ & ... & ... \\
        \noalign{\smallskip}
        \hline
    \end{tabular}
    \tablefoot{
        \tablefoottext{a}{Fixed parameter in the line fit.}
    }
    \tablebib{
        (1)~\citet{chen2025alma};
        (2)~\citet{qin2022atoms};
        (3)~\citet{walsh2007australia};
        (4)~\citet{yu2015molecular};
        (5)~\citet{hsu2022alma};
        (6)~\citet{van2014origin};
        (7)~\citet{de2025hot};
        (8)~\citet{szabo2023effelsberg}.
    }
\end{sidewaystable*}

            \begin{sidewaystable*}
    \centering
    \setlength{\tabcolsep}{3pt}
    \caption{Velocity of the sources with respect to the local standard rest ($v_\text{LSR}$) estimated for every detection with \textsc{MADCUBA}}
    \label{tab:vel1}
    \begin{tabular}{lccccccccccc}
        \hline
        \hline
        \noalign{\smallskip}
        \multicolumn{1}{l}{} & \multicolumn{11}{c}{$v_\text{LSR}$ (km/s)} \\
        \noalign{\smallskip}
        \cline{2-12}
        \noalign{\smallskip}
        Molecule & AFGL 5180 a & AFGL 5180 b & G010.62 & G011.94 & G012.81 & G045.47 & G049.27 & G049.37 & G058.77 & G061.48 a & G061.48 b \\
        \noalign{\smallskip}
        \hline
        \noalign{\smallskip}
        $^{13}$CO & ... & ... & ... & ... & ... & ... & ... & ... & 33.40(0.15) & 21.45(0.03) & 21.58(0.03) \\
        C$^{18}$O & 2.91(0.02) & 2.96(0.03) & -2.87(0.03) & ... & ... & 62.93(0.04) & 67.70(0.06) & ... & 32.70(0.03) & 21.73(0.02) & 21.40(0.05) \\
        H$_{2}$CO & 2.77(0.02) & 3.42(0.02) & -3.47(0.02) & ... & ... & 62.64(0.09) & 67.85(0.03) & 51.280(0.014) & 32.72(0.03) & 21.80(0.02) & 21.57(0.03) \\
        H$_{2}$$^{13}$CO & 2.83(0.06) & 2.86(0.09) & -2.64(0.05) & ... & ... & 63.7(0.1) & 68.4(0.2) & 51.5(0.1) & 32.84(0.09) & 21.81(0.09) & ... \\
        D$_{2}$CO & ... & ... & ... & ... & ... & ... & ... & ... & ... & ... & ... \\
        CH$_{3}$OH & 2.75(0.02) & 3.24(0.05) & -2.92(0.03) & 38.13(0.04) & 37.44(0.02) & 62.94(0.02) & 68.02(0.06) & 51.51(0.02) & 32.59(0.03) & 22.05(0.05) & 21.82(0.04) \\
        $^{13}$CH$_{3}$OH & ... & ... & -2.0(0.2) & ... & 38.0(0.5) & ... & ... & ... & ... & ... & ... \\
        CH$_{2}$DOH & ... & ... & ... & ... & ... & ... & ... & ... & ... & ... & ... \\
        CH$_{3}$$^{18}$OH & ... & ... & ... & ... & ... & ... & ... & ... & ... & ... & ... \\
        CH$_{3}$CHO & 2.41(0.11) & 2.9(0.2) & -1.98(0.14) & ... & 37.5(0.2) & ... & ... & 51.4(0.2) & 32.7(0.1) & ... & ... \\
        CH$_{3}$OCHO & ... & ... & -2.06(0.09) & ... & ... & ... & ... & ... & ... & ... & ... \\
        CH$_{3}$OCH$_{3}$ & 2.47(0.14) & 4.6(0.5) & -2.42(0.08) & ... & 37.8(0.2) & 64.4(0.3) & ... & 52.0(0.3) & 32.8(0.2) & ... & ... \\
        CH$_{3}$COCH$_{3}$ & ... & ... & ... & ... & ... & ... & ... & ... & ... & ... & ... \\
        C$_{2}$H$_{5}$OH & ... & ... & -1.83(0.13) & ... & 37.3(0.2) & ... & ... & ... & ... & ... & ... \\
        $^{13}$CN & ... & ... & ... & 36.27(0.04) & 35.34(0.04) & ... & ... & ... & ... & ... & ... \\
        C$^{15}$N & ... & ... & ... & ... & ... & ... & ... & 50.9(0.2) & 31.0(0.3) & ... & ... \\
        HNCO & ... & ... & ... & ... & ... & ... & ... & ... & ... & ... & ... \\
        CH$_{3}$C$^{15}$N & ... & ... & ... & ... & ... & ... & ... & ... & ... & ... & ... \\
        C$_{2}$H$_{3}$CN & ... & ... & -1.1(0.2) & ... & ... & ... & ... & ... & ... & ... & ... \\
        C$_{2}$H$_{5}$CN & ... & ... & -0.8(0.1) & ... & ... & ... & ... & ... & ... & ... & ... \\
        CH$_{3}$NCO & ... & ... & ... & ... & ... & ... & ... & ... & ... & ... & ... \\
        NH$_{2}$CHO & ... & ... & 0.3(0.2) & ... & ... & ... & ... & ... & ... & ... & ... \\
        NH$_{2}$CN & ... & ... & ... & ... & ... & ... & ... & ... & ... & ... & ... \\
        NH$_{2}$D & ... & ... & ... & ... & ... & ... & ... & ... & ... & ... & ... \\
        HC$_{3}$N & ... & ... & ... & ... & ... & ... & ... & ... & ... & ... & ... \\
        SO & 2.96(0.02) & 3.08(0.04) & -2.46(0.03) & 38.38(0.05) & ... & 63.67(0.03) & 67.67(0.05) & 51.25(0.02) & 32.52(0.06) & 21.79(0.02) & 21.72(0.05) \\
        SO$_{2}$ & ... & ... & -0.93(0.05) & ... & ... & 64.3(0.2) & 67.3(0.5) & ... & ... & ... & ... \\
        $^{34}$SO$_{2}$ & ... & ... & -0.7(0.3) & ... & ... & ... & ... & ... & ... & ... & ... \\
        CCS & ... & ... & -2.4(0.3) & ... & ... & ... & ... & ... & ... & ... & ... \\
        O$^{13}$CS & ... & ... & -1.0(0.2) & ... & ... & ... & ... & ... & ... & ... & ... \\
        HDCS & 2.8(0.2) & 2.7(0.2) & ... & ... & ... & 63.2(0.3) & ... & 51.8(0.2) & ... & ... & ... \\
        H$_{2}$$^{13}$CS & ... & ... & -2.8(0.3) & ... & ... & ... & ... & ... & ... & ... & ... \\
        SiO & 2.83(0.08) & 2.97(0.07) & -2.67(0.06) & ... & 36.77(0.06) & 63.24(0.06) & 67.7(0.3) & 51.43(0.09) & 32.2(0.2) & 21.5(0.2) & 23.86(0.09) \\
        c-C$_{3}$H$_{2}$ & ... & ... & ... & ... & ... & ... & ... & 51.0(0.2) & ... & 21.61(0.11) & ... \\
        \noalign{\smallskip}
        \hline
        \noalign{\smallskip}
        average & $2.75(0.07)$ & $3.19(0.14)$ & $-1.93(0.13)$ & $37.59(0.04)$ & $37.2(0.2)$ & $63.46(0.11)$ & $67.8(0.2)$ & $51.4(0.13)$ & $32.55(0.12)$ & $21.72(0.07)$ & $21.99(0.05)$ \\
        \noalign{\smallskip}
        \hline
        \noalign{\smallskip}
    \end{tabular}
\end{sidewaystable*}

            \addtocounter{table}{-1}
\begin{sidewaystable*}
    \centering
    \setlength{\tabcolsep}{3pt}
    \caption{continued.}
    \label{tab:vel2}
    \begin{tabular}{lccccccccccc}
        \hline
        \hline
        \noalign{\smallskip}
        \multicolumn{1}{l}{} & \multicolumn{11}{c}{$v_\text{LSR}$ (km/s)} \\
        \noalign{\smallskip}
        \cline{2-12}
        \noalign{\smallskip}
        Molecule & G305.20 & G305.80 & G309.92 & G317.40 & G318.95 & G337.40 & MMS-6 & NGC 2071 & OMC-2 & OMC1-S & W51 e2 \\
        \noalign{\smallskip}
        \hline
        \noalign{\smallskip}
        $^{13}$CO & ... & ... & ... & ... & ... & ... & 11.29(0.02) & ... & 10.69(0.04) & ... & ... \\
        C$^{18}$O & ... & ... & ... & ... & ... & ... & 11.274(0.008) & 7.5$^{\tablefootmark{a}}$ & 10.43(0.02) & ... & ... \\
        H$_{2}$CO & -40.16(0.04) & -31.63(0.13) & -59.71(0.03) & ... & -34.18(0.06) & -39.71(0.02) & 11.176(0.012) & 7.40(0.08) & 10.38(0.02) & 6.24(0.02) & ... \\
        H$_{2}$$^{13}$CO & ... & -32.39(0.13) & ... & ... & ... & -39.80(0.08) & 11.36(0.03) & 7.6(0.2) & 10.51(0.13) & 6.27(0.04) & 56.2(0.5) \\
        D$_{2}$CO & ... & ... & ... & ... & ... & ... & 11.25(0.13) & ... & ... & ... & ... \\
        CH$_{3}$OH & -40.12(0.09) & -33.2(0.1) & -59.83(0.04) & ... & -34.18(0.06) & -39.73(0.06) & 11.13(0.04) & 7.3(0.2) & ... & 6.10(0.04) & 57.3(0.1) \\
        $^{13}$CH$_{3}$OH & -40.20(0.13) & -34.2(0.1) & -59.96(0.06) & ... & -34.48(0.08) & -39.75(0.03) & 11.6(0.2) & ... & ... & 4.7(0.4) & ... \\
        CH$_{2}$DOH & ... & ... & ... & ... & -33.7(0.2) & ... & ... & ... & ... & ... & ... \\
        CH$_{3}$$^{18}$OH & -40.0$^{\tablefootmark{a}}$ & -34.4(0.2) & -59.98(0.03) & ... & -34.47(0.11) & -39.41(0.09) & ... & ... & ... & ... & ... \\
        CH$_{3}$CHO & -40.31(0.09) & -32.2(0.4) & -59.38(0.11) & ... & -34.59(0.08) & -39.74(0.06) & ... & ... & ... & 6.14(0.04) & ... \\
        CH$_{3}$OCHO & -40.14(0.06) & -33.46(0.04) & -59.70(0.02) & ... & -34.41(0.03) & -39.83(0.03) & ... & ... & ... & 5.72(0.13) & ... \\
        CH$_{3}$OCH$_{3}$ & -40.2(0.2) & -34.02(0.15) & -59.88(0.14) & ... & -34.61(0.15) & -40.10(0.15) & 11.42(0.08) & 10.0(0.3) & ... & 6.3(0.1) & ... \\
        CH$_{3}$COCH$_{3}$ & -40.10(0.15) & -34.02(0.15) & -59.5(0.1) & ... & -35.01(0.12) & -39.71(0.06) & ... & ... & ... & ... & ... \\
        C$_{2}$H$_{5}$OH & -39.5(0.2) & -34.34(0.14) & -59.93(0.11) & ... & -35.16(0.09) & -39.6(0.1) & ... & ... & ... & ... & ... \\
        $^{13}$CN & ... & ... & ... & ... & ... & ... & ... & ... & ... & ... & 58.3(0.3) \\
        C$^{15}$N & ... & ... & ... & ... & ... & ... & ... & ... & ... & ... & ... \\
        HNCO & ... & -33.6(0.2) & -59.74(0.07) & ... & -34.6(0.2) & -39.85(0.11) & ... & ... & ... & 6.5(0.2) & ... \\
        CH$_{3}$C$^{15}$N & ... & ... & ... & ... & ... & -40.86(0.12) & ... & ... & ... & ... & ... \\
        C$_{2}$H$_{3}$CN & -40.56(0.09) & -32.5(0.2) & ... & ... & ... & -40.90(0.13) & ... & ... & ... & ... & ... \\
        C$_{2}$H$_{5}$CN & -40.49(0.04) & -32.8(0.1) & -59.35(0.09) & ... & -34.28(0.07) & -39.84(0.04) & ... & 6.0(0.2) & ... & 7.74(0.13) & ... \\
        CH$_{3}$NCO & ... & ... & -59.2(0.7) & ... & ... & -39.1(0.2) & ... & ... & ... & ... & ... \\
        NH$_{2}$CHO & -37.6(0.2) & -32.0(0.3) & -59.3(0.2) & ... & -33.58(0.11) & -38.47(0.14) & ... & ... & ... & 3.7(0.2) & ... \\
        NH$_{2}$CN & ... & ... & ... & ... & -35.6(0.2) & ... & ... & ... & ... & ... & ... \\
        NH$_{2}$D & ... & ... & ... & ... & -35.01(0.13) & -41.8(0.3) & ... & ... & ... & ... & ... \\
        HC$_{3}$N & -40.65(0.03) & -31.51(0.06) & -58.94(0.07) & ... & -34.10(0.05) & ... & ... & ... & ... & ... & ... \\
        SO & ... & ... & ... & ... & ... & ... & 11.15(0.02) & 7.5(0.2) & 10.40(0.04) & 6.18(0.05) & ... \\
        SO$_{2}$ & -40.2(0.4) & -31.7(0.2) & -59.6(0.2) & ... & -35.0(0.3) & -39.7(0.2) & ... & ... & ... & 7.62(0.14) & ... \\
        $^{34}$SO$_{2}$ & ... & ... & ... & ... & ... & ... & ... & ... & ... & ... & ... \\
        CCS & ... & ... & ... & ... & ... & ... & ... & ... & ... & 6.1(0.2) & ... \\
        O$^{13}$CS & ... & ... & -59.02(0.07) & ... & ... & -40.00(0.04) & ... & ... & ... & 6.4(0.2) & ... \\
        HDCS & ... & ... & ... & ... & ... & ... & ... & 9.93(0.13) & ... & 6.37(0.08) & ... \\
        H$_{2}$$^{13}$CS & -39.7(0.5) & -37.0(1.0) & -59.31(0.11) & ... & -35.1$^{\tablefootmark{a}}$ & -40.7(0.6) & ... & ... & ... & 7.1(1.0) & ... \\
        SiO & ... & ... & ... & ... & ... & ... & ... & 7.2(0.2) & ... & 5.56(0.11) & 58.2(0.2) \\
        c-C$_{3}$H$_{2}$ & -39.6(0.3) & ... & ... & ... & ... & ... & 11.04(0.04) & 10.3(0.2) & ... & 6.14(0.11) & ... \\
        \noalign{\smallskip}
        \hline
        \noalign{\smallskip}
        average & $-40.0(0.2)$ & $-33.2(0.2)$ & $-59.55(0.12)$ & ... & $-34.53(0.12)$ & $-39.93(0.13)$ & $11.27(0.06)$ & $8.1(0.2)$ & $10.48(0.05)$ & $6.2(0.2)$ & $57.5(0.3)$ \\
        \noalign{\smallskip}
        \hline
        \noalign{\smallskip}
    \end{tabular}
    \tablefoot{
        \tablefoottext{a}{Fixed parameter in the line fit.}
    }
\end{sidewaystable*}

            The best-fit parameters obtained with \textsc{MADCUBA}, through the procedure described in Sect.~\ref{sect:lineidentification}. In Table \ref{tab:dens1}, \ref{tab:fwhm1}, \ref{tab:tex1}, and \ref{tab:vel1}, we present the column densities ($N_\text{tot}$), the line widths (FWHM), the excitation temperatures ($T_\text{ex}$), and the source velocities ($v_\text{LSR}$), respectively.
    
        \section{Spectra of the sample}
            \label{app:spectra_SOMA}
            The observed and the synthetic spectra of our sample (described in Sect.~\ref{sect:detections}) are shown in Figs.~\ref{fig:spectra1}-\ref{fig:spectra20}. 
            \insertplot{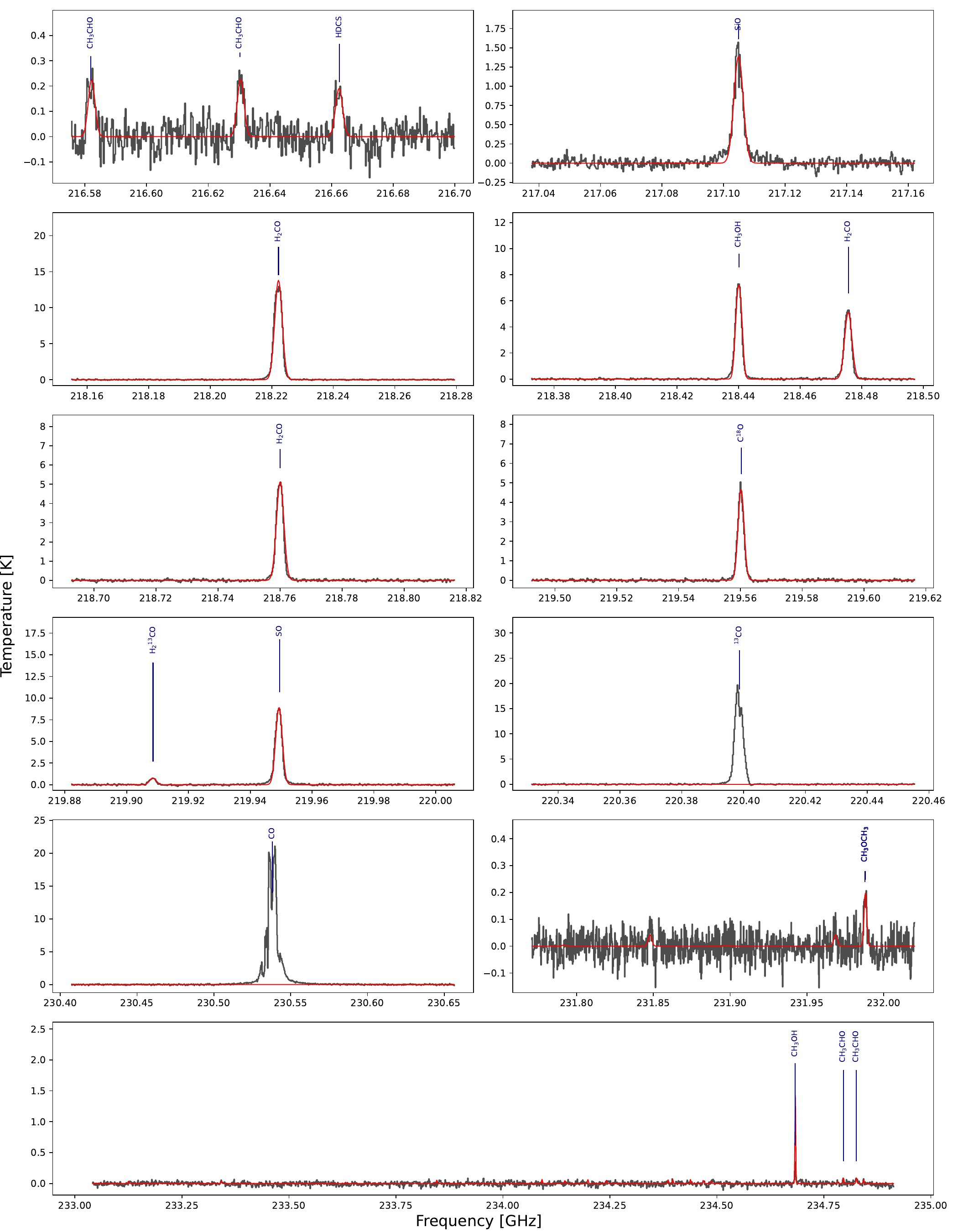}{The continuum-subtracted spectra of AFGL 5180 a (in grey). The synthetic spectra are overplotted (in red). Every detected transition is shown (with a vertical line) and labeled with the respective molecule. The displayed frequencies are in the rest frame (i.e., shifted from the observed frequencies using the $v_\text{LSR}$ shown in Table~\ref{tab:sources}).}{fig:spectra1}
            \insertplot{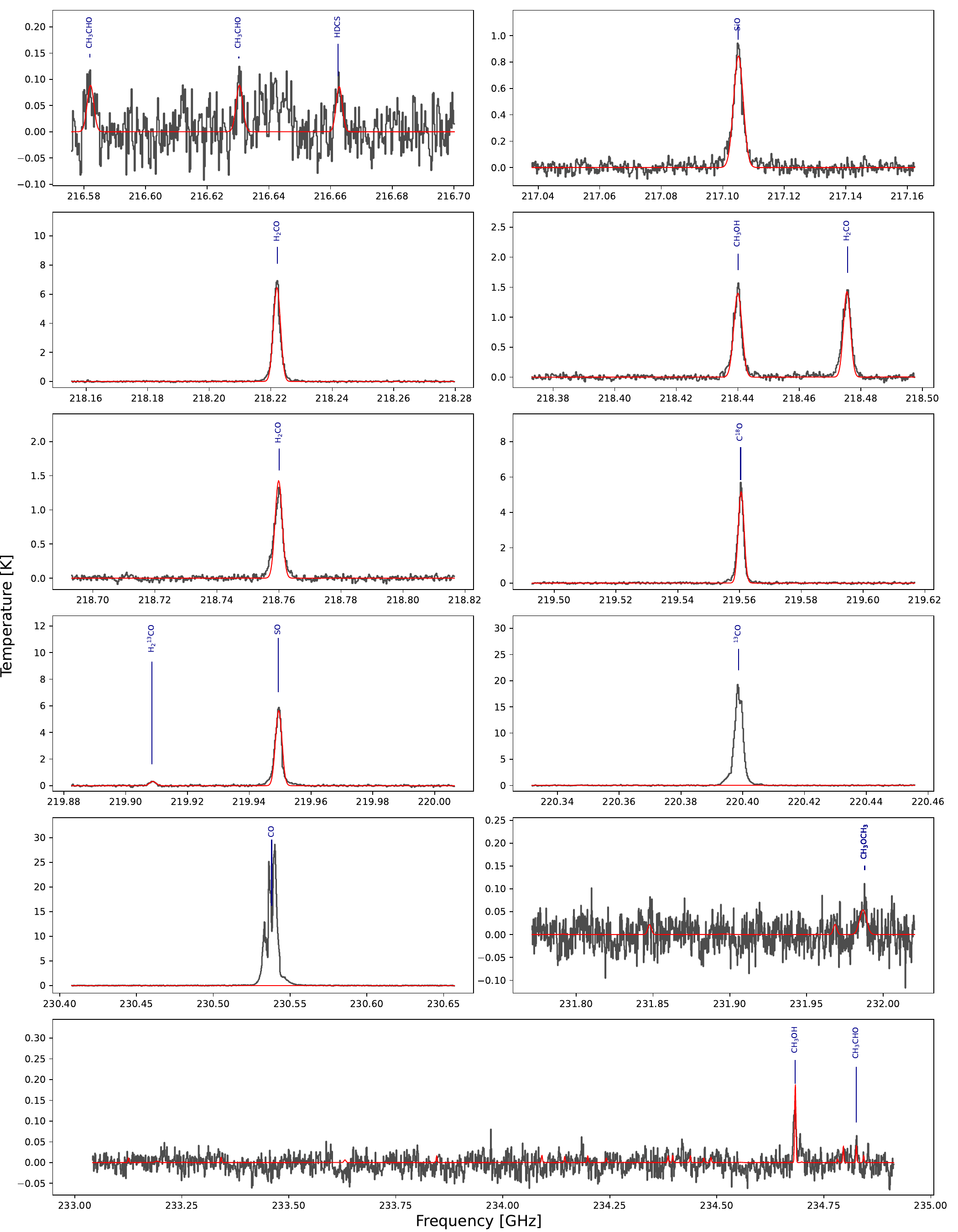}{Same for AFLG 5180 b.}{fig:spectra2}
            \insertplot{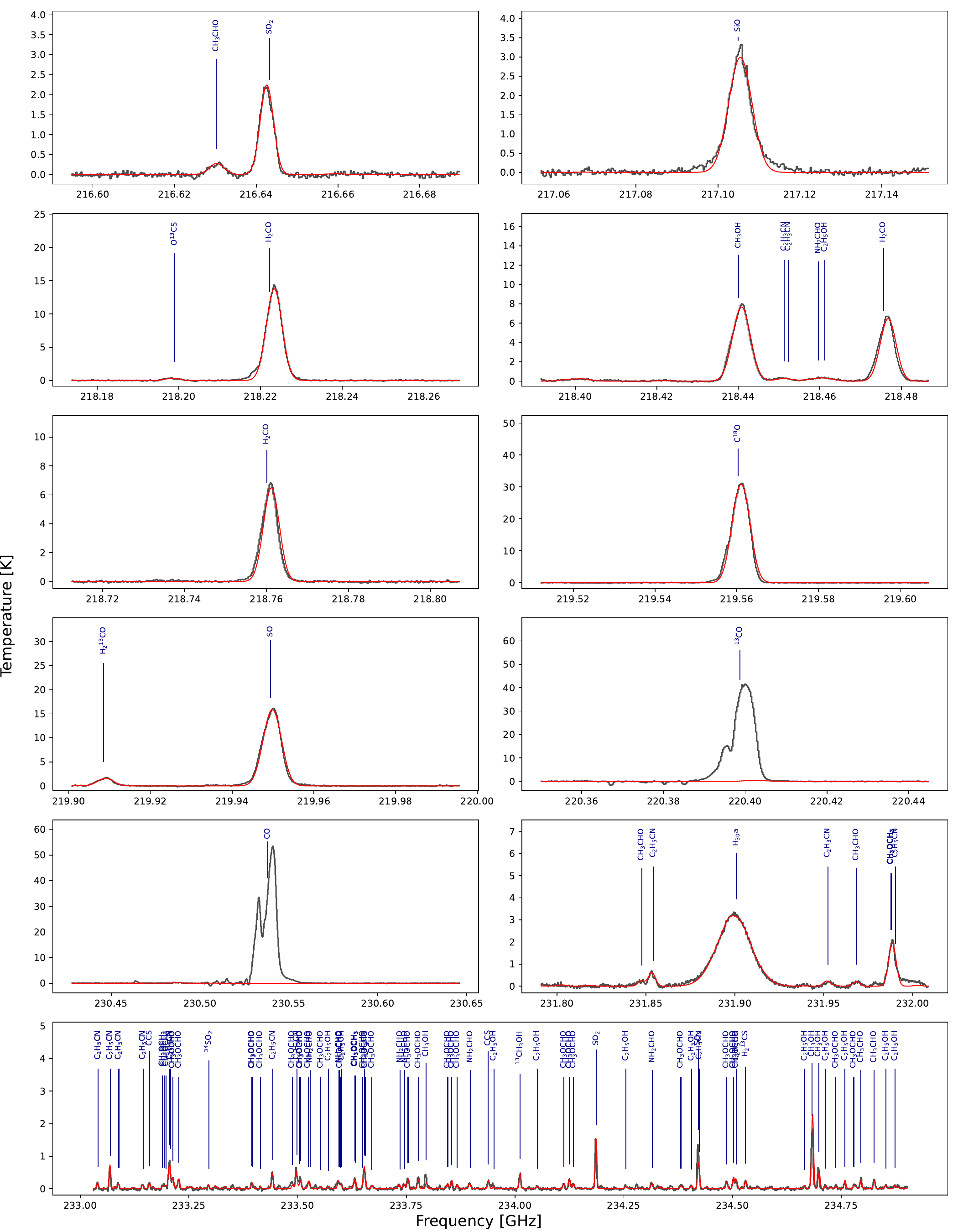}{Same for G010.62.}{fig:spectra3}
            \insertplot{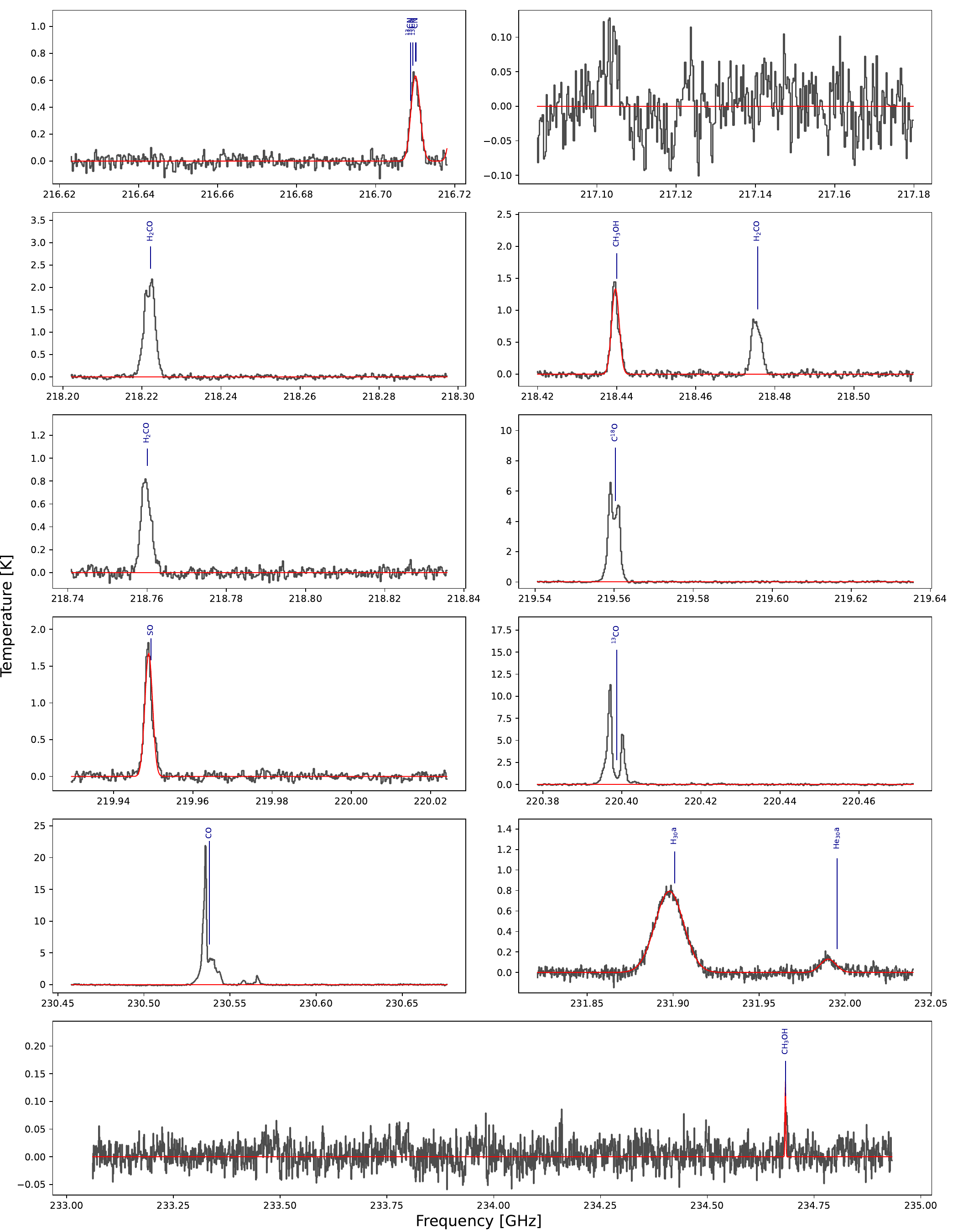}{Same for G011.94.}{fig:spectra4}
            \insertplot{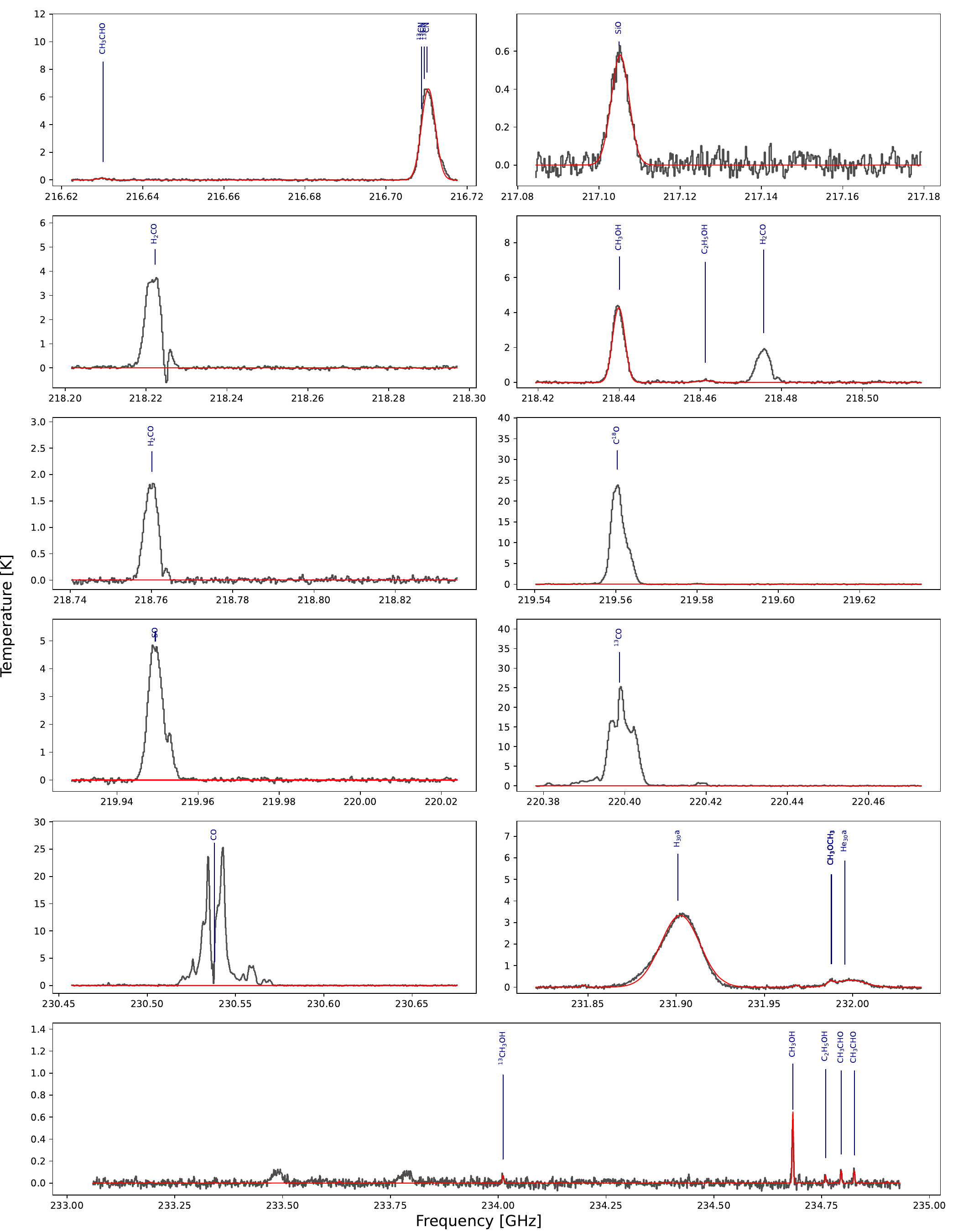}{Same for G012.81.}{fig:spectra5}
            \insertplot{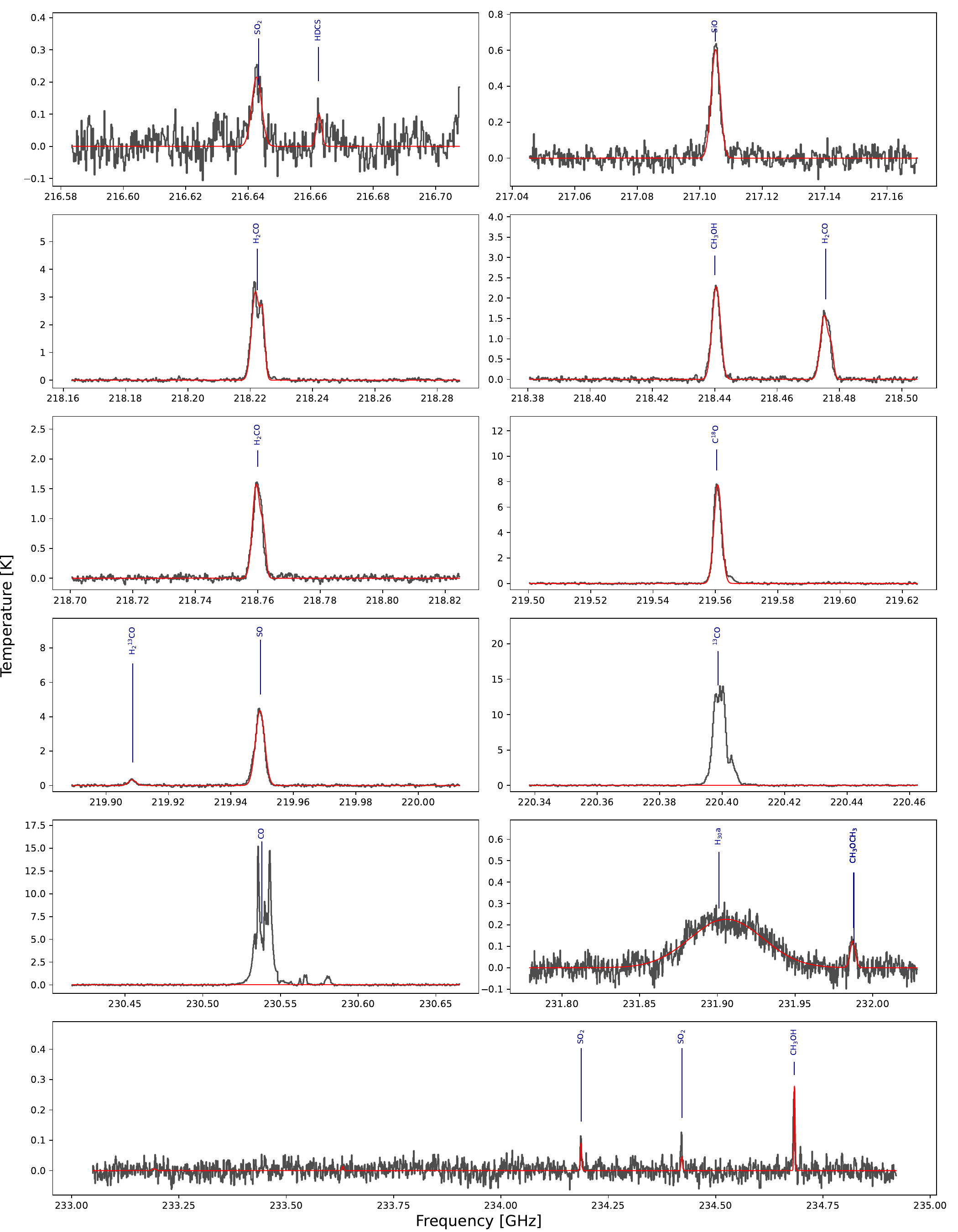}{Same for G045.47.}{fig:spectra6}
            \insertplot{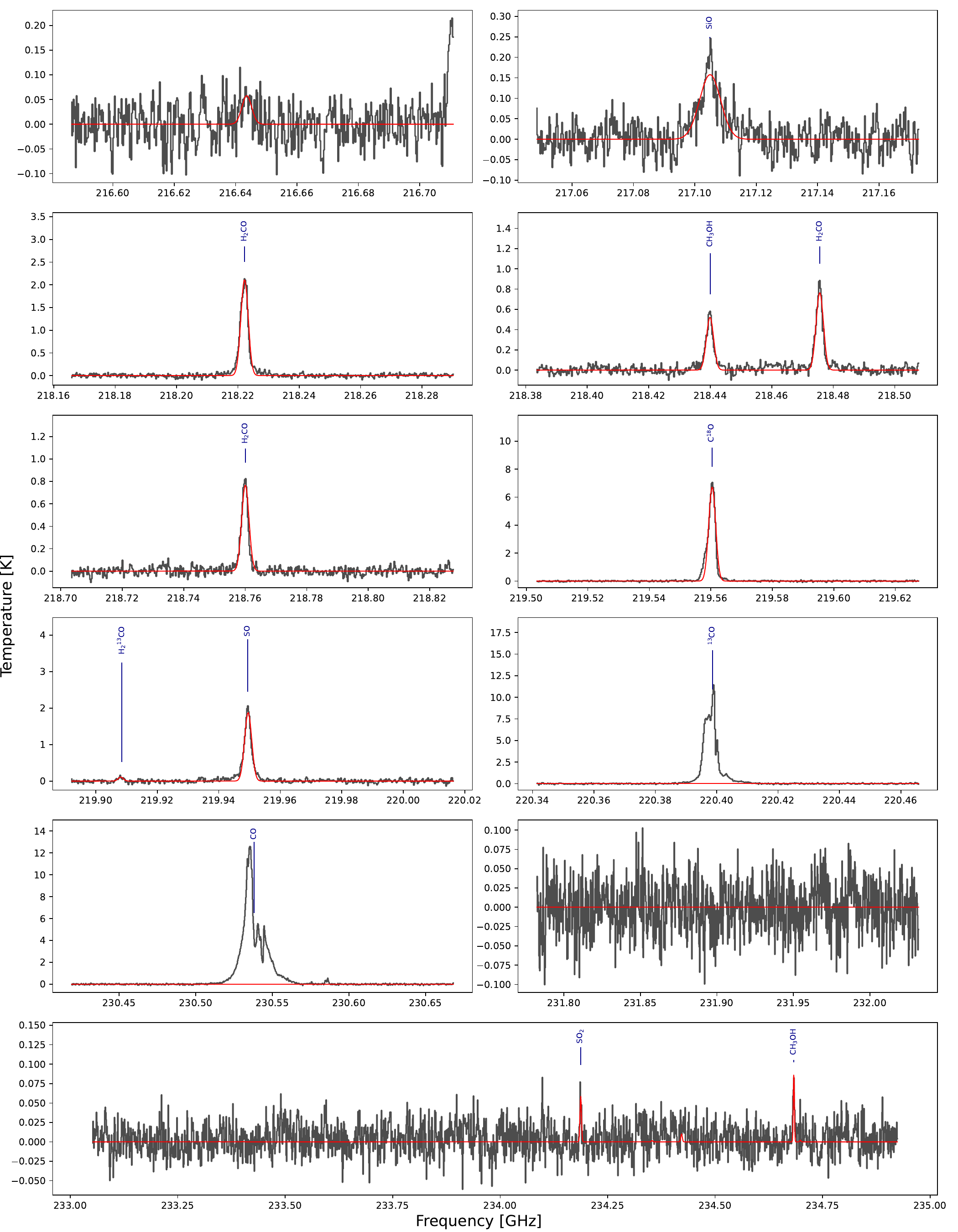}{Same for G049.27.}{fig:spectra7}
            \insertplot{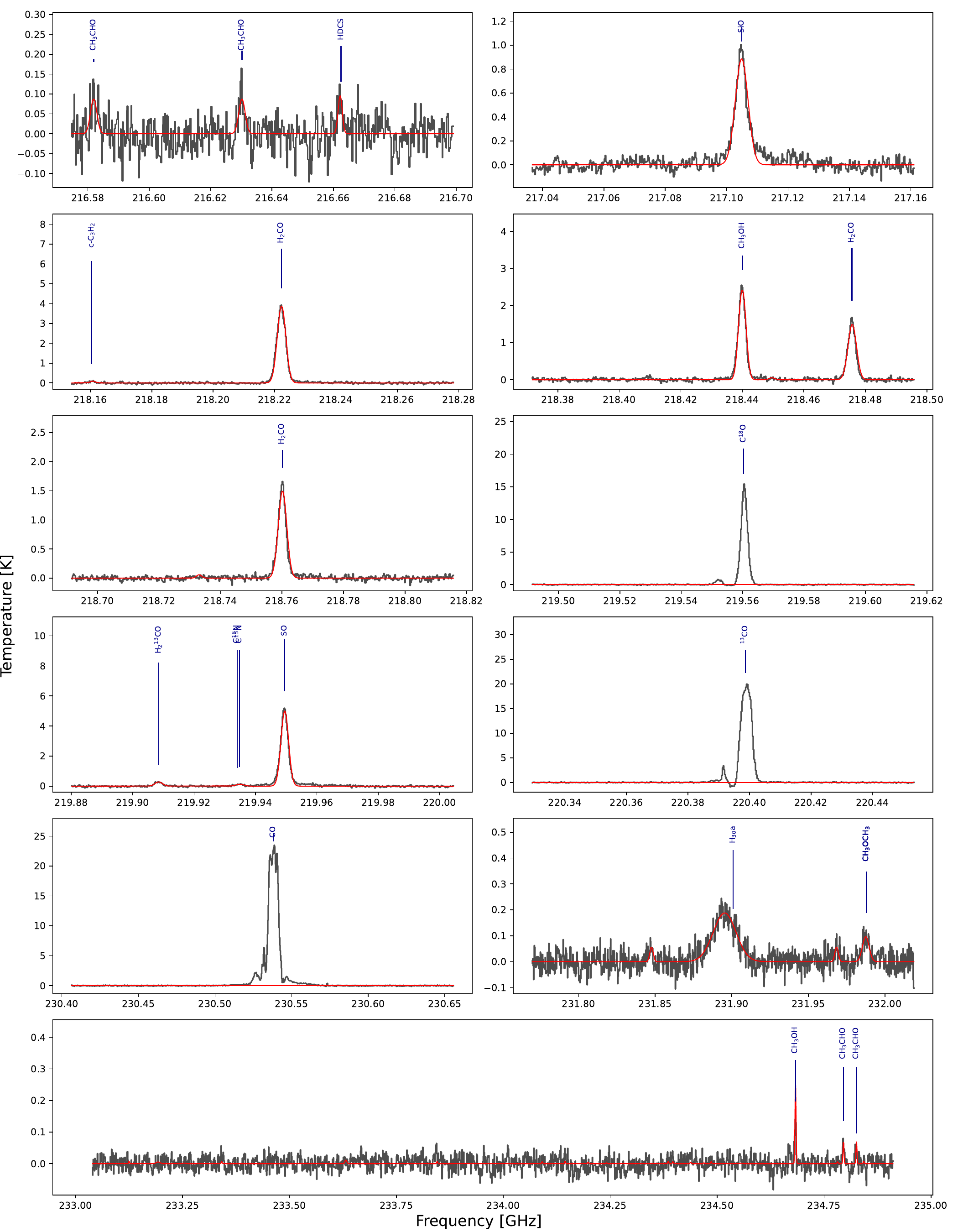}{Same for G049.37.}{fig:spectra8}
            \insertplot{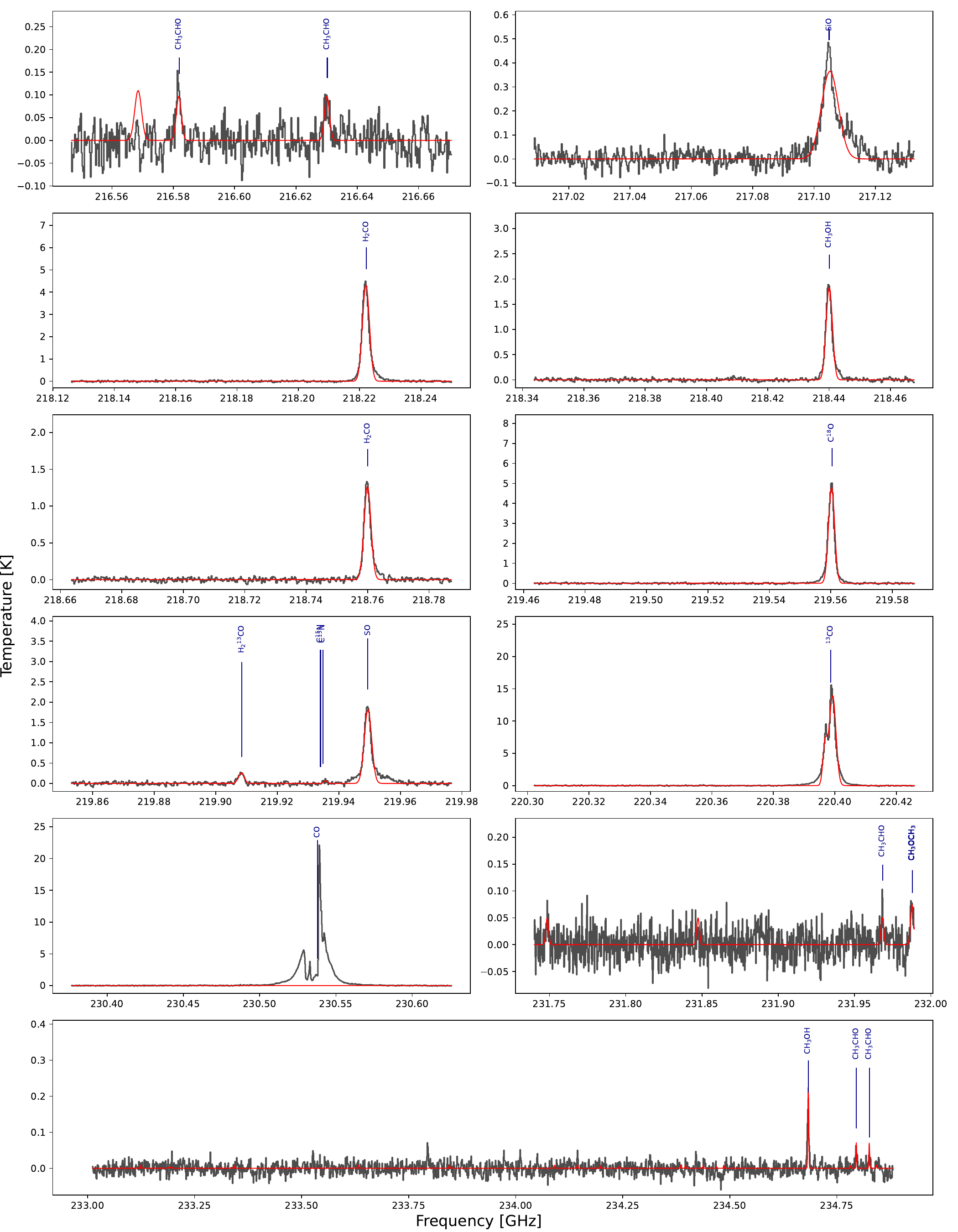}{Same for G058.77.}{fig:spectra9}
            \insertplot{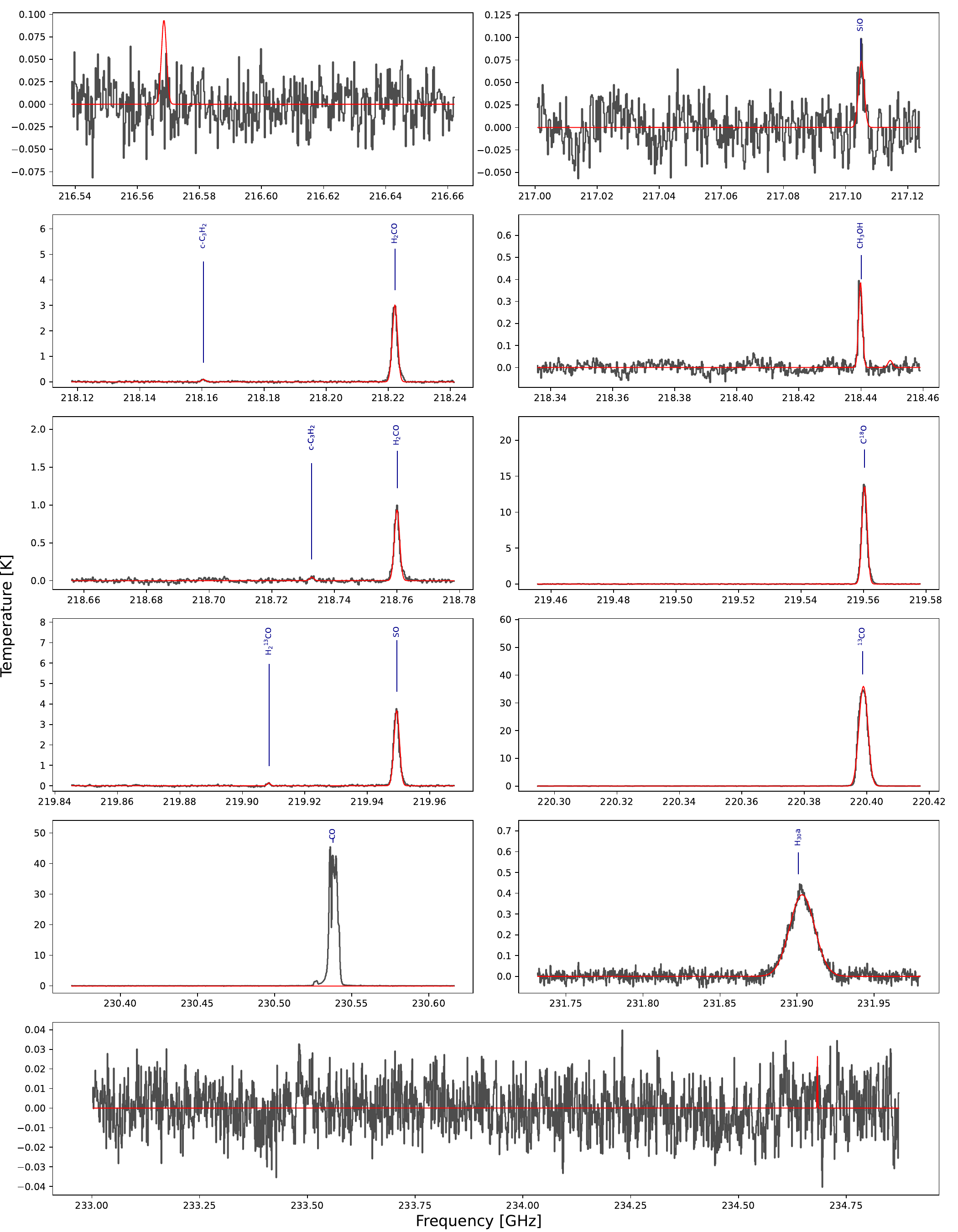}{Same for G061.48 a.}{fig:spectra10}
            \insertplot{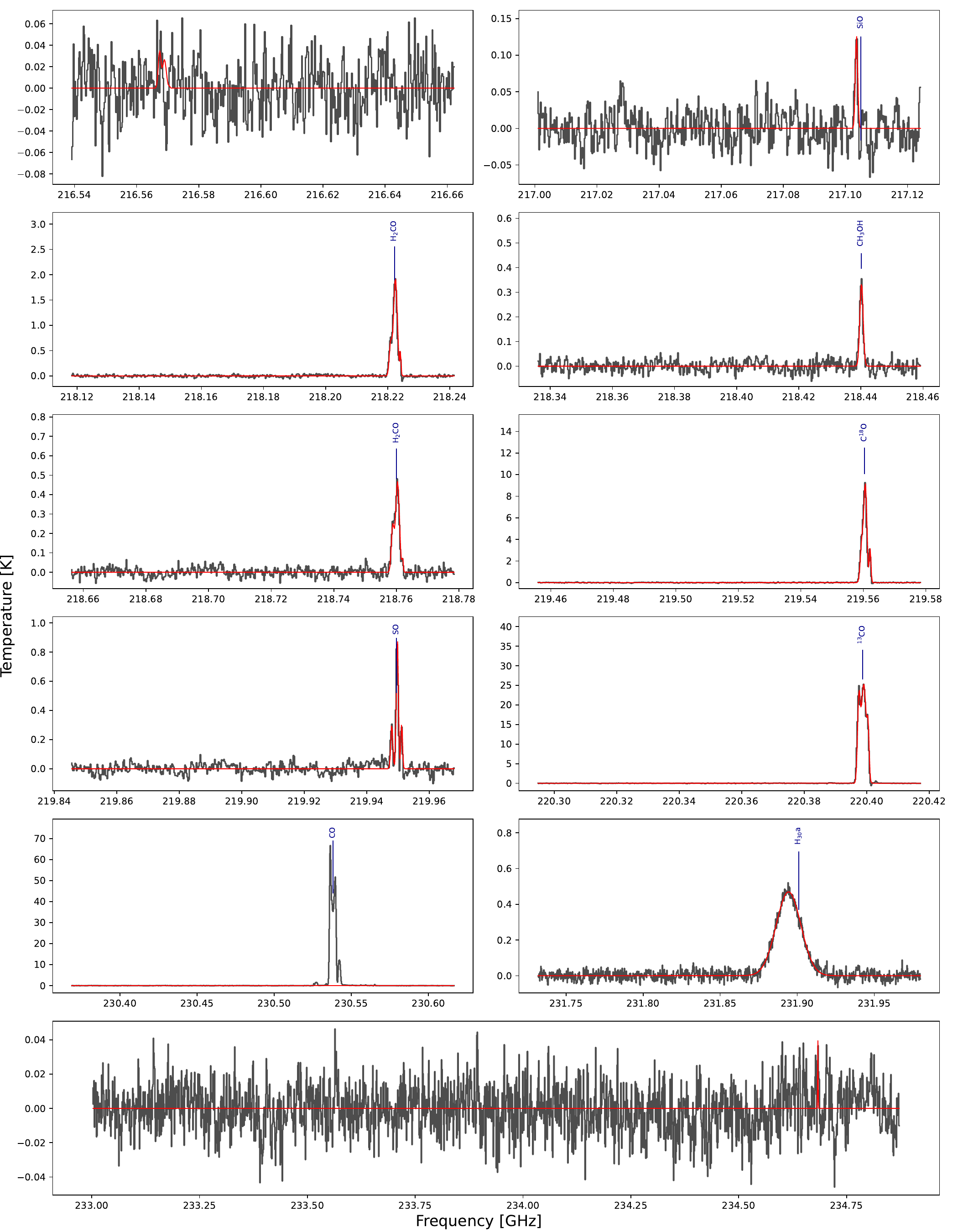}{Same for G061.48 b.}{fig:spectra11}
            \insertplot{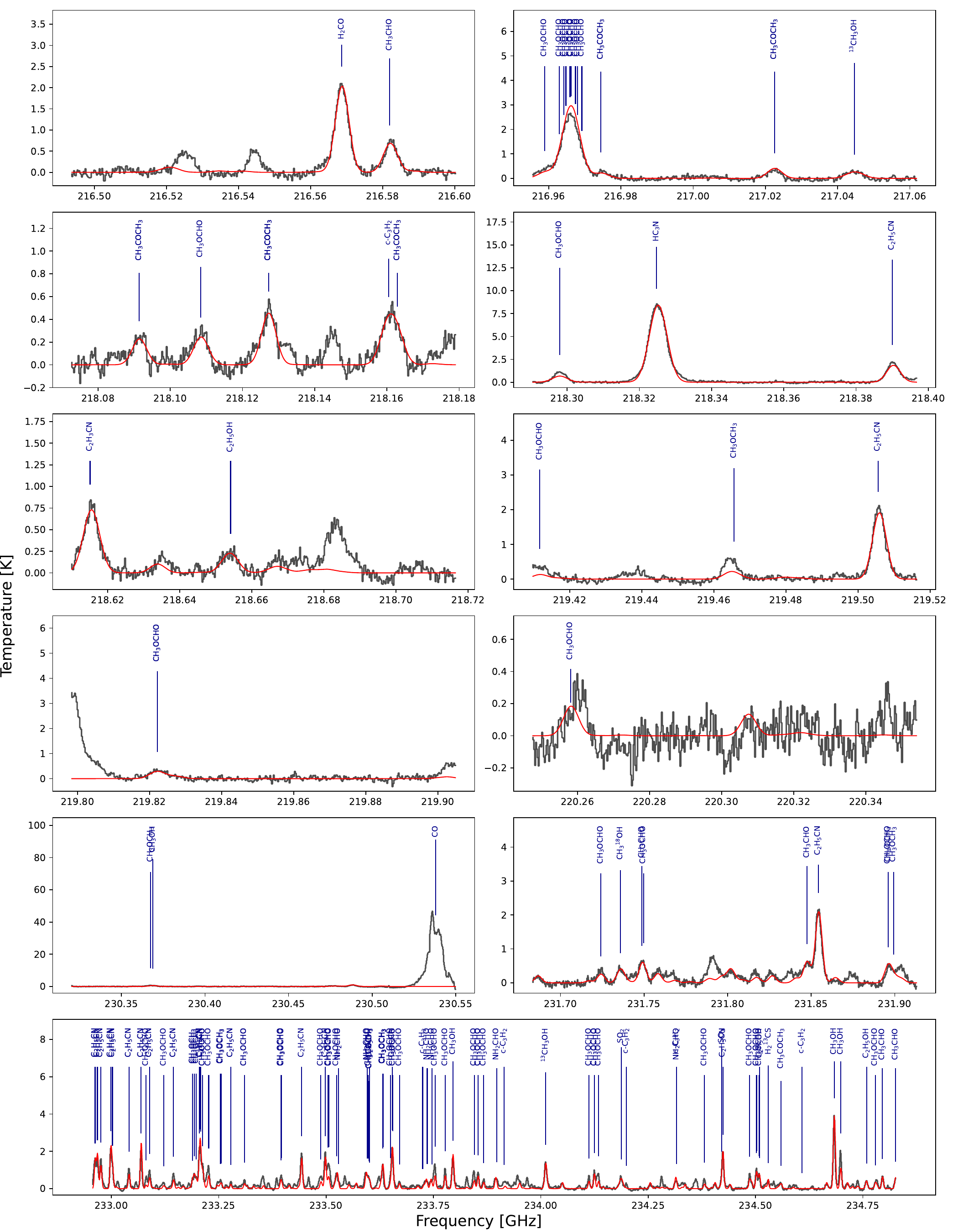}{Same for G305.20.}{fig:spectra12}
            \insertplot{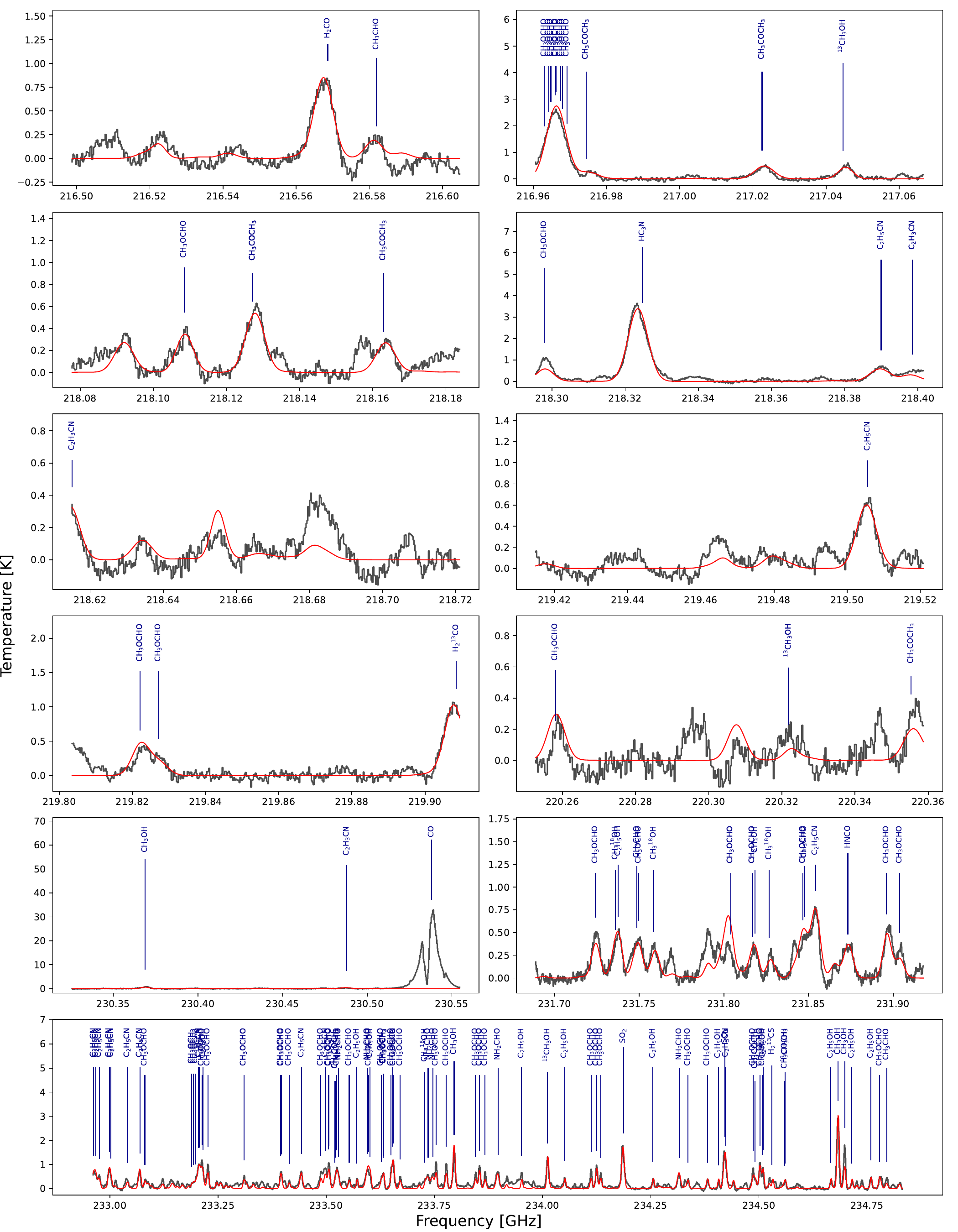}{Same for G305.80.}{fig:spectra13}
            \insertplot{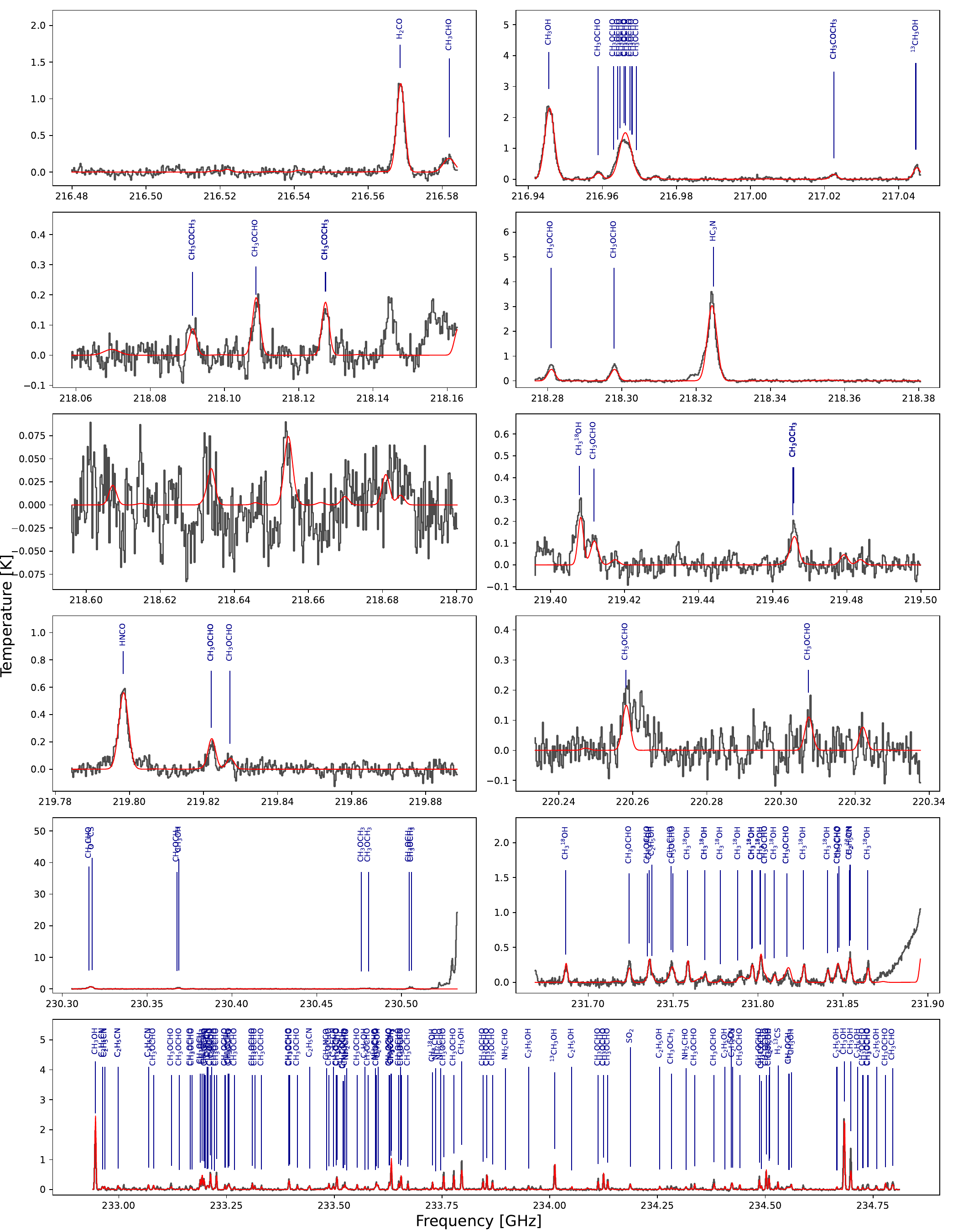}{Same for G309.92.}{fig:spectra14}
            \insertplot{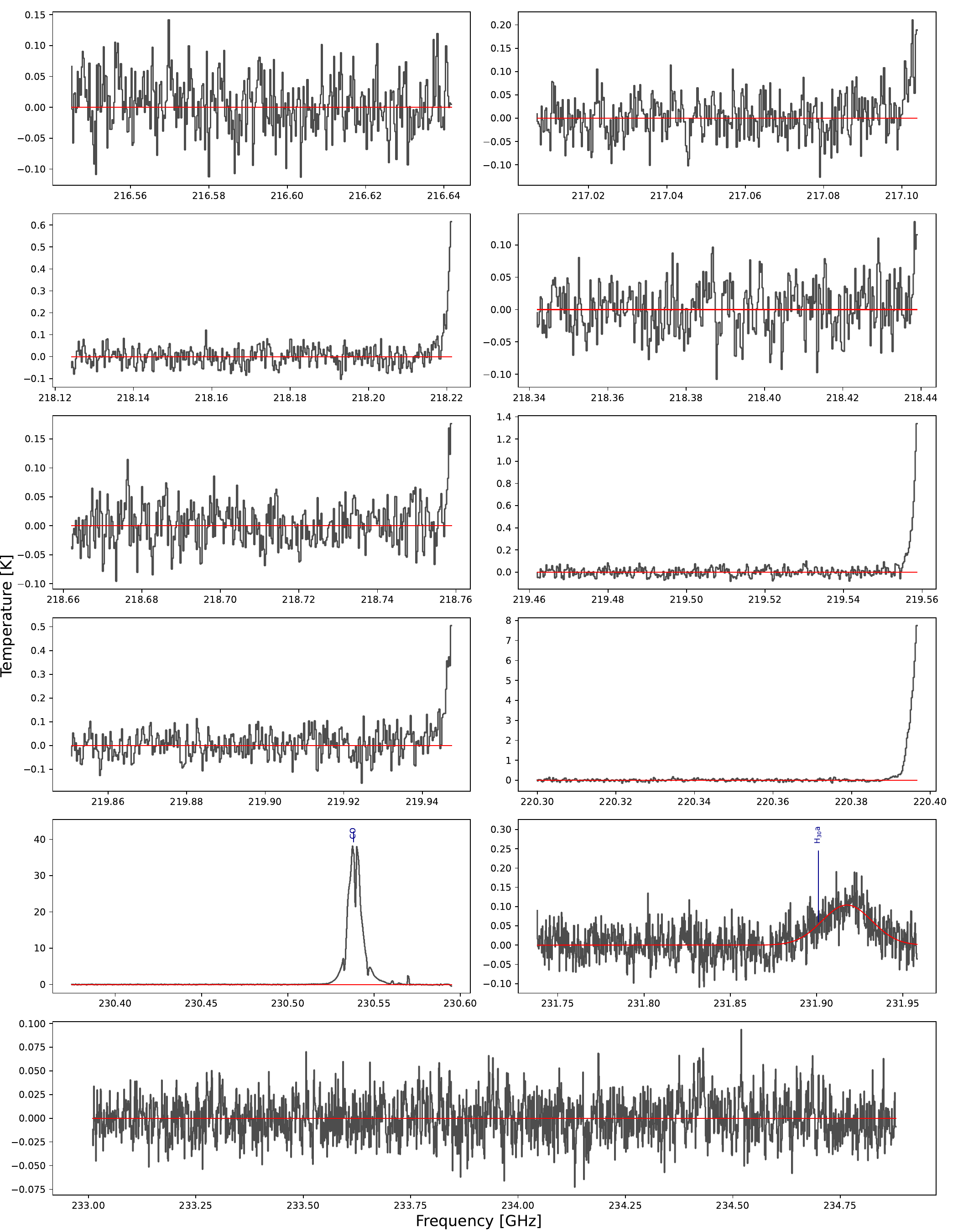}{Same for G317.40.}{fig:spectra15}
            \insertplot{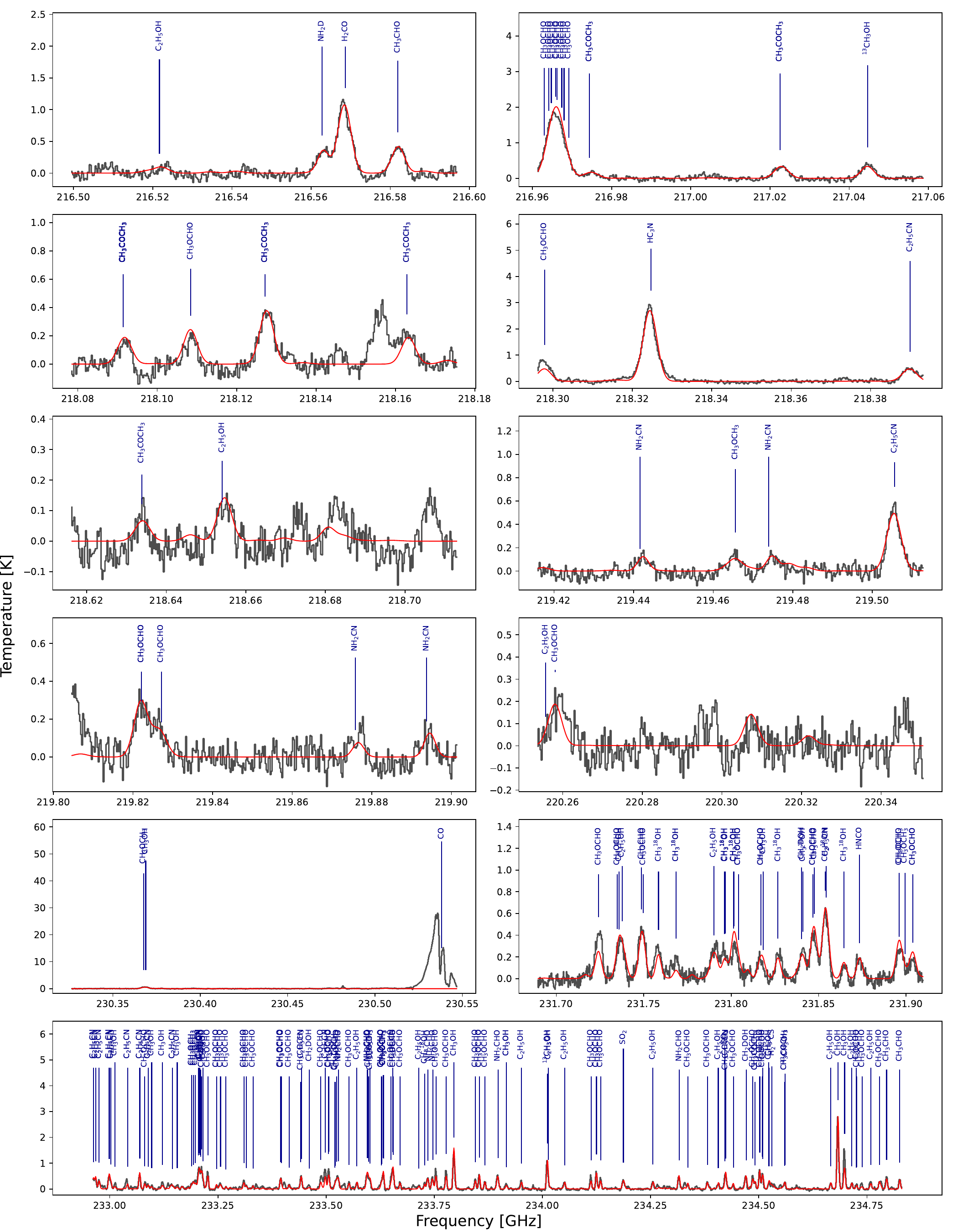}{Same for G318.95.}{fig:spectra16}
            \insertplot{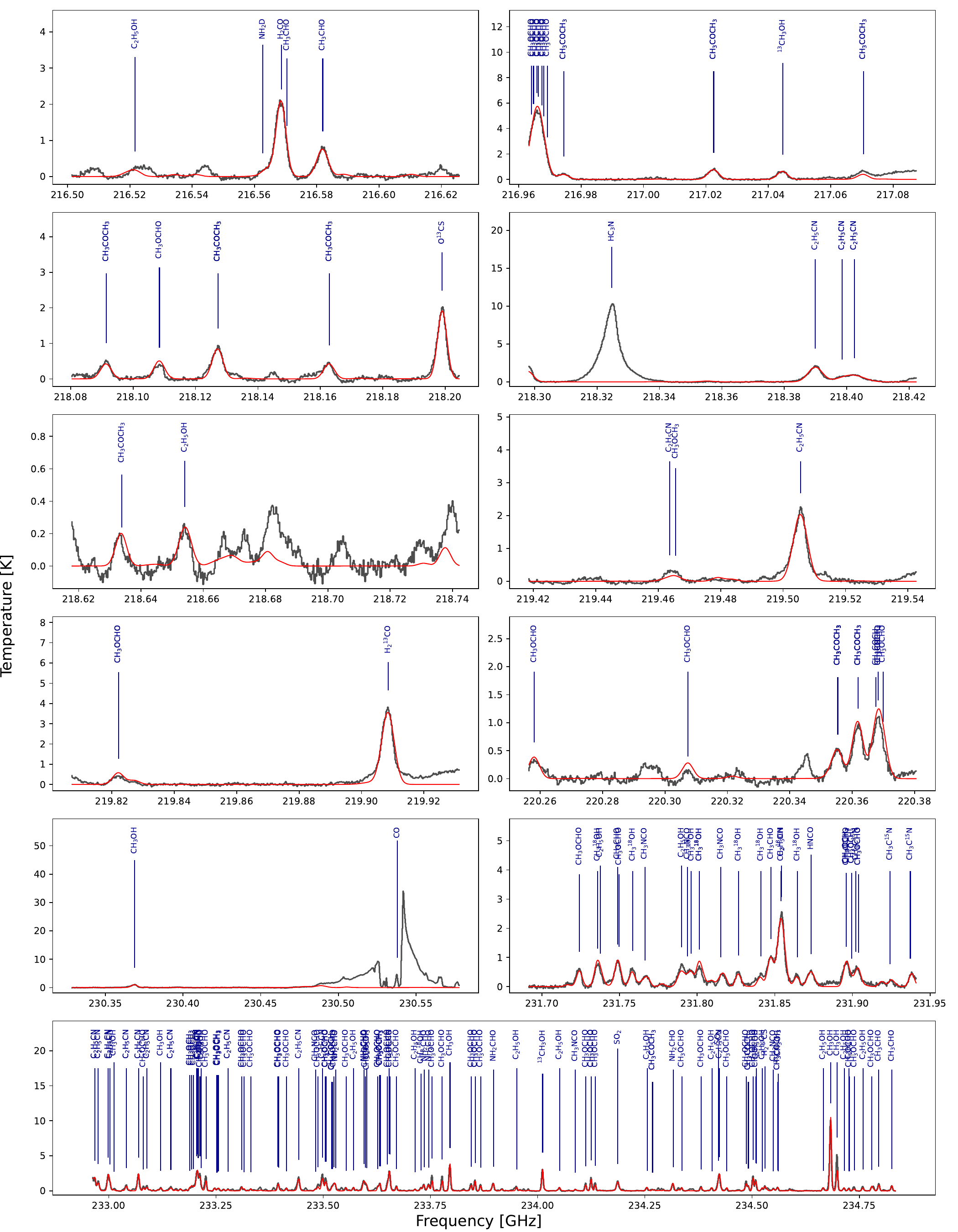}{Same for G337.40.}{fig:spectra17}
            \insertplot{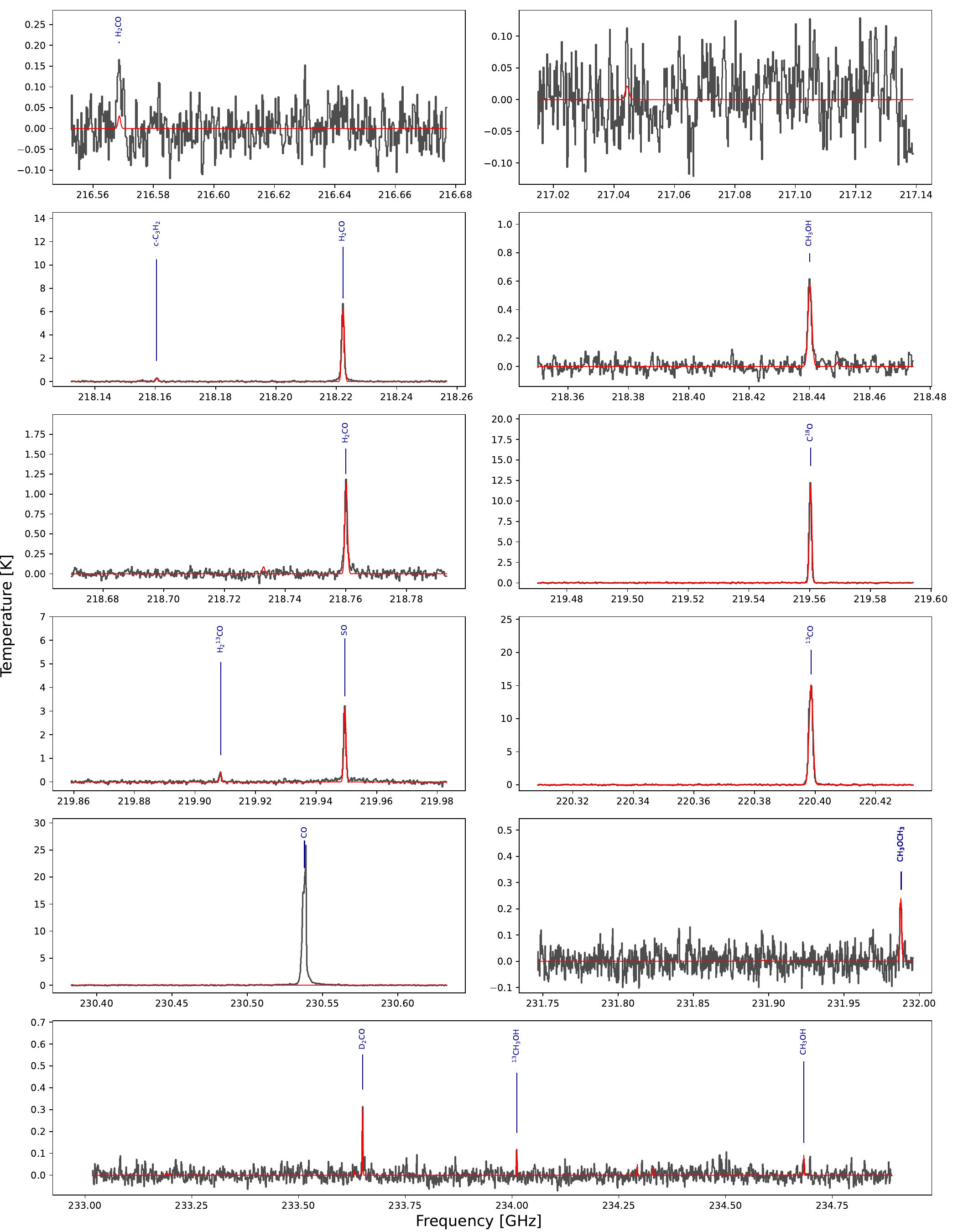}{Same for MMS-6.}{fig:spectra18}
            \insertplot{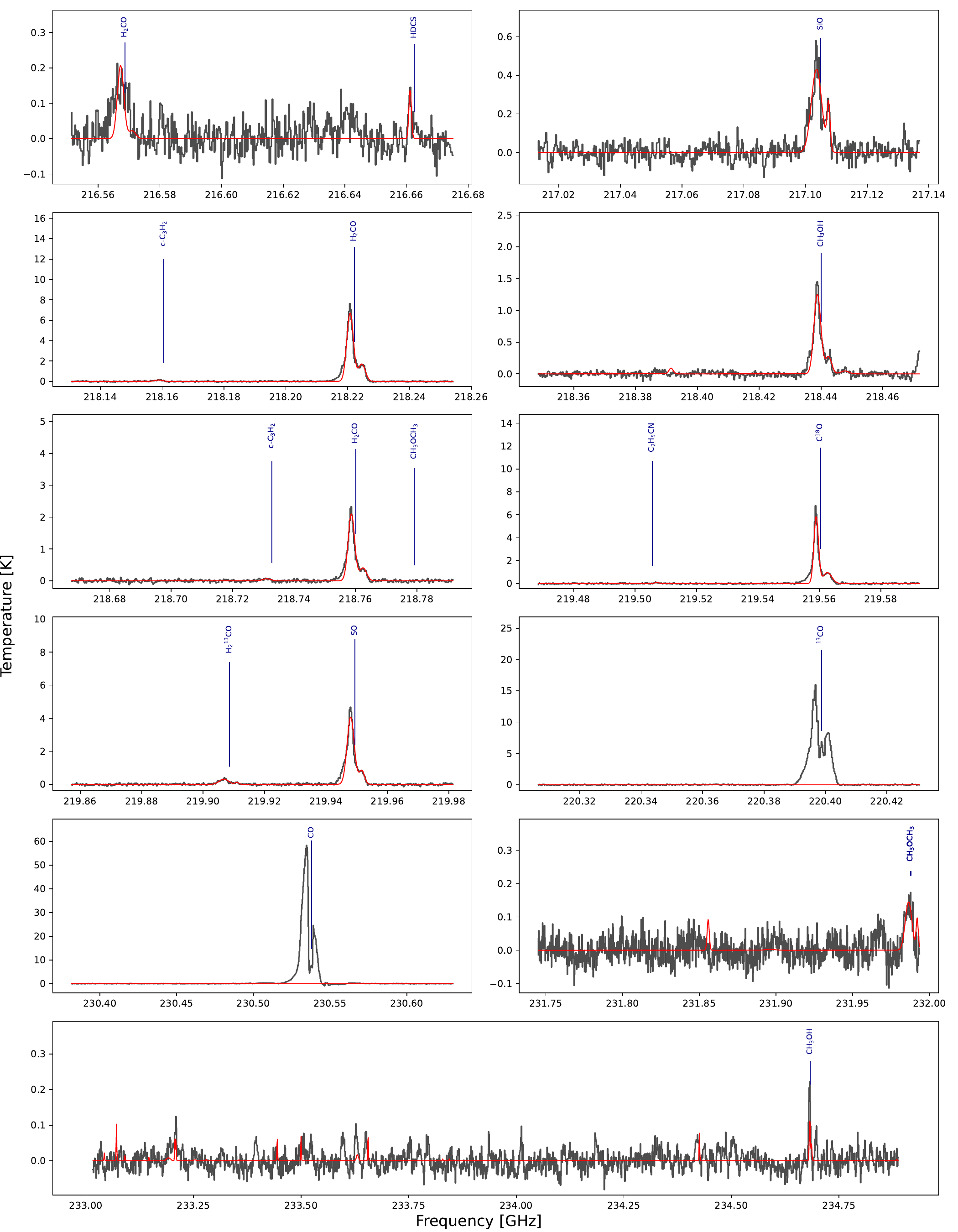}{Same for NGC 2071.}{fig:spectra19}
            \insertplot{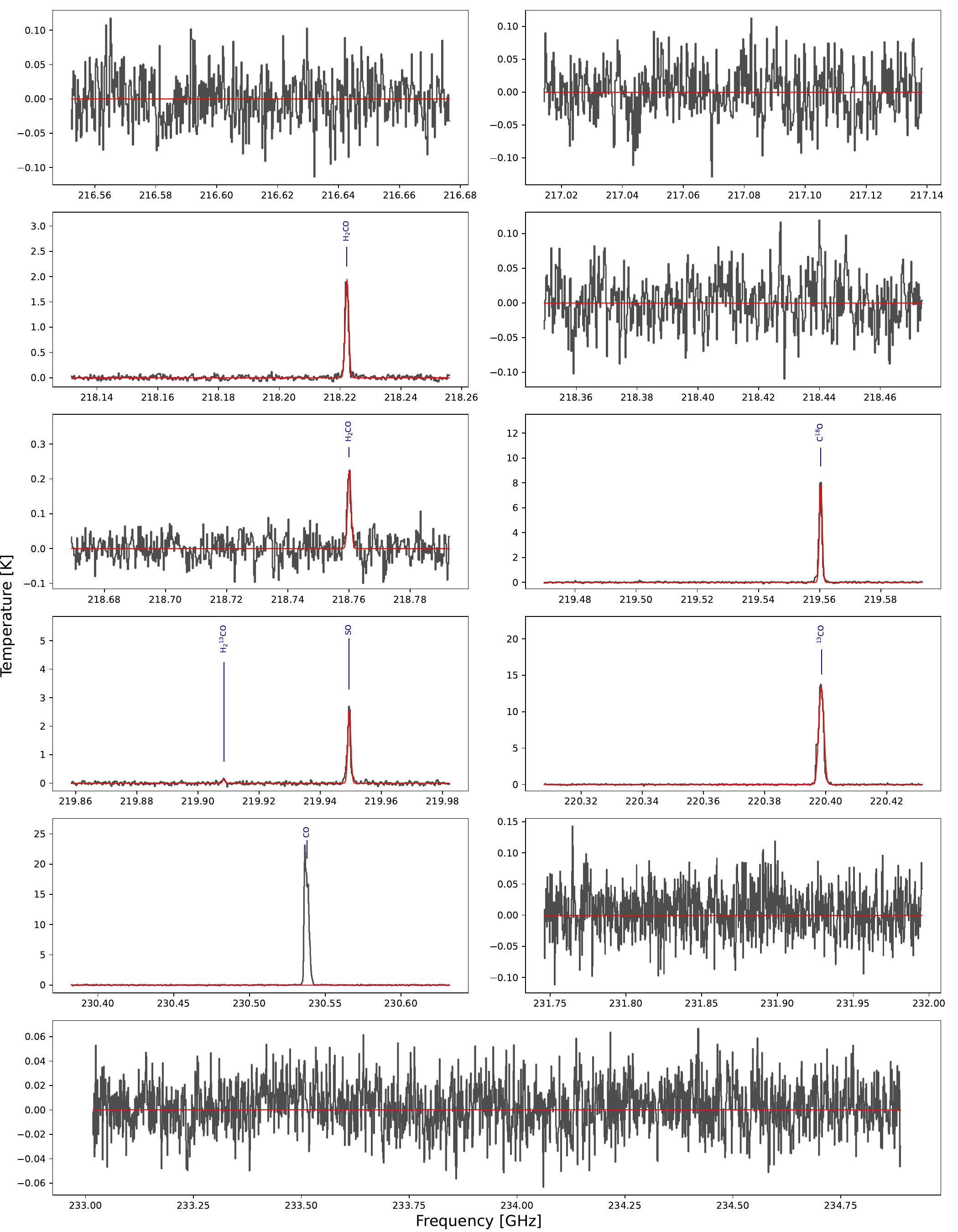}{Same for OMC-2.}{fig:spectra20}
            \insertplot{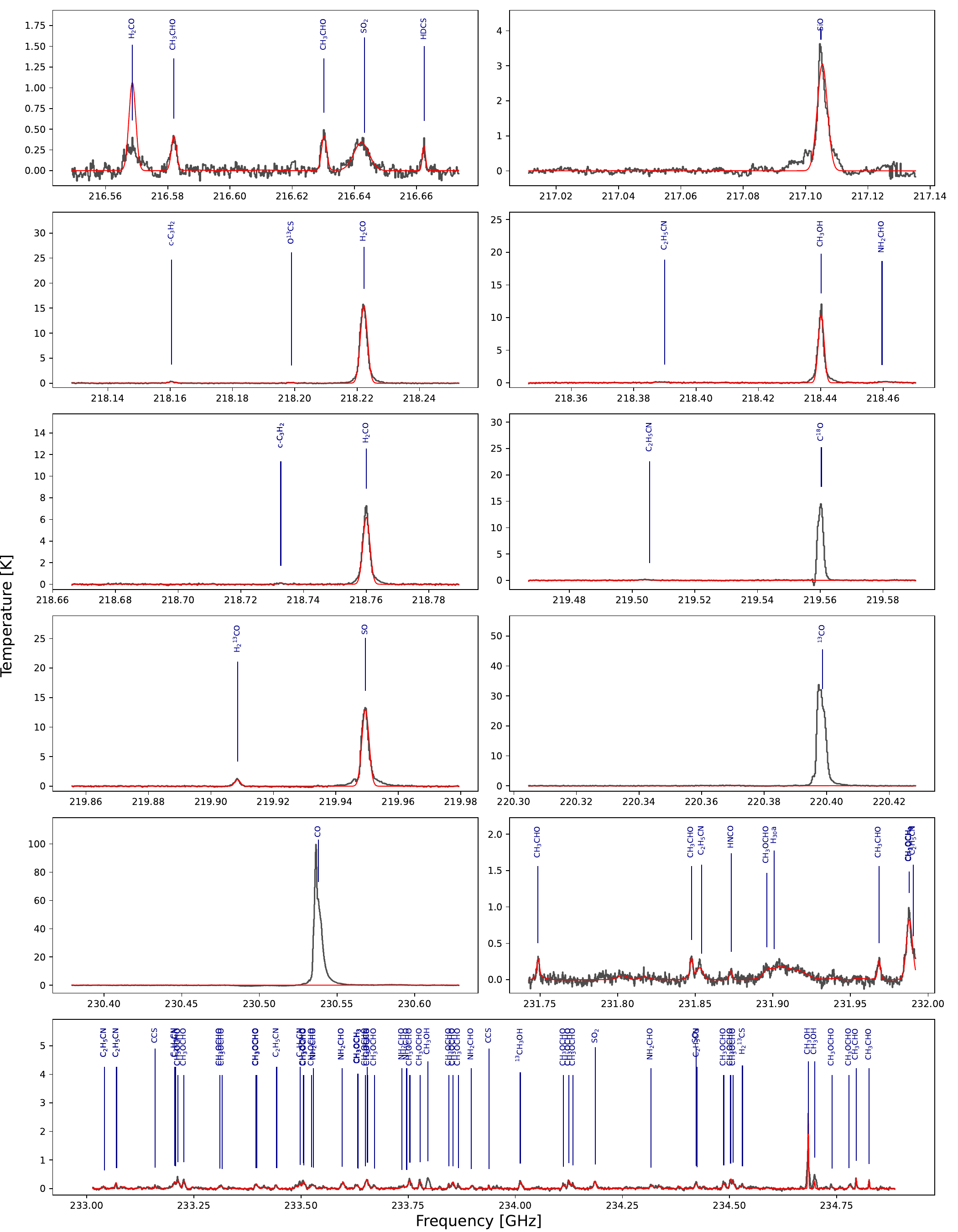}{Same for OMC1-S.}{fig:spectra21}
            \insertplot{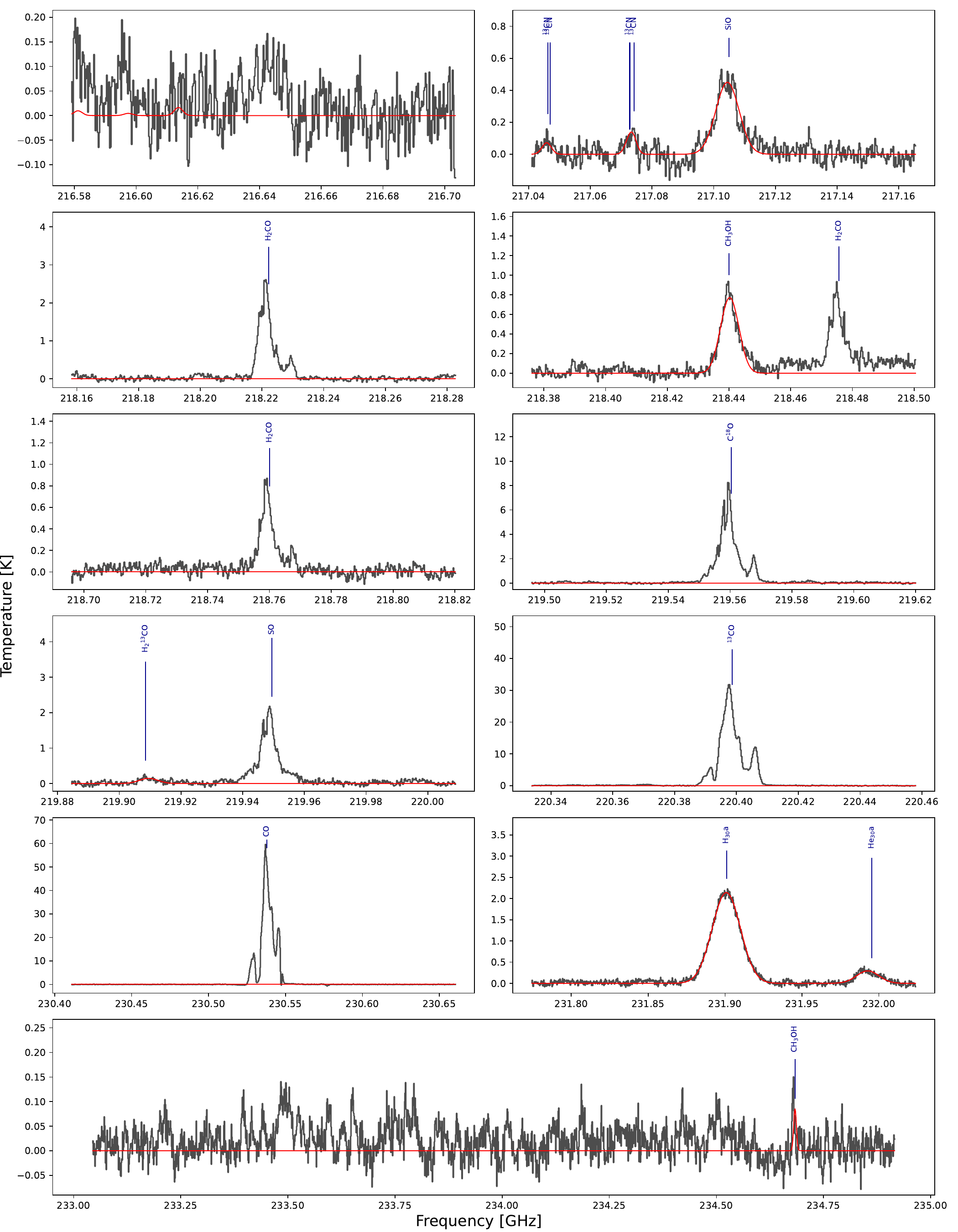}{Same for W51 e2.}{fig:spectra22}                    
    \end{appendix}
\end{document}